\documentclass[9pt,twocolumn,twoside]{opticajnl}
\journal{opticajournal} 

\setboolean{shortarticle}{false}

\usepackage{lineno}

\usepackage{graphicx}
\usepackage{tikz}
\usepackage{multirow}
\usepackage{pgfplots}
\usepackage{booktabs}
\usepackage{amssymb}
\usepackage{oplotsymbl}
\usepackage{amsmath}

\usepackage{rotating} 
\usepackage{tabularx} 
\usepackage{threeparttable} 

\title{Silicon-Integrated Next-Generation Plasmonic Devices for Energy-Efficient Semiconductor Applications}

\author[1]{Nasir Alfaraj}
\author[1,*]{Amr S. Helmy}

\affil[1]{The Edward S.\@ Rogers Sr.\@ Department of Electrical and Computer Engineering, University of Toronto, 10 King's College Road, Toronto, Ontario M5S 3G4, Canada.}

\affil[*]{a.helmy@utoronto.ca}

\begin{abstract}
Silicon-based integrated photonics has demonstrated significant advances in miniaturization and performance, yet critical challenges remain in achieving efficient on-chip communication at high bandwidths.\@ Plasmonic devices on silicon and silicon-on-insulator substrates offer a promising solution, enabling subwavelength light confinement and enhanced light-matter interactions through hybrid modes.\@ However, integrating traditional plasmonic materials like gold and silver into silicon-based platforms presents significant challenges, particularly due to their incompatibility with standard silicon processing techniques and their increased optical losses at longer wavelengths, which can hinder performance in near-infrared applications.\@ Plasmonic devices, leveraging advances in device architectures, have the potential to close these performance gaps and enable the next generation of high-speed, on-chip data communication.\@ This review explores recent progress in silicon‐integrated hybrid-mode plasmonic devices, highlighting the potential of transparent conductive oxides like indium tin oxide for low‐loss and tunable operation.\@ Key device topologies including coupled hybrid plasmonic waveguides and dielectric-based heterostructures are examined, along with fabrication techniques and practical considerations.\@ By critically comparing various plasmonic approaches and identifying their respective advantages and limitations, a path toward realizing the full potential of plasmonics in shaping the future of high-performance, silicon-based integrated photonics is charted.
\end{abstract}

\setboolean{displaycopyright}{false} 

\begin{document}
	
\maketitle

\renewcommand\thefootnote{}
\footnotetext{Preprint of a manuscript submitted to \emph{Advanced Materials Technologies} on February 24, 2025.\@ The version submitted to the journal is currently under review, and this preprint may not reflect the final peer-reviewed article.}

\section{Introduction}\label{intro}

The persistent pursuit of faster data transfer rates \cite{tossoun2024high,ning2024photonic}, higher bandwidth (BW) \cite{tasker2024bi}, and more efficient computation \cite{cheung2024energy,gosciniak2024schottky} is pushing the boundaries of existing electronic and photonic technologies, demanding novel solutions to meet these escalating needs \cite{pelucchi2022potential}.\@ Integrated photonics, specifically leveraging plasmonic devices compatible with silicon (Si) platforms, offers a promising avenue to address these challenges \cite{shekhar2024roadmapping}.\@ This field aims to combine the benefits of photonics---using light for information processing---with the advantages of Si-based device architectures, paving the way for the development of compact, energy-efficient, and high-performance devices integrated on a single chip \cite{bolognesi2023fully,li2023integrated,zanetto2023time}.\@ Plasmonics, the rapidly evolving field dedicated to manipulating light at subwavelength scales, provides a powerful tool for realizing this integration \cite{sharma2024past}.\@ However, achieving this vision while ensuring low power consumption and high performance in increasingly miniaturized devices, all within the constraints of established Si fabrication processes, presents a complex interplay of material properties, device design, and fabrication techniques \cite{alfaraj2023heteroepitaxial,guo2022multi,su2019record,lin2015dynamically}.

Si's prevalence in microelectronics, including integrated photonics, stems from its abundance, mature fabrication processes, and integrability with complementary metal--oxide--semiconductor (CMOS) technology \cite{yuvaraja2024three,pitters2024atomically}.\@ This allows for large-scale integration of diverse optical components on silicon-on-insulator (SOI) platforms, enabling complex photonic circuits.\@ SOI is particularly advantageous as it provides strong optical confinement within the Si waveguide layer due to the underlying insulating layer, minimizing losses and enabling denser integration.\@ While germanium (Ge) offers superior carrier mobilities, Si's cost-effectiveness and established infrastructure make it the preferred choice.\@ Alternative platforms like sapphire, while possessing attractive optical properties \cite{wang2024heterogeneous,alfaraj2022optical}, face challenges related to fabrication complexity and material quality \cite{alfaraj2020epitaxial,sun2017structural}.\@ Si's ability to be nanostructured using well-established techniques further solidifies its position as the leading material for integrated photonics.

This review explores the integration of plasmonics with Si platforms, examining the inherent trade-offs between on-chip area, power consumption, and device performance.\@ Performance is evaluated based on loss-confinement parameters such as extinction ratio (ER), insertion loss (IL), and propagation loss (PL).\@ We critically examine the limitations of traditional Si-based photonic devices and explore how plasmonic device architectures, particularly coupled hybrid plasmonic waveguides (CHPWs), can enhance device miniaturization, efficiency, and functionality.\@ Focusing on CHPWs, we analyze the footprint of various Si-integrated plasmonic-photonic components, including modulators, photodetectors (PDs), and passive waveguides.\@ These are compared to both traditional dielectric counterparts and plasmonic devices implemented with noble metals like gold (Au) and silver (Ag).\@ We then investigate the optoelectronic characteristics of these devices, addressing the inherent energy losses in plasmonic systems and exploring potential solutions through materials and design optimization.\@ Finally, key performance metrics, such as modulation depth (MD), BW, and responsivity (\emph{R}$_\mathrm{ph}$), are evaluated and compared across the different device types.\@ Strategies for overcoming limitations like PL levels and material dispersion are also discussed.

Recent advancements in hybrid mode integration, particularly leveraging CHPWs, are explored to address these challenges and further enhance device performance.\@ The development of hybrid modes in plasmonics \cite{li2023narrow,oulton2008hybrid} integrated with Si is particularly promising, as these modes can combine the benefits of both plasmonic and dielectric waveguides, offering a pathway to achieve both strong confinement and low losses.\@ By examining the current state-of-the-art and highlighting future directions, this review aims to provide a comprehensive overview of the challenges and opportunities in plasmonics integrated with Si photonics, paving the way for developing next-generation integrated photonic devices.

\subsection{The drive for miniaturization}

\begin{figure*}[h]
	\centering
	\includegraphics[width=0.85\linewidth]{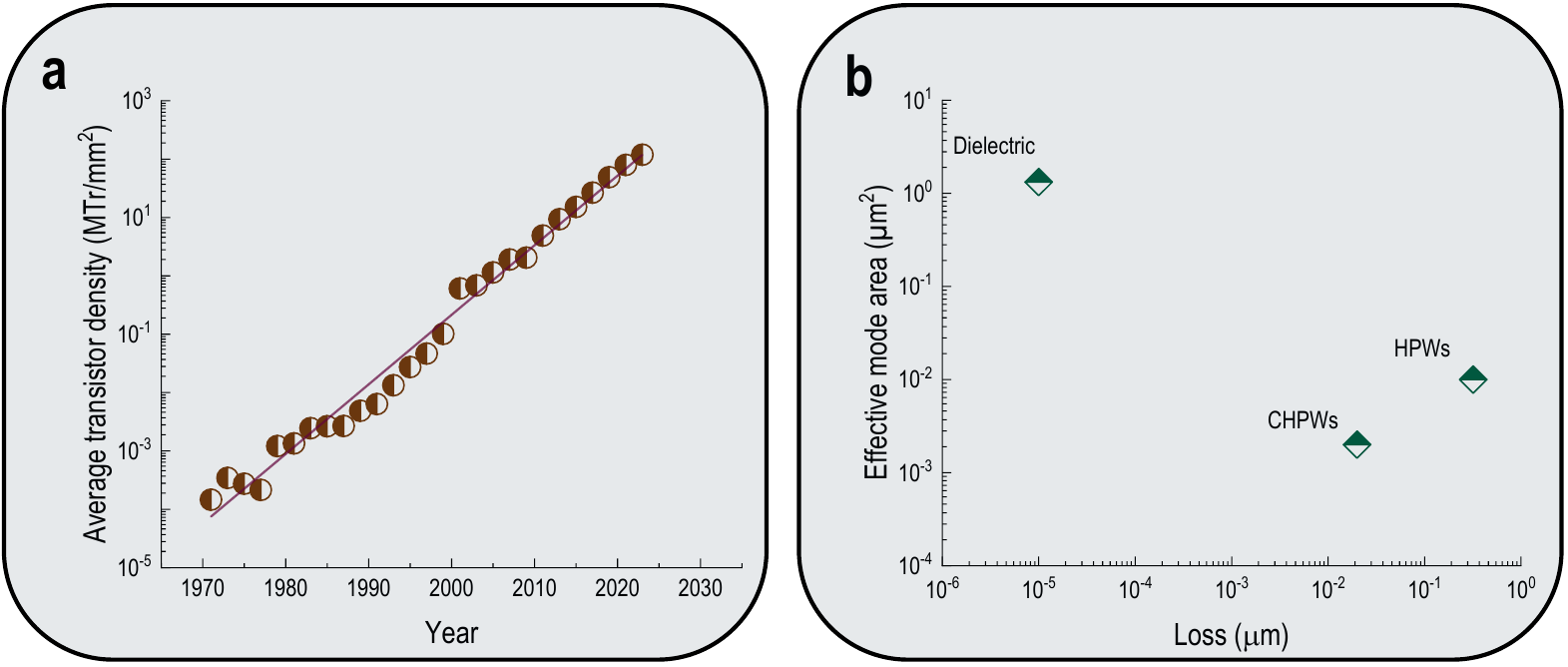}
	\caption{Moore's Law and the potential of plasmonics for increased optical component density.\@ \textbf{a}, The manifestation of Moore's Law.\@ This scatter plot charts the exponential growth in transistor density over time.\@ The red line represents an exponential fit to the data.\@ The plot showcases the average transistor density achieved by industry leaders such as Intel, AMD, Apple, and Nvidia.\@ The data, sourced from publicly available microchip datasheets and technology reports, underscores the remarkable exponential increase in transistor density (measured in mega-transistors per square millimeter) over the years.\@ \textbf{b}, Subwavelength confinement and increased optical component density.\@ This diagram contrasts the mode size of traditional dielectric waveguides based on TIR with that of select plasmonic waveguides (HPWs and CHPWs), highlighting the dramatic size reduction enabled by plasmonics.\@ In traditional dielectric waveguides, the mode size is limited by the diffraction limit of light.\@ Plasmonic waveguides suffer from higher propagation losses due to the absorption of light by the metal.\@ Dielectric waveguides generally have lower losses and can guide light over longer distances (loss-confinement trade-off).}
	\label{miniaturization}
\end{figure*}

As the microelectronics industry grapples with the limitations of Moore's Law, which is an empirical observation that has historically predicted the doubling of transistor density on a chip approximately every two years, the field of photonics is exploring new avenues for miniaturization and performance enhancement \cite{taha2024exploring,burg2021moore,khan2018science}.\@ Plasmonics, with its ability to confine and manipulate light at the nanoscale, has emerged as a key enabler for achieving these goals \cite{burla2019500}.\@ While Moore's original 1965 article projected this trend for the following decade, it has surprisingly held for far longer, driving advancements in computing power for several decades \cite{moore1997microprocessor,moore1965cramming}.\@ This trend is illustrated in \figurename{ \ref{miniaturization}}, which shows the exponential increase in both transistor count per chip and areal density over time.\@ However, in the first two decades of the 21st century, as the 22 nm node milestone approached, researchers began to question the sustainability of this exponential growth due to physical limitations and escalating costs \cite{liao2023integrated,cavin2012science,thompson2006moore}.\@ The resulting uncertainties about the future of Moore's Law have sparked interest in alternative paradigms for continued technological progress.\@ The challenges encountered in further miniaturizing electronic components have, in a way, inspired the formulation of a ``Moore's Law for Photonics'' concept, suggesting a similar trend of increasing integration density for photonic components on a chip.\@ This pursuit has driven extensive research into subwavelength optical structures and novel materials that can overcome the diffraction limit, ultimately enabling denser and more powerful photonic integrated circuits (PICs) \cite{su2019record,naraine2023subwavelength,meng2021optical,alfaraj2019deep}.

The exponential fit to the transistor density in \figurename{ \ref{miniaturization}\textbf{a}} data reveals a remarkable adherence to Moore's Law, with an average two-year increase of 97.13\%, nearing the predicted doubling.\@ This trend is further supported by the observation that out of 51 two-year periods, the transistor density doubled or more in 17 instances, while it decreased in only 4.\@ The graph and the underlying data underscore the rapid pace of technological advancement in the semiconductor industry, driving the continuous improvement in computing power and capabilities.\@ However, it is crucial to acknowledge that this exponential growth may not be sustainable indefinitely due to physical limitations and economic considerations.\@ Future innovations may require novel approaches to maintain this trajectory of progress.\@ This is where device engineers play a crucial role.\@ By exploring innovative architectures like CHPWs, optics engineers can push the boundaries of miniaturization, enabling the development of more compact and efficient photonic devices.\@ \figurename{ \ref{miniaturization}\textbf{b}} illustrates the impact of these architectures on propagation losses and optical mode areas.

\subsection{Si integration challenges in plasmonics}

While plasmonic devices offer significant advantages for miniaturization in integrated photonics, their integration with Si platforms presents several challenges.\@ To address the limitations of traditional electronics, one particularly promising avenue for achieving miniaturization in integrated photonics lies in harnessing plasmonic phenomena, which allow for the confinement and manipulation of light at scales below the diffraction limit \cite{ozbay2006plasmonics}.\@ However, integrating traditional plasmonic materials like Au and Ag directly into Si-based fabrication processes presents significant challenges.\@ While Au and Ag excel at subwavelength confinement due to their high DC conductivity and resulting low ohmic losses \cite{naik2013alternative}, they are not readily compatible with standard Si processing techniques \cite{cheng2023plasmonic}.\@ Their incompatibility stems from several factors, including contamination, electromigration, and adhesion issues.\@ For instance, the high temperatures involved in Si device fabrication can cause Au and Ag atoms to diffuse into Si, acting as deep-level traps and recombination centers that disrupt its carefully engineered electronic properties and degrade chip performance \cite{ocana2022strain,sang2022ultrahigh,lewis2003stability,fahey1989point}.\@ Additionally, the high current densities common in integrated circuits (ICs) can lead to electromigration in Au and Ag interconnects, causing voids and hillocks that compromise device reliability \cite{waliullah2024electromigration,ceric2023statistical,d1971electromigration,berenbaum1969surface}.\@ Moreover, the poor adhesion of Au and Ag to Si and silicon dioxide (SiO$_\mathrm{2}$) \cite{le2019postdeposition,benjamin1960adhesion}, a prevalent dielectric in Si devices, further complicates their integration.

These challenges increase fabrication complexity and cost, limiting the scalability of plasmonic devices within the Si photonics ecosystem.\@ Furthermore, Au and Ag exhibit significant losses at optical frequencies, particularly in the near-infrared (NIR, approximately 155 THz and above) due to interband transitions \cite{west2010searching,johnson1972optical}.\@ These losses can severely limit the performance and efficiency of plasmonic devices for optical communication and other applications.\@ While integrating Au and Ag at later fabrication stages remains a possibility, albeit challenging \cite{li2024enhancing,lu2017through}, the search for alternative plasmonic materials compatible with standard CMOS technology and Si foundry processes has become crucial in integrated photonics.

Promising avenues to overcome these challenges include exploring alternative plasmonic materials such as transparent conductive oxides (TCOs) \cite{gosciniak2023transparent} or developing hybrid-mode plasmonic structures that combine the benefits of both plasmonic and dielectric materials \cite{ma2014asymmetric}.\@ These approaches aim to achieve a balance between the desirable optical properties of plasmonic materials and the compatibility and manufacturability requirements of Si-based platforms.

\section{CMOS Technology and Advanced Logic Foundries}

CMOS technology, the cornerstone of modern ICs, relies on Si as the core semiconductor material.\@ Advanced logic foundries specialize in fabricating these CMOS-based ICs, often pushing the boundaries of Si technology to manufacture complex chips like central processing units (CPUs), graphics processing units (GPUs), and system-on-chip (SoC) ICs \cite{shalf2020future}.\@ These foundries excel in processes related to creating transistors with complementary p-type and n-type semiconductors for low-power Boolean logic functions, forming the foundation of CMOS logic \cite{park2016neuromorphic}.\@ Furthermore, these foundries often work with materials that integrate well with Si processes, such as silicon germanium (SiGe) \cite{goley2021zero} used for high-speed transistors in applications like wireless communication, and silicon carbide (SiC) \cite{nielsen2024high,rhee2024one} and gallium nitride (GaN) \cite{lu2024hybrid,udabe2023gallium,alfaraj2017thermodynamic}, which are wide-bandgap semiconductors used for power electronics and high-frequency applications.

The inherent compatibility of Si with various materials has driven the mass production of ICs, with foundries adapting to incorporate new materials and processes to support advancements in CMOS design \cite{sell2022intel,hill2021cmos}.\@ However, the integration of plasmonic components presents unique challenges and opportunities.\@ Traditional plasmonic materials, such as Au and Ag, face compatibility issues with Si processing, particularly with the high temperatures involved in CMOS front-end-of-line (FEOL) processes.\@ This limitation has spurred research into alternative plasmonic materials that can seamlessly integrate into existing Si fabrication workflows.\@ These materials must not only exhibit excellent optical properties for subwavelength light manipulation but also align with the manufacturing infrastructure of Si technology, ensuring cost-effectiveness and scalability \cite{lin2020monolithic,alfaraj2015functional,babicheva2013towards}.

A prime example of this adaptation is the transition from SiO$_\mathrm{2}$ to hafnium oxide (HfO$_\mathrm{2}$) as the gate dielectric.\@ HfO$_\mathrm{2}$'s higher dielectric constant enables the use of a physically thinner layer while maintaining the same effective oxide thickness (EOT) as a thicker SiO$_\mathrm{2}$ layer.\@ This thinner physical layer reduces leakage current due to quantum tunneling, a significant issue with ultra-thin SiO$_\mathrm{2}$ layers, while still providing the required capacitance \cite{noheda2023lessons}.\@ Similarly, the evolution of interconnect materials from aluminum (Al) to copper (Cu), along with the use of metals like tungsten (W), titanium (Ti), and cobalt (Co) for specific purposes, demonstrates the flexibility of Si foundries in accommodating evolving technological needs.\@ Titanium nitride (TiN), a promising refractory material \cite{gadalla2020imaging,alfaraj2019deep_}, further exemplifies this versatility, serving as both a diffusion barrier and metal gate electrode \cite{briggs2016fully}.

Successful Si device fabrication hinges on close collaboration between design and manufacturing \cite{reynolds2023single,han2023advances,ghoneim2015enhanced}.\@ Advanced logic foundries provide design rules and process parameters that must be strictly followed to ensure compatibility and reliability \cite{jamil2023reliability,quader1994hot,lee1992design}.\@ CMOS fabrication involves an even more complex sequence of steps, including lithography, etching, deposition, and ion implantation, all meticulously optimized for high-volume production \cite{hasan2018promising,cheng2017ifinfet}.\@ However, variations in foundry processes can impact device performance \cite{yuan2022100,vincent2020process}.\@ Therefore, CMOS designs incorporate techniques to mitigate these variations and ensure consistency across different foundries.\@ The continuous evolution of CMOS technology, with newer nodes offering smaller feature sizes and enhanced performance, presents ongoing challenges.\@ Advanced logic foundries may specialize in specific technology nodes, requiring CMOS designs to target compatible facilities.\@ CMOS designs, often incorporating specialized processes or non-standard materials, may further limit foundry options.\@ Ultimately, selecting a foundry for CMOS fabrication requires careful consideration of factors such as technology node compatibility, process complexity, foundry capabilities, cost, and production capacity.\@ \tablename{ \ref{CMOS_SI}} summarizes the key aspects of advanced logic foundries and their role in Si-based CMOS device fabrication.

\begin{table*}[h]
	\centering
	\caption{Aspects of advanced logic foundries and their role in Si-based CMOS device fabrication.\label{CMOS_SI}}
    \renewcommand{\arraystretch}{1} 
	\begin{tabular*}{\textwidth}{@{\extracolsep\fill}l >{\raggedright\arraybackslash}m{0.6\textwidth} @{\extracolsep\fill}}
		\toprule
		\textbf{Aspect} & \textbf{Description}\\
		\midrule
		\addlinespace[0.8em]
		Focus & Fabrication of advanced CMOS-based ICs, pushing the limits of Si technology to manufacture complex chips like CPUs, GPUs, and SoCs \cite{shalf2020future}.\\
		\addlinespace[0.8em]
		Core competency &  Processes related to creating transistors with complementary p-type and n-type semiconductors for low-power, high-performance logic functions \cite{park2016neuromorphic}.\\
		\addlinespace[0.8em]
		Materials & Utilize Si as the core material, but also work with Si-compatible materials like SiGe, SiC, and GaN to enhance device performance and functionality \cite{iucolano2019gan,kazior2014beyond}.\\
		\addlinespace[0.8em]
		Manufacturing expertise & Possess advanced fabrication facilities and expertise in processes like lithography, etching, deposition, and ion implantation, all optimized for high-volume production of complex chips \cite{hasan2018promising,cheng2017ifinfet}.\\
		\addlinespace[0.8em]
		Design and manufacturing collaboration & Provide design rules and process parameters to ensure compatibility between CMOS designs and their manufacturing processes \cite{jamil2023reliability,quader1994hot,lee1992design}.\\
		\bottomrule
	\end{tabular*}
\end{table*}

The exploration of Si-compatible plasmonic materials has gained significant momentum, with TCOs like indium tin oxide (ITO).\@ As shown in \tablename{ \ref{optical_properties}}, these materials offer low optical losses, tunable optical properties, and compatibility with CMOS back-end-of-line (BEOL) processes \cite{jaffray2024nonlinear,kinsey2019near}, making them particularly suitable for on-chip optical communication.\@ ITO, in particular, exhibits epsilon-near-zero (ENZ) behavior at telecommunication wavelengths, facilitating enhanced light-matter interactions and enabling subwavelength light confinement crucial for miniaturized photonic devices \cite{alfaraj2023facile,li2023invertible}.\@ Furthermore, its lower carrier concentration compared to traditional plasmonic metals like Au and Ag translates to reduced optical losses in the NIR regime, a critical advantage for Si-based plasmonic devices.

\begin{table*}[h]
	\centering
	\caption{Optoelectronic properties of selected plasmonic materials at different wavelengths (visible: 0.532 $\mu$m, NIR: 1.55 $\mu$m).\@ The table presents approximate values of the optical properties for various plasmonic materials at these wavelengths.\@ Actual values may vary depending on factors such as synthesis method, film thickness, and the underlying substrate.\label{optical_properties}}
    \renewcommand{\arraystretch}{1} 
	\begin{tabular*}{\textwidth}{@{\extracolsep\fill}lllllll@{\extracolsep\fill}}
		\toprule
		\textbf{Material} & \textbf{$\boldsymbol{\lambda}$ ($\mu$m)} & \textbf{$\boldsymbol{\varepsilon'}$} & \textbf{$\boldsymbol{\varepsilon''}$} & \textbf{$\boldsymbol{\ell}$ ($\mu$m)$^\bot$} & \textbf{FOM}$^\bot$ & \textbf{CMOS integrability} \\
		\midrule
		\multirow{2}{*}[0.5em]{Au \cite{mcpeak2015plasmonic}} 
		& 0.532 & $-$5.23 & 1.98 & 1.17 & 2.65 & \multirow{2}{*}[0.5em]{Potentially BEOL$^\dagger$} \\
		& 1.55 & $-$127 & 5.37 & 739 & 23.6 & \\[0.5em]
		\multirow{2}{*}[0.5em]{Ag \cite{mcpeak2015plasmonic}} 
		& 0.532 & $-$11.9 & 0.29 & 41 & 40.6 & \multirow{2}{*}[0.5em]{Potentially BEOL$^\dagger$} \\
		& 1.55 & $-$134 & 3.62 & 1219 & 37 & \\[0.5em]
		\multirow{2}{*}[0.5em]{ITO \cite{minenkov2024monitoring}} 
		& 0.532 & 3.67 & 0.03 & 43 & 138 & \multirow{2}{*}[0.5em]{MEOL, BEOL} \\
		& 1.55 & $-$2 & 0.64 & 1.53 & 3.11 & \\[0.5em]
		\multirow{2}{*}[0.5em]{Al \cite{mcpeak2015plasmonic}} 
		& 0.532 & $-$31.5 & 8.24 & 10.2 & 3.83 & \multirow{2}{*}[0.5em]{FEOL, MEOL, BEOL} \\
		& 1.55 & $-$198 & 38.1 & 254 & 5.2 & \\[0.5em]
		\multirow{2}{*}[0.5em]{TiN \cite{beliaev2023optical}} 
		& 0.532 & $-$1.61 & 3.32 & 0.07 & 0.48 & \multirow{2}{*}[0.5em]{FEOL, MEOL, BEOL} \\
		& 1.55 & $-$59.3 & 34.8 & 25 & 1.71 & \\[0.5em]
		\multirow{2}{*}[0.5em]{Cu \cite{mcpeak2015plasmonic}} 
		& 0.532 & $-$5.86 & 4.92 & 0.59 & 1.19 & \multirow{2}{*}[0.5em]{MEOL, BEOL$^\ddagger$} \\
		& 1.55 & $-$119 & 4.24 & 826 & 28.1 & \\[0.5em]
		\multirow{2}{*}[0.5em]{Graphene$^\S$ \cite{song2018broadband}} 
		& 0.532 & 5.61 & 7.27 & 0.37 & 0.77 & \multirow{2}{*}[0.5em]{Potentially MEOL \& BEOL$^\S$} \\
		& 1.55 & 4.75 & 15.8 & 0.35 & 0.3 & \\[0.5em]
		\multirow{2}{*}[0.5em]{MoS$_\mathrm{2}$$^\S$ \cite{islam2021plane}} 
		& 0.532 & 8.7 & 4.43 & 1.45 & 1.97 & \multirow{2}{*}[0.5em]{Potentially MEOL \& BEOL$^\S$} \\
		& 1.55 & 7.85 & 0.61 & 24.9 & 12.9 & \\[0em]
		\bottomrule
	\end{tabular*}
	\begin{tablenotes}
		\item $^\bot$The propagation length $\ell_\mathrm{p}$ is calculated using equation \ref{prop_length}.\@ The Figure of Merit (FOM) is calculated as the ratio of the absolute value of the real permittivity to the imaginary permittivity, $\mathrm{FOM} = \frac{|\varepsilon'|}{\varepsilon''}$.
		\item $^\dagger$Potential integration, although challenging due to diffusion and high-temperature incompatibility.
		\item $^\ddagger$Special barrier layers and processes are employed to prevent Cu diffusion into the underlying FEOL structures \cite{kuo2023mos2}.
		\item $^\S$CMOS integrability of 2D materials is an active area of research and depends heavily on the specific integration approach and fabrication process.\@ While they show promise for MEOL and BEOL integration, their compatibility with FEOL processes remains infeasible due to potential high-temperature degradation (MoS$_\mathrm{2}$) \cite{zhu2023low,lemme2023low} and the need for careful control of growth and transfer processes (graphene) \cite{elbanna20232d}.
	\end{tablenotes}
\end{table*}

\tablename{ \ref{optical_properties}} highlights the optical properties and CMOS compatibility of various plasmonic materials.\@ While noble metals like Au and Ag exhibit superior performance in the visible and NIR ranges \cite{steinberger2006dielectric}, their integration, particularly within FEOL processes, remains challenging due to potential diffusion and high-temperature incompatibility.\@ ITO, with its lower optical losses and CMOS BEOL compatibility, presents a compelling alternative, especially for applications in the NIR regime.\@ Beyond TCOs, research is exploring hybrid plasmonic materials, combining metals and dielectrics to tailor optical responses, and two-dimensional (2D) materials like graphene and molybdenum disulfide (MoS$_\mathrm{2}$), which offer unique electronic and optical properties \cite{yoshioka2024chip,zampa2024strong,lan2024enhanced,wen2022pathways}.\@ However, achieving full compatibility of 2D materials with CMOS processes, especially FEOL, is an ongoing research effort due to potential high-temperature degradation (MoS$_\mathrm{2}$) \cite{zhu2023low,lemme2023low} and the need for careful control of growth and transfer processes (graphene) \cite{elbanna20232d}.

While standard Si foundries may excel at CMOS fabrication, the unique demands of integrating optical functionality necessitate specialized processes often found in PIC foundries.\@ These foundries cater to the specific material and fabrication requirements of PICs, enabling the integration of optical components alongside Si electronics.\@ Examples include AIM Photonics (US), a public-private partnership focused on advancing the manufacturing of PICs \cite{nierenberg2023manufacturability,mcdonough2023aim}, and LioniX International (Netherlands) \cite{nikbakht2023asymmetric}, focusing on silicon nitride (Si$_\mathrm{3}$N$_\mathrm{4}$ or SiN) platforms.\@ While CMOS compatibility remains crucial for many applications, the broader context of Si integration encompasses a wider range of materials and processes, highlighting the importance of both standard Si foundries and specialized PIC foundries for advancing hybrid photonic devices.

\section{SPP Guided Waves}\label{SPP_guided_waves}

Before exploring the specifics of small-footprint plasmonic waveguides compatible with Si processing, it is crucial to establish a foundational understanding of surface plasmon polaritons (SPPs) and the inherent trade-offs associated with their utilization.\@ SPPs are evanescent electromagnetic (EM) waves that propagate along the interface between a metal and a dielectric.\@ The term ``surface plasmon polariton'' reflects the hybrid nature of these waves, combining the collective oscillations of free electrons at the metal surface (surface plasmons) with the EM fields extending into the dielectric (polaritons).\@ They are characterized by several key parameters:\@ their effective wavelength, which is shorter than that of light in free space or the dielectric, leading to strong field confinement under their resulting higher effective refractive indices;\@ their propagation length $\ell_\mathrm{p}$, which is limited by losses in the metal;\@ and their penetration depth $\delta_\mathrm{d}$ into the surrounding media \cite{barnes2006surface}.

\begin{figure}
	\centering
	\includegraphics[width=0.93\linewidth]{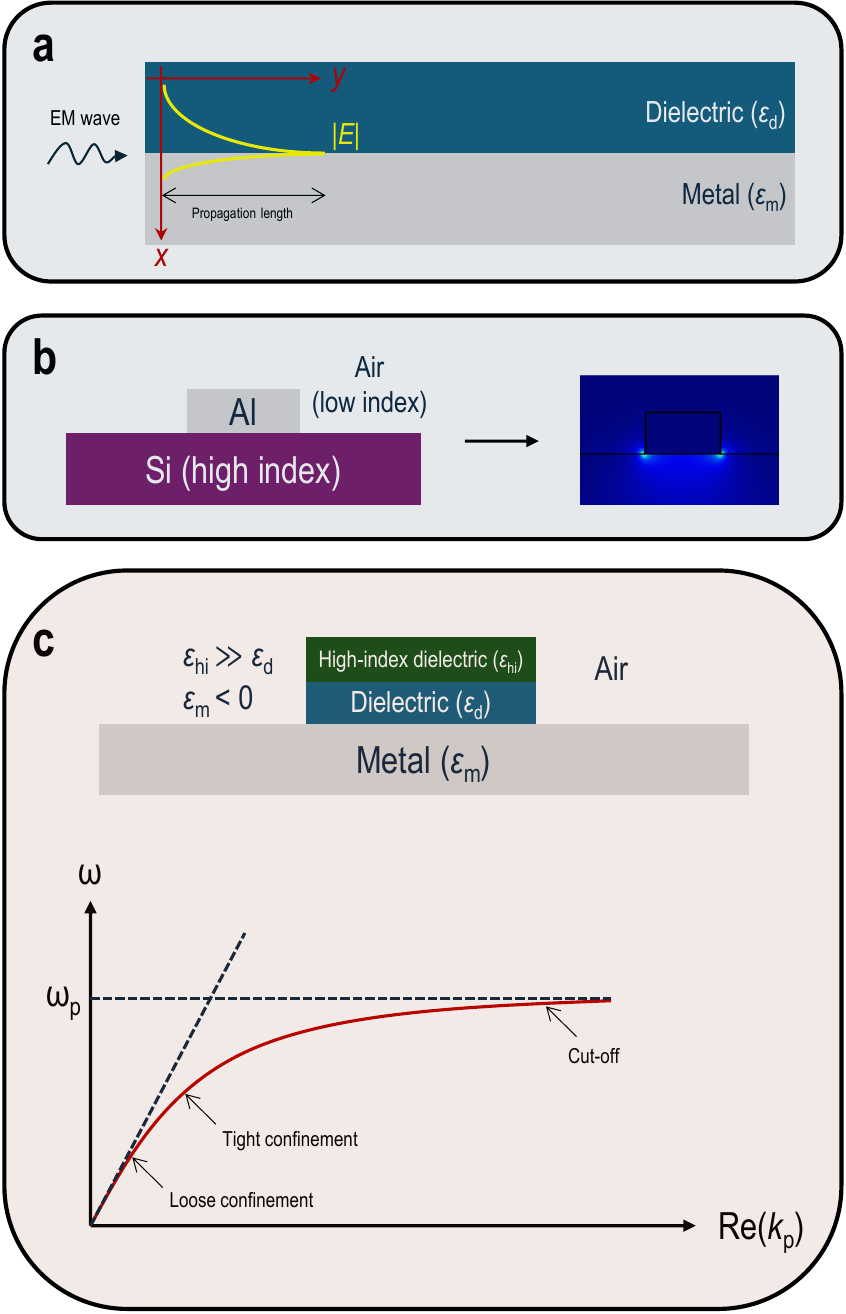}
	\caption{SPP propagation.\@ \textbf{a}, Cross-sectional schematic of SPP propagation at a metal-dielectric interface, illustrating the exponential decay of the E-field into both materials.\@ \textbf{b}, Cross-sectional illustration of a basic plasmonic waveguide structure, with a metal layer on top of a high-index dielectric layer, demonstrating the intensified E-field due to boundary conditions and confinement within the high-index material (i.e., Si with \emph{n}$_\mathrm{Si} \approx$ 3.48 at $\lambda$ = 1.55 $\mu$m) as shown in the adjacent E-field distribution plot.\@ Because SPPs result from a resonant phenomenon, SPP decay is a challenging issue that impedes the commercialization potential of plasmonic devices.\@ Hence, more research is needed to advance the performance of SPP-based devices while optimizing field overlap and modal losses.\@ \textbf{c}, Cross-section of a fundamental plasmonic supermode waveguide design with an accompanying diagram indicating the different regions of operation for achieving optimal mode confinement.\@ Panel \textbf{c} is adapted with permission from Alam et al.\@ \cite{alam2007super}.\@ Copyright 2007, Optica Publishing Group.}
	\label{SPP_propagation}
\end{figure}

Plasmonic waveguides have emerged as a promising solution to overcome the diffraction limit of light, a fundamental constraint in traditional optics that restricts the miniaturization of photonic components.\@ These waveguides exploit the unique properties of SPPs, which are hybrid EM modes formed at the interface between a metal and a dielectric.\@ As illustrated in \figurename{ \ref{SPP_propagation}\textbf{a}}, SPPs exhibit strong field confinement at this interface, enabling the manipulation of light at scales significantly smaller than its free-space wavelength.\@ The effective refractive index of an SPP mode, denoted as \emph{n}$_\mathrm{SPP}$, is higher than that of light in free space or the dielectric, and can be described by equation \ref{SPP_eff}:
\begin{equation}\label{SPP_eff}
	n_\mathrm{SPP} = \sqrt{\frac{\varepsilon_\mathrm{m} \varepsilon_\mathrm{d}}{\varepsilon_\mathrm{m} + \varepsilon_\mathrm{d}}},
\end{equation}

\noindent{where $\varepsilon_\mathrm{m}$ and $\varepsilon_\mathrm{d}$ are the permittivities of the metal and dielectric, respectively.\@ This higher effective refractive index leads to a shorter effective wavelength for SPPs, further contributing to their exceptional confinement capabilities.\@ It is important to note that the degree of confinement for the guided mode depends on the frequency of operation.\@ At frequencies much lower than the surface plasmon resonance frequency, the mode will be loosely confined.\@ As the frequency increases but remains below the resonance, the mode becomes more tightly confined.\@ However, at frequencies near the surface plasmon resonance, the plasmonic mode becomes tightly bound to the metallic interface, and coupling with other modes becomes negligible, leading to a cutoff condition \cite{alam2007super}}

In traditional total internal reflection (TIR) waveguides, modal areas are typically on the order of 0.2 $\mu$m$^\mathrm{2}$ \cite{chrostowski2015silicon,lin2012efficient,maker2011low}, while plasmonic waveguides can achieve modal areas as small as 0.002 $\mu$m$^\mathrm{2}$ \cite{su2019record}.\@ This significant reduction in the modal area enables the design of much more compact photonic components, making them ideal for ICs.\@ However, TIR waveguides primarily suffer from scattering losses due to fabrication imperfections, which are usually minimal compared to the losses in plasmonic waveguides, which can reach up to 1.5 dB/$\mu$m \cite{wong2018performance}.\@ The propagation of SPPs involves a rapid decay of the electric field (E-field) into both the dielectric and the metal.\@ At the metal-dielectric interface, the E-field is significantly intensified due to the imposed boundary conditions and the higher refractive index of the dielectric.\@ This is analogous to the field enhancement observed in a stack of Al (metal) and Si (dielectric) as shown in \figurename{ \ref{SPP_propagation}\textbf{b}}.\@ The strong field confinement is advantageous for enhancing light-matter interactions, but it also contributes to increased absorption losses.

Conventionally, the diffraction limit of light restricts the size of an optical mode propagating in a dielectric waveguide to about half of the light's wavelength $\lambda$ in the dielectric material.\@ To overcome this limitation and achieve higher on-chip integration density, plasmonic waveguides leverage SPPs.\@ These hybrid modes arise from the coupling of EM fields to resonant oscillations of electrons at the metal-dielectric interface, enabling subwavelength confinement of light \cite{oulton2008hybrid}.\@ Alternatively stated, the higher effective refractive index of SPPs compared to light in other media results in the strong optical field confinement observed at metal-dielectric interfaces.\@ This confinement allows for waveguiding through highly localized plasmon polaritons, which are essentially plasmons coupled with photons, propagating as EM fields coupled to resonant oscillations of the electron plasma.\@ Ma et al.\@ explore the concept of long-range (LR) SPP modes in asymmetric hybrid plasmonic waveguides, demonstrating that field symmetry on either side of the metal layer is crucial for achieving LR propagation, even in the presence of structural asymmetry \cite{ma2014asymmetric}.

However, this confinement comes at the cost of increased propagation losses and potential crosstalk.\@ To ensure minimal crosstalk within a dense array of plasmonic waveguides, the penetration depth $\delta_\mathrm{d}$ of an SPP's tangential component into the surrounding medium should be carefully controlled.\@ In a typical scenario where the waveguide is exposed to air, the longest operating wavelength is often chosen such that the emanating field penetrates no greater than three wavelengths on the air-exposed side \cite{blaber2010review}.\@ A more conservative criterion, as proposed by Blaber et al.\@ \cite{blaber2010review}, restricts the penetration depth to no more than half a wavelength (assuming a low-loss Drude metal), as follows \cite{barnes2006surface,blaber2010review}:
\begin{equation}
	\frac{\delta_\mathrm{d}}{\lambda} = \frac{1}{\mathrm{Re}\{\sqrt{-1/(\varepsilon_\mathrm{m} + 1)}\}} = \frac{1}{2\pi}\Big|\frac{\varepsilon'_\mathrm{m}+\varepsilon_\mathrm{d}}{\varepsilon^2_\mathrm{d}}\Big|^{\frac{1}{2}} \leq 0.5,
\end{equation}

\noindent{While a dielectric cladding layer can further reduce the penetration depth, it is important to consider its impact on the propagation length $\ell_\mathrm{p}$ of the SPP mode, which can be approximated as:}
\begin{equation}\label{prop_length}
	\ell_\mathrm{p} = \frac{1}{2\mathrm{Im}\{\lambda\sqrt{\varepsilon_\mathrm{m}/(\varepsilon_\mathrm{m}+1)}\}} \approx \lambda \frac{(\varepsilon'_\mathrm{m})^{2}}{2\pi\varepsilon''_\mathrm{m}}.
\end{equation}

\noindent{These equations highlight the fundamental trade-off between field confinement (and thus miniaturization potential) and propagation losses in plasmonic waveguides.\@ Thus, for optimal high-density integration in air-cladded waveguides, we have calculated that the real part of the metal's permittivity should ideally lie within the range $-$10.87 $\leq$ $\varepsilon'_\mathrm{m}$ $\leq$ 8.87 (i.e., $\varepsilon'_\mathrm{m} \in [-\pi^\mathrm{2}+\mathrm{1}, \, \pi^\mathrm{2}-\mathrm{1}]$).\@ This trade-off between confinement and loss, coupled with the challenges of integrating traditional plasmonic materials like Au and Ag with CMOS technology, is further exacerbated by the resonant nature of SPPs, which makes their decay a critical hurdle to overcome for practical applications.\@ Consequently, extensive research efforts are directed towards enhancing the performance of SPP-based devices by optimizing the delicate balance between field overlap ($\Phi$) and modal losses, particularly within the constraints imposed by integrating materials compatible with Si processing.}

Alam et al.\@ highlighted the importance of minimizing the overlap of the mode with the lossy dielectric (e.g., Si) to reduce propagation losses, even when operating near the bandgap of the dielectric \cite{alam2014marriage,alam2007super}.\@ While the structure shown in \figurename{ \ref{SPP_propagation}\textbf{b}} is similar to the one investigated by these researchers, they proposed a structure where a high-index dielectric (e.g., Si) is placed adjacent to a metal with a thin, low-index dielectric spacer, as shown in \figurename{ \ref{SPP_propagation}\textbf{c}} \cite{alam2007super}.\@ This waveguide design utilizes a coupled mode configuration, where an SPP mode interacts with a dielectric waveguide mode to create a \emph{supermode}.\@ The structure consists of a high-refractive-index dielectric slab positioned near a metallic surface, with a lower-index medium separating the two.\@ This arrangement, depicted in \figurename{ \ref{SPP_propagation}\textbf{c}}, facilitates coupling between the SPP mode at the metal surface and the guided mode within the dielectric slab.\@ The resulting supermode exhibits strong field confinement in the region between the metal and the dielectric.\@ Although the dielectric slab can support both transverse electric (TE) and transverse magnetic (TM) modes, the metal-dielectric interface selectively supports only TM-polarized SPPs.\@ Consequently, the coupled mode guided by this structure will also possess TM polarization.

The confinement of the guided mode in this coupled waveguide structure depends on the operational frequency ($\omega$).\@ At lower frequencies ($\omega \ll \omega_\mathrm{SPP}$), where $\omega_\mathrm{SPP}$ is the surface plasmon frequency, the propagation constant of the supermode is similar to that of light in the surrounding dielectric medium.\@ This leads to a loosely confined mode between the dielectric slab and the metal plane as can be observed in \figurename{ \ref{SPP_propagation}\textbf{c}}.\@ As $\omega$ increases, the supermode becomes more confined within the low-index SiO$_\mathrm{2}$ spacer layer.\@ However, when $\omega$ approaches the $\omega_\mathrm{SPP}$ ($\omega \simeq \omega_\mathrm{SPP}$), the coupling between the plasmon mode and the dielectric mode weakens.\@ This is because the plasmonic mode becomes tightly bound to the metal interface, and the supermode reaches its cutoff point \cite{alam2007super}.

It is important to note that $\omega_\mathrm{SPP}$ is directly related to the cutoff frequency of the supermode.\@ Simulations conducted over a wavelength range of 0.5 to 1.2 $\mu$m (illustrated in Figure 3 in Alam et al.'s paper) confirm these theoretical predictions.\@ The simulations were conducted using a 2D finite-element method (FEM) with the following parameters: a dielectric ridge of 100 nm width and 100 nm height, a low-index spacer layer of 50 nm thickness, and a metal thickness of 100 nm.\@ The material parameters used were:\@ \emph{n$_\mathrm{m}$} = 0.134 + \emph{i}3.19 for the metal, \emph{n}$_\mathrm{d}$ = 1.444 for the dielectric, and \emph{n}$_\mathrm{hi}$ = 3.45 for the high-index dielectric.\@ The results demonstrate that the inclusion of Si losses does not significantly impact the propagation distance, even near the band edge.\@ This observation underscores the effective confinement of the mode within the low-index spacer layer, thereby minimizing interaction with the lossy Si material.\@ Furthermore, the mode exhibits tight confinement at shorter wavelengths, with its size approaching the width of the dielectric ridge.\@ As the wavelength increases, the confinement weakens, and the mode size expands \cite{alam2007super}.

The inherent trade-off between confinement and loss in plasmonic waveguides necessitates innovative design approaches to achieve LR propagation while maintaining subwavelength dimensions.\@ One such approach, as explored by Ma et al., involves the utilization of asymmetric hybrid plasmonic waveguides, which leverage the coupling between two dissimilar hybrid plasmonic waveguides to achieve field symmetry across the metal layer, a critical condition for enabling the existence of LR modes even in asymmetric structures \cite{ma2014asymmetric}.

\begin{figure*}[h]
	\begin{minipage}{0.48\textwidth} 
		\begin{tikzpicture}
			\begin{axis}[
				title={Penetration depth vs.\@ metal permittivity},
				xlabel={$\varepsilon'_\mathrm{m}$},
				ylabel={$\delta_d / \lambda$},
				zlabel={$\varepsilon''_\mathrm{m}$},
				colormap/viridis,
				]
				\addplot3[
				surf,
				domain=-20:10,
				domain y=0.1:6,
				]
				{1/(2*pi) * sqrt(abs( (x + 1) / 1 ))}; 
			\end{axis}
		\end{tikzpicture}
	\end{minipage}\hfill 
	\begin{minipage}{0.48\textwidth}
		\begin{tikzpicture}
			\begin{axis}[
				title={Propagation length vs.\@ metal permittivity},
				xlabel={$\varepsilon'_\mathrm{m}$},
				ylabel={$\ell_p / \lambda$},
				zlabel={$\varepsilon''_\mathrm{m}$},
				colormap/viridis,
				]
				\addplot3[
				surf,
				domain=-20:10,
				domain y=0.1:6,
				]
				{ (x^2) / (2*pi*y) };
			\end{axis}
		\end{tikzpicture}
	\end{minipage}
	\caption{Trade-off between confinement and loss in plasmonic waveguides.\@ The 3D plots illustrate the dependence of \textbf{a}, wavelength-normalized penetration depth $\delta_\mathrm{d} / \lambda$ and \textbf{b}, wavelength-normalized propagation length $\ell_\mathrm{p} / \lambda$ of SPPs on the real ($\varepsilon'_\mathrm{m}$) and imaginary ($\varepsilon''_\mathrm{m}$) parts of the metal's permittivity in an air-cladded waveguide.}
	\label{fig:tradeoff}
\end{figure*}
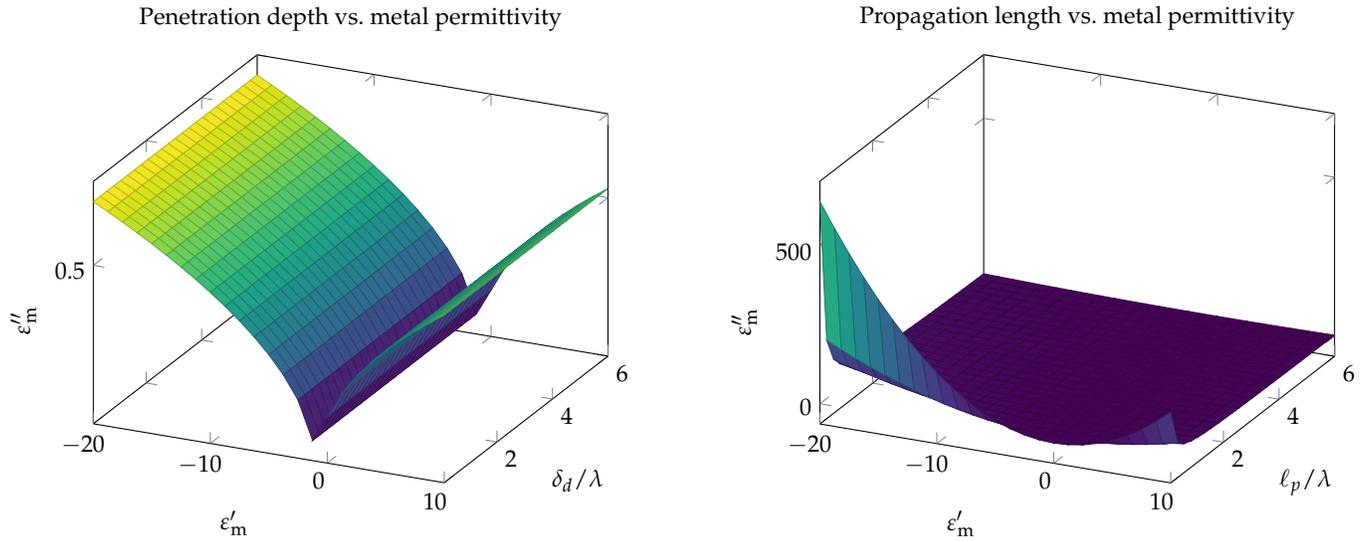

\figurename{ \ref{fig:tradeoff}} visually captures the inherent trade-off between field confinement and propagation loss in plasmonic waveguides, as predicted by the analytical expressions for penetration depth $\delta_\mathrm{d}$ and propagation length $\ell_\mathrm{p}$ derived from the low-loss Drude model.\@ The 3D plots illustrate the dependence of these parameters on the real ($\varepsilon'_\mathrm{m}$) and imaginary ($\varepsilon''_\mathrm{m}$) parts of the metal's permittivity, assuming an air-cladded waveguide (i.e., $\varepsilon_\mathrm{d} = 1$).\@ As shown in \figurename{ \ref{fig:tradeoff}\textbf{a}}, a more negative real permittivity, which typically occurs near the plasmon resonance frequency, leads to stronger confinement of the SPP mode, evident in the decreasing wavelength-normalized penetration depth $\delta_\mathrm{d} / \lambda$.\@ However, this enhanced confinement often comes at the cost of increased losses and shorter propagation lengths, as illustrated in \figurename{ \ref{fig:tradeoff}\textbf{b}}.\@ The wavelength-normalized propagation length $\ell_\mathrm{p} / \lambda$ initially increases with a more negative real permittivity but then starts to decrease as the imaginary permittivity, representing the losses, becomes dominant.\@ The choice of metal and waveguide design must therefore carefully balance these competing factors to achieve optimal performance for specific applications.

\section{Emerging Plasmonic Devices for Si}

By integrating plasmonic devices with mature Si technologies, such as those based on SOI wafers, this approach seeks to combine the unique light manipulation capabilities of SPP-guided waves at the nanoscale with the advantages of established Si and CMOS manufacturing processes.\@ This section delves into two key plasmonic device categories, modulators, and PDs, highlighting their potential for revolutionizing on-chip data communication through enhanced speed, sensitivity, and miniaturization.\@ We will explore recent advancements in these areas, emphasizing their compatibility with Si integration and their potential to address the escalating BW demands of modern communication networks.

\subsection{Modulators}

Plasmonic modulators are a promising technology for encoding information onto optical signals with high speed and in a compact form factor, which is critical for on-chip data communication applications.\@ These devices leverage the interaction of light with surface plasmons to enable dynamic control of optical properties such as amplitude, phase, and polarization \cite{crabb2022amplitude,smolyaninov2019programmable}.\@ Plasmonic structures offer exceptional light confinement capabilities, significantly enhancing light-matter interactions.\@ This characteristic enables optical modulation with a remarkably small footprint and broader BW compared to conventional Si-based modulators.\@ Si optical cavities, for example, typically exhibit quality factors (\emph{Q}s) in the thousands, which inherently limit achievable BWs \cite{eppenberger2023resonant,winzer2017scaling}.\@ While techniques like optical peaking can enhance BW, even advanced Si microring resonator modulators (MRMs) have not exceeded 42.5--54 GHz BW range \cite{hu2023300,sakib2022240}.\@ Consequently, data transmission rates remain limited to approximately 240 Gbps for four-level pulse amplitude modulation (4PAM) \cite{sakib2022240} and up to 302 Gbps using a discrete-multitone format (DMT) \cite{hu2023300}.\@ In the context of Si integration, plasmonic modulators can be seamlessly integrated with Si photonic components, leveraging mature Si fabrication processes and enabling the development of high-density photonic circuits \cite{zemtsov2023plasmon}.\@ \tablename{ \ref{modulators}} presents a performance comparison of various modulator structures, including both plasmonic and non-plasmonic modulators, highlighting their compatibility with Si integration for optical on-chip data communication.

\begin{sidewaystable}
	\def\d{\hphantom{0}}
	\caption{Performance comparison of various modulator structures, highlighting their compatibility with Si integration.\@ The table includes both plasmonic and non-plasmonic modulators, evaluated across key metrics.\@ Si integration potential is classified as 'Seamless,' 'Strategic,' or 'Challenging,' reflecting the ease of incorporating each structure into Si foundry and CMOS processes.\label{modulators}}%
    \renewcommand{\arraystretch}{0.82} 
	\begin{tabular*}{\textwidth}{@{\extracolsep{\fill}}lllllllll@{\extracolsep{\fill}}}
		\toprule
		\begin{footnotesize}Device structure\end{footnotesize} & \begin{footnotesize}Substrate\end{footnotesize} & \begin{footnotesize}Si integ.\end{footnotesize} & \begin{footnotesize}Type\end{footnotesize} & \begin{footnotesize}\emph{A} ($\mu$m$^\mathrm{2}$)\end{footnotesize} & \begin{footnotesize}PL/IL\end{footnotesize} & \begin{footnotesize}BW$_\mathrm{3-dB}$ (GHz)\end{footnotesize} & \begin{footnotesize}ER/MD\end{footnotesize} & \begin{footnotesize}$\lambda$ ($\mu$m)\end{footnotesize} \\
		\midrule
		\begin{footnotesize}Au/PZT/LN/Pt/TiO$_\mathrm{2}$/YSZ/SiO$_\mathrm{2}$ coupler \cite{yezekyan2024plasmonic} \end{footnotesize}& \begin{footnotesize}Si\end{footnotesize} & \begin{footnotesize}Challenging\end{footnotesize} & \begin{footnotesize}Plasmonic\end{footnotesize} & \begin{footnotesize}---\end{footnotesize} & \begin{footnotesize}0.36 dB/$\mu$m (IL)\end{footnotesize} & \begin{footnotesize}2\end{footnotesize} \begin{footnotesize}$\times$ 10$^{-\mathrm{4}}$\end{footnotesize} & \begin{footnotesize}> 40\% (MD)\end{footnotesize} & \begin{footnotesize}1.55\end{footnotesize} \\
		\begin{footnotesize}OEO/Au/SiO$_\mathrm{2}$ MZ \cite{bisang2024plasmonic}\end{footnotesize} & \begin{footnotesize}SOI\end{footnotesize} & \begin{footnotesize}Challenging\end{footnotesize} & \begin{footnotesize}Plasmonic\end{footnotesize} & \begin{footnotesize}1.58\end{footnotesize} & \begin{footnotesize}$-$5.9 (on-chip), $-$34 dB (IL)\end{footnotesize} & \begin{footnotesize}> 100\end{footnotesize} & \begin{footnotesize}---\end{footnotesize} & \begin{footnotesize}1.528\end{footnotesize} \\
		\begin{footnotesize}OEO/Au/SiO$_\mathrm{2}$ RRM \cite{bisang2024plasmonic}\end{footnotesize} & \begin{footnotesize}SOI\end{footnotesize} & \begin{footnotesize}Challenging\end{footnotesize} & \begin{footnotesize}Plasmonic\end{footnotesize} & \begin{footnotesize}1.05\end{footnotesize} & \begin{footnotesize}---\end{footnotesize} & \begin{footnotesize}> 67\end{footnotesize} & \begin{footnotesize}---\end{footnotesize} & \begin{footnotesize}1.285\end{footnotesize} \\
		\begin{footnotesize}Au (grating)/HfO$_\mathrm{2}$/Si MOSCAP \cite{naeini2024high}\end{footnotesize} & \begin{footnotesize}Si\end{footnotesize} & \begin{footnotesize}Strategic\end{footnotesize} & \begin{footnotesize}Plasmonic\end{footnotesize} & \begin{footnotesize}20--616\end{footnotesize} & \begin{footnotesize}2--4.2 dB\end{footnotesize} & \begin{footnotesize}22\end{footnotesize} & \begin{footnotesize}$\leq$ 2.3\% (MD)\end{footnotesize} & \begin{footnotesize}1.55\end{footnotesize} \\
		\begin{footnotesize}OEO/Au/SiO$_\mathrm{2}$ microring-racetrack \cite{eppenberger2023resonant}\end{footnotesize} & \begin{footnotesize}SOI\end{footnotesize} & \begin{footnotesize}Challenging\end{footnotesize} & \begin{footnotesize}Plasmonic\end{footnotesize} & \begin{footnotesize}814\end{footnotesize} & \begin{footnotesize}0.049 dB/$\mu$m\end{footnotesize} & \begin{footnotesize}176\end{footnotesize} & \begin{footnotesize}5.2 dB (ER)\end{footnotesize} & \begin{footnotesize}1.55\end{footnotesize} \\
		\begin{footnotesize}a-Si/Al/SiO$_\mathrm{2}$/ITO/Si CHPW \cite{alfaraj2023facile}\end{footnotesize} & \begin{footnotesize}SOI\end{footnotesize} & \begin{footnotesize}Seamless\end{footnotesize} & \begin{footnotesize}Plasmonic\end{footnotesize} & \begin{footnotesize}2\end{footnotesize} & \begin{footnotesize}0.128 dB/$\mu$m (IL)\end{footnotesize} & \begin{footnotesize}26--600\end{footnotesize} & \begin{footnotesize}10.63 dB (ER)\end{footnotesize} & \begin{footnotesize}1.55\end{footnotesize} \\
		\begin{footnotesize}SiN-TFLN MZ \cite{valdez2023100}\end{footnotesize} & \begin{footnotesize}LNOI/SiN\end{footnotesize} & \begin{footnotesize}Strategic\end{footnotesize} & \begin{footnotesize}EO\end{footnotesize} & \begin{footnotesize}7.2 $\times$ 10$^\mathrm{3}$\end{footnotesize} & \begin{footnotesize}0.0015 dB/$\mu$m (IL)\end{footnotesize} & \begin{footnotesize}100\end{footnotesize} & \begin{footnotesize}31 dB (ER)\end{footnotesize} & \begin{footnotesize}0.784\end{footnotesize} \\
		\begin{footnotesize}SiN-TFLN MZ \cite{valdez2023buried}\end{footnotesize} & \begin{footnotesize}LNOI/SOI\end{footnotesize} & \begin{footnotesize}Strategic\end{footnotesize} & \begin{footnotesize}EO\end{footnotesize} & \begin{footnotesize}---\end{footnotesize}& \begin{footnotesize}0.003--0.004 dB/$\mu$m (IL)\end{footnotesize} & \begin{footnotesize}> 110\end{footnotesize} & \begin{footnotesize}28 dB (ER)\end{footnotesize} & \begin{footnotesize}1.31\end{footnotesize} \\
		\begin{footnotesize}Au-BTO MZ \cite{kohli2023plasmonic}\end{footnotesize} & \begin{footnotesize}SiN\end{footnotesize} & \begin{footnotesize}Challenging\end{footnotesize} & \begin{footnotesize}Plasmonic\end{footnotesize} & \begin{footnotesize}---\end{footnotesize} & \begin{footnotesize}0.5 dB/$\mu$m\end{footnotesize} & \begin{footnotesize}> 70\end{footnotesize} & \begin{footnotesize}---\end{footnotesize} & \begin{footnotesize}1.557\end{footnotesize} \\
		\begin{footnotesize}Au NW/Ti/LN/air/Au NW/Ti MZ \cite{thomaschewski2022plasmonic}\end{footnotesize} & \begin{footnotesize}LN\end{footnotesize} & \begin{footnotesize}Challenging\end{footnotesize} & \begin{footnotesize}Plasmonic\end{footnotesize} & \begin{footnotesize}< 10$^\mathrm{6}$\end{footnotesize} & \begin{footnotesize}1.3 dB/$\mu$m (PL)\end{footnotesize} & \begin{footnotesize}10--900\end{footnotesize} & \begin{footnotesize}2.5 dB (ER)\end{footnotesize} & \begin{footnotesize}1.55\end{footnotesize} \\
		\begin{footnotesize}Au/Al$_\mathrm{2}$O$_\mathrm{3}$/ITO MZ \cite{amin2021heterogeneously}\end{footnotesize} & \begin{footnotesize}SOI\end{footnotesize} & \begin{footnotesize}Strategic\end{footnotesize} & \begin{footnotesize}Plasmonic\end{footnotesize} & \begin{footnotesize}---\end{footnotesize} & \begin{footnotesize}6.7--10.1 dB/$\mu$m (IL)\end{footnotesize} & \begin{footnotesize}1.1\end{footnotesize} & \begin{footnotesize}3--8 dB (ER)\end{footnotesize} & \begin{footnotesize}1.55\end{footnotesize} \\
		\begin{footnotesize}Au/graphene-SiO$_\mathrm{2}$ Si rib \cite{alaloul2021low}\end{footnotesize} & \begin{footnotesize}SOI\end{footnotesize} & \begin{footnotesize}Challenging\end{footnotesize} & \begin{footnotesize}Plasmonic\end{footnotesize} & \begin{footnotesize}5.52\end{footnotesize} & \begin{footnotesize}0.52 (IL), 0.43 (PL) dB/$\mu$m\end{footnotesize} & \begin{footnotesize}> 100\end{footnotesize} & \begin{footnotesize}3.5 dB (ER)\end{footnotesize} & \begin{footnotesize}1.55\end{footnotesize} \\
		\begin{footnotesize}a-Si/Al/SiO$_\mathrm{2}$/ITO/Si CHPW \cite{lin2020monolithic}\end{footnotesize} & \begin{footnotesize}SOI\end{footnotesize} & \begin{footnotesize}Seamless\end{footnotesize} & \begin{footnotesize}Plasmonic\end{footnotesize} & \begin{footnotesize}1--3\end{footnotesize} & \begin{footnotesize}< 0.1 dB/$\mu$m (IL)\end{footnotesize} & \begin{footnotesize}26--636\end{footnotesize} & \begin{footnotesize}> 10 dB (ER)\end{footnotesize} & \begin{footnotesize}1.55\end{footnotesize} \\
        \begin{footnotesize}Si$_\mathrm{3}$N$_\mathrm{4}$/VO$_\mathrm{2}$ rib \cite{wong2019broadband}\end{footnotesize} & \begin{footnotesize}Si\end{footnotesize} & \begin{footnotesize}Strategic\end{footnotesize} & \begin{footnotesize}All-optical\end{footnotesize} & \begin{footnotesize}2--10\end{footnotesize} & \begin{footnotesize}0.98 (IL), 1.33 (PL) dB/$\mu$m\end{footnotesize} & \begin{footnotesize}---\end{footnotesize} & \begin{footnotesize}$\leq$ 10 dB (ER)\end{footnotesize} & \begin{footnotesize}1.55\end{footnotesize} \\
		\begin{footnotesize}Ag/SiO$_\mathrm{2}$/Si MIS ring \cite{haffner2018low}\end{footnotesize} & \begin{footnotesize}SOI\end{footnotesize} & \begin{footnotesize}Strategic\end{footnotesize} & \begin{footnotesize}Plasmonic\end{footnotesize} & \begin{footnotesize}---\end{footnotesize} & \begin{footnotesize}---\end{footnotesize} & \begin{footnotesize}> 110\end{footnotesize} & \begin{footnotesize}> 10 dB (ER)\end{footnotesize} & \begin{footnotesize}1.54--1.55\end{footnotesize} \\
		\begin{footnotesize}VO$_\mathrm{2}$/Si/Ag/SiO$_\mathrm{2}$/Si HPW \cite{wong2018performance}\end{footnotesize} & \begin{footnotesize}SiO$_\mathrm{2}$\end{footnotesize} & \begin{footnotesize}Strategic\end{footnotesize} & \begin{footnotesize}Plasmonic\end{footnotesize} & \begin{footnotesize}0.4\end{footnotesize} & \begin{footnotesize}1.4 dB/$\mu$m (IL)\end{footnotesize} & \begin{footnotesize}---\end{footnotesize} & \begin{footnotesize}3.8 dB (ER)\end{footnotesize} & \begin{footnotesize}1.55\end{footnotesize} \\
		\begin{footnotesize}Graphene-SiO$_\mathrm{2}$ Si rib \cite{hu2016broadband}\end{footnotesize} & \begin{footnotesize}Si\end{footnotesize} & \begin{footnotesize}Strategic\end{footnotesize} & \begin{footnotesize}Electroabsorptive\end{footnotesize} & \begin{footnotesize}50\end{footnotesize} & \begin{footnotesize}0.06--0.11 (PL) dB/$\mu$m\end{footnotesize} & \begin{footnotesize}5.9\end{footnotesize} & \begin{footnotesize}5.2 dB (ER)\end{footnotesize} & \begin{footnotesize}1.55\end{footnotesize} \\
		\begin{footnotesize}Si MZ \cite{haffner2015all}\end{footnotesize} & \begin{footnotesize}SOI\end{footnotesize} & \begin{footnotesize}Challenging\end{footnotesize} & \begin{footnotesize}Plasmonic\end{footnotesize} & \begin{footnotesize}10\end{footnotesize} & \begin{footnotesize}0.4--0.5 dB/$\mu$m (PL)\end{footnotesize} & \begin{footnotesize}70--100\end{footnotesize} & \begin{footnotesize}6 dB (ER)\end{footnotesize} & \begin{footnotesize}1.534--1.55\end{footnotesize} \\
		\begin{footnotesize}Graphene/Al$_\mathrm{2}$O$_\mathrm{3}$--SiO$_\mathrm{2}$--Si$_\mathrm{3}$N$_\mathrm{4}$--Au/Pd/Ti \cite{phare2015graphene}\end{footnotesize} & \begin{footnotesize}Si\end{footnotesize} & \begin{footnotesize}Strategic\end{footnotesize} & \begin{footnotesize}Electroabsorptive\end{footnotesize} & \begin{footnotesize}10$^\mathrm{4}$\end{footnotesize} & \begin{footnotesize}---\end{footnotesize} & \begin{footnotesize}30\end{footnotesize} & \begin{footnotesize}> 28 dB (ER)\end{footnotesize} & \begin{footnotesize}1.555\end{footnotesize} \\
		\begin{footnotesize}Ag/SiO$_\mathrm{2}$/Si MIS \cite{dionne2009plasmostor}\end{footnotesize} & \begin{footnotesize}SOI\end{footnotesize} & \begin{footnotesize}Strategic\end{footnotesize} & \begin{footnotesize}Plasmonic\end{footnotesize} & \begin{footnotesize}4--75\end{footnotesize} & \begin{footnotesize}0.207--2.37 dB/$\mu$m (PL)\end{footnotesize} & \begin{footnotesize}3--15\end{footnotesize} & \begin{footnotesize}$\leq$ 4.56 dB (ER)\end{footnotesize} & \begin{footnotesize}1.55\end{footnotesize} \\
		\bottomrule
	\end{tabular*}
\end{sidewaystable}

As shown in \tablename{ \ref{modulators}}, plasmonic modulators have been realized on various substrates, including Si, SOI, SiN, lithium niobate (LiNbO$_\mathrm{3}$ or LN), and lithium niobate on insulator (LNOI).\@ Modulators on SOI substrates are considered to have the most seamless integration potential, while those on SiN and LNOI substrates are strategically compatible.\@ The type of modulator also plays a crucial role in its performance.\@ Plasmonic modulators are generally smaller and faster than their non-plasmonic counterparts, making them ideal for applications where footprint and BW are critical factors.\@ However, plasmonic modulators can also suffer from higher IL levels due to the inherent losses associated with plasmonic materials.\@ The performance of the modulators is evaluated across several key metrics, including footprint, propagation loss/insertion loss (PL/IL), 3-dB speed/BW (BW$_\mathrm{3-dB}$), extinction ratio/modulation depth (ER/MD), and operating wavelength ($\lambda$).\@ These metrics provide a comprehensive overview of the capabilities of each modulator structure, allowing for a direct comparison of their strengths and weaknesses.

Bisang et al.\@ demonstrated a plasmonic Mach--Zehnder modulator (MZM) housed on an SOI substrate \cite{bisang2024plasmonic}.\@ The MZM is based on Au, SiO$_\mathrm{2}$, and organic electro-optic (OEO) materials operating at cryogenic temperatures (4 K) at $\lambda$ = 1.528 $\mu$m and is 15 $\mu$m-long and 105 nm-wide.\@ The MZM device achieved BW$_\mathrm{3-dB}$ levels above 100 GHz at low drive voltages: electrical energy consumption (\emph{E}$_\mathrm{bit}$) of the 16 GBd 2PAM modulator with a peak-to-peak voltage (\emph{V}$_\text{PP, 50-$\Omega$}$) of 96 mV is as low as 230 aJ/bit.\@ For the 128 GBd 2PAM modulator with a \emph{V}$_\text{PP, 50-$\Omega$}$ of 1 V, the \emph{E}$_\mathrm{bit}$ is 25 fJ/bit, and for 80 GBd 4PAM modulator with a \emph{V}$_\text{PP, 50-$\Omega$}$ of 897 mV, the \emph{E}$_\mathrm{bit}$ is 5.6 fJ/bit, making it suitable for cryogenic data transmission with low energy consumption.\@ In addition, the authors demonstrated a plasmonic ring-resonator modulator (RRM) based on the same materials (Au, SiO$_\mathrm{2}$, and OEO) on SOI.\@ The RRM operated at $\lambda$ = 1.285 $\mu$m and achieved a BW$_\mathrm{3-dB}$ of 67 GHz.\@ The RRM device had dimensions of 30 $\times$ 40 $\mu$m and a plasmonic slot length and width of 10 $\mu$m and 105 nm, respectively.\@ The \emph{E}$_\mathrm{bit}$ for the 128 GBd 2PAM signal with the RRM modulator with a \emph{V}$_\text{PP, 50-$\Omega$}$ of 1 V was 13 fJ/bit, and 3.6 fJ/bit for the 128 GBd 4PAM signal.\@ Furthermore, for the 180 GBd 2PAM signal with the RRM modulator with a \emph{V}$_\text{PP, 50-$\Omega$}$ of 1 V, the \emph{E}$_\mathrm{bit}$ was 17 fJ/bit.\@ The authors also showed a significant reduction (over 40\%) in plasmonic propagation losses at cryogenic temperatures (4 K) compared to room temperature for different plasmonic slot widths, verifying reduced losses in a plasmonic waveguide at low temperatures.\@ However, the MZM device exhibited high optical IL levels ($-$34 dB, for data transmission of up to 160 Gbit/s, plasmonic MZM operated at 4 K at around $\lambda$ = 1.528 $\mu$m), attributed mainly to poor fiber-to-chip coupling.\@ The integrability of these devices with standard CMOS and Si processes is challenging due to the incorporation of Au and OEOs.\@ Another approach to realizing high-speed plasmonic modulators is to use ferroelectric materials, as demonstrated by Kohli et al.\@ \cite{kohli2023plasmonic}.

Kohli et al.\@ demonstrated a high-speed plasmonic ferroelectric modulator monolithically integrated on a foundry-produced SiN platform \cite{kohli2023plasmonic}.\@ The modulator utilizes barium titanate (BaTiO$_\mathrm{3}$ or BTO) as the active electro-optic (EO) material, leveraging its large Pockels coefficient (\emph{r}$_\mathrm{42}$ up to 1300 pm/V in the strained case) for enhanced modulation efficiency.\@ The device structure consists of a Mach--Zehnder interferometer (MZI) with a 15 $\mu$m long plasmonic phase shifter in each arm.\@ The phase shifter features a metal-BTO-metal slot waveguide, enabling strong confinement of both the optical and electric fields for efficient modulation.\@ The modulator was fabricated on a SiN platform with the BTO integrated into a BEOL process.\@ The photonic-to-plasmonic converter, responsible for coupling light from the SiN waveguide to the plasmonic slot, was identified as a major contributor to the device's high IL level (29 dB).\@ This discrepancy was attributed to fabrication defects, particularly in the metallization of the coupling section.\@ Optimization of the fabrication process, specifically the metallization, is expected to reduce IL levels by more than 10 dB.\@ The modulator, operating at $\lambda$ = 1.557 $\mu$m, exhibited a flat EO frequency response up to 70 GHz, indicating a BW$_\mathrm{3-dB}$ beyond this value.\@ A half-wave-voltage length product (\emph{V}$_\pi$\emph{L}) of 144 \emph{V}-$\mu$m was achieved.\@ Data experiments demonstrated a symbol rate of 216 GBd with a 2PAM signal and 256 Gbit/s with a 4PAM signal, both with bit-error-ratios (BER) below the soft-decision forward error correction (SD-FEC) limit.\@ While the authors do not explicitly discuss the compatibility of their BTO-plasmonic modulator with standard CMOS and Si processes, the monolithic integration of the device on a foundry-produced SiN platform suggests potential compatibility.\@ However, the use of Au and the integration process of BTO may require further investigation to ensure full compatibility with standard CMOS and Si fabrication technologies.\@ In addition to plasmonic-organic and plasmonic-ferroelectric modulators, hybrid integration of LN with SiN has also been explored for high-speed modulation \cite{valdez2023100}.

Valdez et al.\@ presented two hybrid integrated MZMs based on thin-film lithium niobate (TFLN) bonded to a SiN layer \cite{valdez2023buried}.\@ The first modulator design featured metal coplanar waveguide electrodes buried in the SOI chip, while the second design had Au electrodes on top of the TFLN layer.\@ Both MZM configurations, operating at a wavelength of 1310 nm, demonstrated a BW$_\mathrm{3-dB}$ greater than 110 GHz and a voltage-driven optical ER greater than 28 dB.\@ The devices with buried electrodes had a length of 0.5 cm and a \emph{V}$_\pi$\emph{L} of 2.8 V-cm, while the devices with top electrodes had a length of 0.4 cm and a \emph{V}$_\pi$\emph{L} of 2.5 V-cm.\@ The authors used edge couplers to couple light onto the chip.\@ The IL was approximately 16 dB for both designs, with a PL of 0.34 dB/cm.\@ The edge coupling loss was 2.7 dB/facet.\@ The majority of the loss was from the optical mode traversing the edge of the bonded TFLN section.\@ Compared to hybrid Si-LN devices, where the feeder waveguide in the underlying high-index Si layer confines most of the light away from the index discontinuity at the unbonded-to-bonded edge interface, the low refractive index of SiN incurs high transition losses.\@ The SiN layer must be close to the surface for the hybrid mode design in the phase-shift EO segment.\@ This hybrid integration approach combines the high EO effect of TFLN with the low loss and design flexibility of the SiN platform.\@ The use of a buried metal layer in the SOI chip for the electrodes is compatible with CMOS fabrication processes.\@ However, the use of edge couplers for optical I/O limits the scalability and integration of this approach.\@ While LN is a well-established material for EO modulation \cite{xu2023attojoule,xu2022dual}, other materials, such as lead zirconate titanate (Pb[Zr$_\mathrm{\emph{x}}$Ti$_{\mathrm{1}-\mathrm{\emph{x}}}$]O$_\mathrm{3}$ (0 $\leq$ \emph{x} $\leq$ 1) or PZT), are also being investigated for their potential in plasmonic modulators.

Yezekyan et al.\@ investigated the potential of PZT as a platform on Si substrate for plasmonic EO modulators \cite{yezekyan2024plasmonic}.\@ The authors fabricated thin-film PZT platforms using a chemical solution deposition technique.\@ They then designed and fabricated plasmonic directional coupler modulators on these platforms.\@ The modulator design, based on previous work with LN \cite{thomaschewski2020plasmonic}, consisted of two identical Au stripes acting as both SPP waveguides and electrodes.\@ The devices were characterized at $\lambda$ = 1.55 $\mu$m.\@ The fabricated PZT platforms exhibited in-plane domain orientations, as revealed by second harmonic generation (SHG) microscopy.\@ Despite this, a high modulation depth (> 40\%) was achieved with a 15 $\mu$m-long modulator, exceeding the performance observed with similar modulators on LN.\@ However, the modulator's frequency response showed an unexpected cutoff at around 200 kHz.\@ This was attributed to domain reorientation effects in the PZT film, confirmed by additional experiments using a dedicated electrode structure.\@ The PZT platforms were fabricated on high-resistivity Si substrate with a 2.4 $\mu$m-thick SiO$_\mathrm{2}$ layer.\@ The PZT film itself was 2 $\mu$m-thick.\@ The plasmonic modulators were formed using Au stripes of 50 nm thickness and 350 nm width, separated by a 300 nm gap.\@ The fabrication process involved photolithography for the connecting electrodes and electron-beam lithography (EBL) for the modulator circuitry.\@ While the use of PZT on Si substrates suggests potential integration possibilities, the compatibility of PZT with standard CMOS and Si processes requires further investigation.\@  Although PZT is a ferroelectric material, and some ferroelectrics like HfO$_\mathrm{2}$ have shown CMOS compatibility, the specific integration process used in this work, including the chemical solution deposition technique and the potential need for high-temperature processing, may present challenges for integration with standard CMOS fabrication technologies.

These examples highlight the diverse approaches and materials being explored in the development of plasmonic modulators for Si integration.\@ The ongoing research in this field is paving the way for high-performance, compact, and energy-efficient modulators that can meet the growing demands of optical on-chip data communication networks.

\subsection{Photodetectors}

Plasmonic PDs offer significant advantages for optical on-chip data communication receivers due to their enhanced sensitivity and high-speed operation \cite{zhang2024plasmonic}.\@ By leveraging plasmonic absorption, these PDs enhance light capture and promote efficient carrier generation, leading to improved responsivities (\emph{R}$_\mathrm{ph}$) and faster response times compared to conventional PDs.\@ This enhanced sensitivity enables efficient detection of weak optical signals, crucial for applications like long-haul fiber optic communication where reliable data reception is paramount.\@ Furthermore, their compact footprint makes them ideal for integration into Si photonic circuits, paving the way for high-performance, miniaturized photodetection systems.\@ This integration allows combining plasmonic PDs with other Si-based components, such as optical waveguides and amplifiers, to create fully integrated photonic receivers capable of handling the increasing data rates and BW demands of modern communication networks \cite{zhang2024can,su2023scalability}.

However, conventional Si-based PDs, often relying on Ge for telecom wavelength detection, face limitations.\@ While achieving \emph{R}$_\mathrm{ph}$ levels and BWs up to 1 A/W and 100 GHz respectively \cite{salamin2018100,thomson2016roadmap,chen20161,novack2013germanium,vivien2012zero,vivien200942}, Ge PDs have a restricted optical BW range and require complex fabrication to address lattice mismatch and ion implantation issues \cite{vivien2012zero,michel2010high,vivien200942}.\@  Additionally, the epitaxial growth process of Ge on Si requires temperatures beyond 700 $^\circ$C, exceeding the thermal budget of typical BEOL processing (450--550 $^\circ$C), hindering integration with some BEOL photonic platforms \cite{ye2014germanium,lee2012back}.

An alternative approach utilizes internal photoemission (IPE) within Schottky diode structures fabricated on Si \cite{lin2020supermode,muehlbrandt2016silicon,goykhman2016chip}.\@ In these devices, thin metal films (typically 50--200 nm-thick) on Si nanowires absorb incident photons, generating ``hot'' carriers with sufficient kinetic energy to overcome the Schottky barrier height ($\phi_\mathrm{B}$) and be emitted.\@ Schottky PDs offer broadband operation (0.2--0.8 eV) and simplified fabrication, further miniaturized by employing plasmonic nanostructures to concentrate incident light.\@ While historically limited to \emph{R}$_\mathrm{ph}$ of a few mA/W and internal quantum efficiency (IQE) around 1\% \cite{brongersma2015plasmon}, recent advancements have yielded significant improvements.\@ For example, in 2020, our IPE-based Schottky CHPW PD demonstrated IQE and external quantum efficiency (EQE) of 3.1\% and 0.1\%, respectively, for a 5 $\mu$m-long CHPW (improvable to $\sim$1\% EQE with optimized light coupling) \cite{lin2020supermode}.\@ This device achieved a maximum  \emph{R}$_\mathrm{ph}$ of 38 mA/W at room temperature, potentially increased to 59 mA/W with a 10 $\mu$m length.

\tablename{ \ref{PDs}} presents a performance comparison of various plasmonic and conventional PD structures, emphasizing their potential for Si integration.

\begin{sidewaystable}
	\def\d{\hphantom{0}}
	\caption{Performance comparison of various PD structures, emphasizing their potential for Si integration.\@ The table encompasses plasmonic, PEC, and Schottky PDs, assessed based on footprint (\emph{A}), \emph{R}$_\mathrm{ph}$, Sens./\emph{D}$^\mathrm{*}$/NEP, speed/BW, and operating wavelength.\@ Si integration compatibility is classified as 'Seamless,' 'Strategic,' or 'Challenging,' reflecting the feasibility of integrating each structure with Si foundry and CMOS processes.\label{PDs}}%
    \renewcommand{\arraystretch}{1} 
	\begin{tabular*}{\textwidth}{@{\extracolsep{\fill}}lllllllll@{\extracolsep{\fill}}}
		\toprule
		\begin{footnotesize}Device structure\end{footnotesize} & \begin{footnotesize}Substrate\end{footnotesize} & \begin{footnotesize}Si integ.\end{footnotesize} & \begin{footnotesize}Type\end{footnotesize} & \begin{footnotesize}\emph{A} ($\mu$m$^\mathrm{2}$)\end{footnotesize} & \begin{footnotesize}\emph{R}$_\mathrm{ph}$\end{footnotesize} & \begin{footnotesize}Sens./$D^\mathrm{*}$/NEP\end{footnotesize} & \begin{footnotesize}BW$_\mathrm{3-dB}$\end{footnotesize} & \begin{footnotesize}$\lambda$ ($\mu$m)\end{footnotesize} \\
		\midrule
		\begin{footnotesize}Au/MoS$_\mathrm{2}$/Al$_\mathrm{2}$O$_\mathrm{3}$/HfN \cite{syong2024enhanced}\end{footnotesize} & \begin{footnotesize}Sapphire\end{footnotesize} & \begin{footnotesize}Challenging\end{footnotesize} & \begin{footnotesize}Plasmonic\end{footnotesize} & \begin{footnotesize}400\end{footnotesize} & \begin{footnotesize}4.45 $\times$ 10$^\mathrm{3}$ A/W\end{footnotesize} & \begin{footnotesize}2.58 $\times$ 10$^\mathrm{12}$ Jones (650 nm)\end{footnotesize} & \begin{footnotesize}729, 146 ms ($\tau_\mathrm{r}$, $\tau_\mathrm{d}$)\end{footnotesize} & \begin{footnotesize}0.63--0.65\end{footnotesize} \\
		\begin{footnotesize}ITO/SiO$_\mathrm{2}$/AuAg nanoalloy \cite{okamoto2024facilely}\end{footnotesize} & \begin{footnotesize}Si\end{footnotesize} & \begin{footnotesize}Challenging\end{footnotesize} & \begin{footnotesize}Plasmonic\end{footnotesize} & \begin{footnotesize}---\end{footnotesize} & \begin{footnotesize}7.3 mA/W ($\lambda$ = 1.31 $\mu$m)\end{footnotesize} & \begin{footnotesize}8.8 $\times$ 10$^\mathrm{8}$ Jones ($\lambda$ = 1.31 $\mu$m)\end{footnotesize} & \begin{footnotesize}---\end{footnotesize} & \begin{footnotesize}1.31\end{footnotesize}, \begin{footnotesize}1.55\end{footnotesize} \\
		\begin{footnotesize}Ge with integ.\@ TiN NHA \cite{mai2024towards}\end{footnotesize} & \begin{footnotesize}Si\end{footnotesize} & \begin{footnotesize}Challenging\end{footnotesize} & \begin{footnotesize}Plasmonic\end{footnotesize} & \begin{footnotesize}1.6 $\times$ 10$^\mathrm{3}$\end{footnotesize} & \begin{footnotesize}114 mA/W\end{footnotesize} & \begin{footnotesize}---\end{footnotesize} & \begin{footnotesize}---\end{footnotesize} & \begin{footnotesize}1.31\end{footnotesize} \\
		\begin{footnotesize}Graphene-Au NGs \cite{fan2023wafer}\end{footnotesize} & \begin{footnotesize}Quartz\end{footnotesize} & \begin{footnotesize}---\end{footnotesize} & \begin{footnotesize}Plasmonic\end{footnotesize} & \begin{footnotesize}2.5 $\times$ 10$^\mathrm{5}$\end{footnotesize} & \begin{footnotesize}2.95 mA/W ($\lambda$ = 1.31 $\mu$m)\end{footnotesize} & \begin{footnotesize}---\end{footnotesize} & \begin{footnotesize}39, 32.1 ms ($\tau_\mathrm{r}$, $\tau_\mathrm{d}$)\end{footnotesize} & \begin{footnotesize}635--1550\end{footnotesize} \\
		\begin{footnotesize}AlGaN/Pt-GaN \cite{liu2022photovoltage}\end{footnotesize} & \begin{footnotesize}Si\end{footnotesize} & \begin{footnotesize}Challenging\end{footnotesize} & \begin{footnotesize}PEC\end{footnotesize} & \begin{footnotesize}7 $\times$ 10$^\mathrm{6}$\end{footnotesize} & \begin{footnotesize}11.39 mA/W ($\lambda$ = 254 nm)\end{footnotesize} & \begin{footnotesize}---\end{footnotesize} & \begin{footnotesize}---\end{footnotesize} & \begin{footnotesize}0.254--0.365\end{footnotesize} \\
		\begin{footnotesize}Au/$\beta$-Ga$_\mathrm{2}$O$_\mathrm{3}$/TiN \cite{alfaraj2021silicon}\end{footnotesize} & \begin{footnotesize}Si\end{footnotesize} & \begin{footnotesize}Strategic\end{footnotesize} & \begin{footnotesize}Schottky PD\end{footnotesize} & \begin{footnotesize}2.5 $\times$ 10$^\mathrm{3}$\end{footnotesize} & \begin{footnotesize}249.84 A/W ($\lambda$ = 250 nm)\end{footnotesize} & \begin{footnotesize}6.38 $\times$ 10$^\mathrm{13}$ Jones ($D^\mathrm{*}$)\end{footnotesize} & \begin{footnotesize}---\end{footnotesize} & \begin{footnotesize}0.25\end{footnotesize} \\
		\begin{footnotesize}Au/$\beta$-Ga$_\mathrm{2}$O$_\mathrm{3}$/TiN \cite{alfaraj2021heteroepitaxial}\end{footnotesize} & \begin{footnotesize}MgO\end{footnotesize} & \begin{footnotesize}---\end{footnotesize} & \begin{footnotesize}Schottky PD\end{footnotesize} & \begin{footnotesize}2.75 $\times$ 10$^\mathrm{5}$\end{footnotesize} & \begin{footnotesize}276.72 A/W ($\lambda$ = 250 nm)\end{footnotesize} & \begin{footnotesize}5.31 $\times$ 10$^\mathrm{13}$ Jones ($D^\mathrm{*}$)\end{footnotesize} & \begin{footnotesize}500 ms\end{footnotesize} & \begin{footnotesize}0.25\end{footnotesize} \\
		\begin{footnotesize}TiN/graphene/Si \cite{alaloul2021plasmon}\end{footnotesize} & \begin{footnotesize}SOI\end{footnotesize} & \begin{footnotesize}Strategic\end{footnotesize} & \begin{footnotesize}Plasmonic\end{footnotesize} & \begin{footnotesize}1.61\end{footnotesize} & \begin{footnotesize}0.6--1.4 A/W\end{footnotesize} & \begin{footnotesize}< 20 pW/Hz$^\mathrm{(1/2)}$ (NEP)\end{footnotesize} & \begin{footnotesize}> 100 GHz\end{footnotesize} & \begin{footnotesize}1.55\end{footnotesize} \\
		\begin{footnotesize}Ru/n-AlGaN NWs \cite{wang2021highly}\end{footnotesize} & \begin{footnotesize}Si\end{footnotesize} & \begin{footnotesize}Challenging\end{footnotesize} & \begin{footnotesize}PEC\end{footnotesize} & \begin{footnotesize}2.5 $\times$ 10$^\mathrm{7}$\end{footnotesize} & \begin{footnotesize}48.8 mA/W (1.5 mW cm$^{-\mathrm{2}}$)\end{footnotesize} & \begin{footnotesize}---\end{footnotesize} & \begin{footnotesize}83, 19 ms ($\tau_\mathrm{r}$, $\tau_\mathrm{d}$)\end{footnotesize} & \begin{footnotesize}0.254\end{footnotesize} \\
		\begin{footnotesize}a-Si/Al/SiO$_\mathrm{2}$-Si CHPW \cite{lin2020supermode}\end{footnotesize} & \begin{footnotesize}SOI\end{footnotesize} & \begin{footnotesize}Seamless\end{footnotesize} & \begin{footnotesize}Plasmonic\end{footnotesize} & \begin{footnotesize}3--12\end{footnotesize} & \begin{footnotesize}59 mA/W\end{footnotesize} & \begin{footnotesize}$-$55 dBm (sens.)\end{footnotesize} & \begin{footnotesize}26--265 GHz\end{footnotesize} & \begin{footnotesize}1.5--1.6\end{footnotesize} \\
		\begin{footnotesize}a-Si/Al/SiO$_\mathrm{2}$-Si CHPW \cite{lin2020monolithic}\end{footnotesize} & \begin{footnotesize}SOI\end{footnotesize} & \begin{footnotesize}Seamless\end{footnotesize} & \begin{footnotesize}Plasmonic\end{footnotesize} & \begin{footnotesize}1--4\end{footnotesize} & \begin{footnotesize}82 mA/W\end{footnotesize} & \begin{footnotesize}$-$54 dBm (sens.)\end{footnotesize} & \begin{footnotesize}26--265 GHz\end{footnotesize} & \begin{footnotesize}1.53--1.57\end{footnotesize} \\
		\begin{footnotesize}Au/$\beta$-Ga$_\mathrm{2}$O$_\mathrm{3}$ NWs MSM \cite{xie2019catalyst}\end{footnotesize} & \begin{footnotesize}Al$_\mathrm{2}$O$_\mathrm{3}$\end{footnotesize} & \begin{footnotesize}Challenging\end{footnotesize} & \begin{footnotesize}PCE\end{footnotesize} & \begin{footnotesize}21\end{footnotesize} & \begin{footnotesize}233 A/W (0.68 $\mu$W/cm$^\mathrm{2}$)\end{footnotesize} & \begin{footnotesize}8.16 $\times$ 10$^\mathrm{12}$ Jones ($D^\mathrm{*}$)\end{footnotesize} & \begin{footnotesize}480, 40 ms ($\tau_\mathrm{r}$, $\tau_\mathrm{d}$)\end{footnotesize} & \begin{footnotesize}0.25\end{footnotesize} \\
		\begin{footnotesize}Au/graphene/SiO$_\mathrm{2}$ \cite{cakmakyapan2018gold}\end{footnotesize} & \begin{footnotesize}Si\end{footnotesize} & \begin{footnotesize}Challenging\end{footnotesize} & \begin{footnotesize}Plasmonic\end{footnotesize} & \begin{footnotesize}900\end{footnotesize} & \begin{footnotesize}0.6 A/W ($\lambda$ = 800 nm)\end{footnotesize} & \begin{footnotesize}1--20 pW/Hz$^\mathrm{(1/2)}$ (NEP)\end{footnotesize} & \begin{footnotesize}> 50 GHz\end{footnotesize} & \begin{footnotesize}0.8--20\end{footnotesize} \\
		\begin{footnotesize}a-Si/Al/SiO$_\mathrm{2}$/Si CHPW \cite{su2017highly}\end{footnotesize} & \begin{footnotesize}SOI\end{footnotesize} & \begin{footnotesize}Seamless\end{footnotesize} & \begin{footnotesize}Plasmonic\end{footnotesize} & \begin{footnotesize}3.1\end{footnotesize} & \begin{footnotesize}5 mA/W\end{footnotesize} & \begin{footnotesize}$-$35 dBm (sens.)\end{footnotesize} & \begin{footnotesize}400 GHz\end{footnotesize} & \begin{footnotesize}1.2--1.8\end{footnotesize} \\
		\bottomrule
	\end{tabular*}
\end{sidewaystable}

While plasmonic PDs offer advantages in sensitivity and speed, it is important to consider other PD technologies and their unique strengths.\@ Photoelectrochemical (PEC) PDs, for example, can operate without an external bias voltage, making them well-suited for low-power applications.\@ Schottky PDs, on the other hand, are known for their simple structure and fabrication, high-speed operation, high sensitivity, and low dark current levels, making them a cost-effective option for many applications \cite{ouyang2024electrode,shi2021status}.\@ Evaluating PD performance involves considering key metrics such as footprint, \emph{R}$_\mathrm{ph}$, sensitivity/detectivity (\emph{D}$^\mathrm{*}$)/noise-equivalent power (NEP), speed/BW, and operating wavelength.\@ These metrics provide a comprehensive overview of each PD structure's capabilities, allowing for a direct comparison of their strengths and weaknesses.\@ Among the diverse approaches being explored, plasmonic PDs have shown promising results, as demonstrated by Syong et al.\@ \cite{syong2024enhanced}.

Syong et al.\@ demonstrated a large-area (400 $\mu$m$^\mathrm{2}$) monolayer MoS$_\mathrm{2}$ PD integrated with a hafnium nitride (HfN) resonant plasmonic metasurface on a sapphire substrate \cite{syong2024enhanced}.\@ The device achieved a record-\emph{I}$_\mathrm{d}$ of 8 pA and a high \emph{D}$^\mathrm{*}$ of 2.58 $\times$ 10$^\mathrm{12}$ Jones.\@ The authors used an Au-assisted exfoliation method and a dry transfer technique to fabricate the large-area monolayer MoS$_\mathrm{2}$ PD.\@ The HfN plasmonic metasurfaces were created through standard nanofabrication techniques involving EBL and dry etching.\@ A 10 nm-thick layer of Al$_\mathrm{2}$O$_\mathrm{3}$ was deposited using atomic layer deposition (ALD) to form a dielectric layer.\@ The monolayer MoS$_\mathrm{2}$ was then transferred onto the fabricated HfN plasmonic metasurfaces.\@ Thermal evaporation and wet etching methods were employed to fabricate the Au contacts.\@ The PD exhibited a significant enhancement in photocurrent compared to the pristine MoS$_\mathrm{2}$ PD.\@ The PD, operating at a wavelength of 660 nm and under a drain voltage of 15 V, exhibited a significant enhancement in photocurrent compared to the pristine MoS$_\mathrm{2}$ PD.\@ The authors attributed this enhancement to the strong localized EM field generated by the resonant plasmonic metasurface.\@ The device also showed a long-term stability of over 40 days.\@ While the use of sapphire substrates in this work is compatible with silicon-on-sapphire (SOS) CMOS technology \cite{kaya2006low,culurciello20068,culurciello2005isolation,culurciello200416}, the incorporation of Au in the fabrication process may pose challenges for integration with standard Si CMOS processes due to its high diffusivity in Si.\@ However, the potential compatibility of 2D semiconductors like MoS$_\mathrm{2}$ with Si CMOS processes, along with the scalability of these materials and HfN across wafer sizes, makes them promising for future optoelectronic applications.\@ Another approach to plasmonic photodetection involves the use of metallic nanoalloys, as shown by Okamoto et al.\@ \cite{okamoto2024facilely}.

Okamoto et al.\@ fabricated a series of plasmonic PDs based on Au-Ag nanoalloys on an n-type Si substrate with varying Au-Ag alloy compositions \cite{okamoto2024facilely}.\@ The devices were fabricated using a cathodic arc plasma deposition method, enabling large-scale manufacturing at a low cost.\@ The process involved depositing Au-Ag nanoalloys on a (111)-oriented Si substrate, followed by the deposition of a 10 nm-thick Ag film on the backside of the substrate as an ohmic contact.\@ A 50 nm-thick Al film was then deposited on the Ag film for protection.\@ Finally, a 220 nm-thick ITO film was deposited on the front side of the substrate using sputtering.\@ The composition of the nanoalloys was controlled by adjusting the Au-Ag ratio in the target material.\@ The fabricated devices exhibited a unique nanostructure consisting of Au-Ag nanoparticles (NPs) and a uniform Au-Ag layer on the Si substrate, separated by a thin SiO$_\mathrm{2}$ layer.\@ The Schottky barrier height at the Si-nanoalloy interface was controlled by adjusting the work function of the nanoalloys through changes in the alloy composition.\@ The device with the Au$_\mathrm{40}$Ag$_\mathrm{60}$ nanoalloy composition showed the highest \emph{R}$_\mathrm{ph}$, achieving 7.3 mA/W at $\lambda$ = 1.31 $\mu$m and 1.9 mA/W at $\lambda$ = 1.55 $\mu$m without any external bias.\@ The enhanced \emph{R}$_\mathrm{ph}$ was attributed to the properly decreased Schottky barrier height and the increased area of the Schottky interface for electron transfer provided by the uniform Au-Ag layer.\@ The IQE of the device with the Au$_\mathrm{40}$Ag$_\mathrm{60}$ nanoalloy composition was 3.3\% at $\lambda$ = 1.31 $\mu$m and 0.52\% at $\lambda$ = 1.55 $\mu$m, which was 4.6 and 6.5 times higher than those of previously reported devices, respectively.\@ The corresponding EQE was 0.69\% at $\lambda$ = 1.31 $\mu$m and 0.152\% at $\lambda$ = 1.55 $\mu$m.\@ The \emph{D}$^\mathrm{*}$ of the device was 8.8 $\times$ 10$^\mathrm{8}$ Jones at $\lambda$ = 1.31 $\mu$m and 2.3 $\times$ 10$^\mathrm{8}$ Jones at $\lambda$ = 1.55 $\mu$m.\@ The authors also investigated the role of the Au-Ag layer in enhancing the efficiency of hot electron injection and found that the large Schottky interface area provided by the uniform layer significantly improved the device performance.\@ While the authors do not explicitly mention CMOS and Si integrability, the fabrication process, which involves standard deposition techniques like cathodic arc plasma deposition and sputtering, suggests potential integrability with CMOS processes.\@ However, the use of Au and Ag NPs in the device structure impedes full integration with standard CMOS fabrication technologies.\@ In addition to plasmonic enhancement, researchers are also exploring CMOS-integrable materials and fabrication processes for PD integration, as exemplified by the work of Mai et al.\@ \cite{mai2024towards}.

Mai et al.\@ presented a CMOS-integrable refractive index sensor based on the monolithic integration of TiN nanohole arrays (NHAs) and Ge photodetectors fabricated on a 200 mm Si substrate \cite{mai2024towards}.\@ The Ge PDs were fabricated using a selective epitaxial growth process, likely involving temperatures around 550 $^\circ$C or higher, as reported in similar processes.\@ Mai et al.\@ did not explicitly state the Ge growth temperature in their work.\@ However, they reference a previous work \cite{lischke2015high} which used a similar Ge epitaxy process.\@ In that paper, the Ge deposition occurs at 550 $^\circ$C with annealing cycles at 800 $^\circ$C to reduce defect density.\@ The PDs, with a 450 nm-thick Ge layer, achieved an \emph{I}$_\mathrm{d}$ density of 129 mA/cm$^\mathrm{2}$.\@ The TiN NHAs were patterned using deep-ultraviolet (DUV) lithography and reactive-ion etching (RIE), placed at a distance of 370 nm from the Ge photodetectors.\@ The integration process did not impact the electrical contact behavior or the yield, which was around 90\%.\@ The devices, operating at $\lambda$ = 1.31 $\mu$m, were characterized by top illumination with both TE and TM polarized light.\@ The photocurrent measurements showed a strong dependence on the polarization of the incident light and the NHA design, confirming the impact of the TiN NHAs on the optical behavior of the PDs.\@ The devices exhibited a \emph{R}$_\mathrm{ph}$ of 0.114 A/W at 1 V reverse bias.\@ The authors also discussed the potential for further device optimization, such as increasing the Ge layer thickness or the angle of incidence for improved \emph{R}$_\mathrm{ph}$.\@ While the integration of Ge PDs with TiN NHAs demonstrates a promising approach for Si and CMOS-integrable refractive index sensors, the high temperatures involved in the Ge epitaxial growth process may require careful management within the thermal budget constraints of the CMOS fabrication flow.\@ AlAloul et al.\@ offer yet another approach to achieving high-performance photodetectors with CMOS integrability \cite{alaloul2021plasmon}.

A recent study by AlAloul et al.\@ introduced a novel graphene-based PD that leverages plasmonic enhancement and CMOS-integrable TiN on an SOI substrate \cite{alaloul2021plasmon}.\@ The device features a Si rib waveguide topped with a graphene sheet and a TiN stripe.\@ The waveguide, with dimensions of 460 nm height and 460 nm width, sits atop a 2 $\mu$m-thick buried oxide layer.\@ The TiN stripe, measuring 20 nm in thickness, spans the width of the waveguide.\@ This PD functions via the photothermoelectric (PTE) effect, where an asymmetric distribution of photoexcited hot carriers in the graphene layer generates a voltage.\@ The TiN stripe amplifies the interaction between the propagating TM mode and the graphene, thereby boosting the optical absorption of graphene and enhancing the photoresponse.\@ The device, operating at $\lambda$ = 1.55 $\mu$m and under zero bias voltage, exhibits a BW$_\mathrm{3-dB}$ exceeding 100 GHz and maintains a consistent photoresponse across the telecom C-band.\@ Notably, it achieves a high \emph{R}$_\mathrm{ph}$ of 1.4 A/W (1.1 A/W external) within a compact 3.5 $\mu$m length.\@ Furthermore, it boasts zero energy consumption and an ultra-low intrinsic NEP of less than 20 pW/Hz$^\mathrm{1/2}$.\@ The fabrication process, utilizing Si and TiN along with standard CMOS and Si fabrication techniques, ensures compatibility with CMOS processes.\@ The TiN film, deposited via DC reactive magnetron sputtering at 150 $^\circ$C \cite{gosciniak2019plasmonic}, further enhances the device's CMOS compatibility across FEOL, MEOL, and BEOL levels.\@ Despite the potential limitations on BW due to the reliance on the PTE effect in graphene, the researchers achieved a high BW by leveraging the unique properties of TiN and the plasmonic enhancement to efficiently extract the photo-generated carriers.

These examples illustrate the ongoing research towards high-performance, compact, and energy-efficient PDs for optical on-chip data communication networks.

\section{CHPWs and Overcoming the Diffraction Limit}

At the telecom wavelength of 1.55 $\mu$m, the diffraction limit for light in Si (\emph{n}$_\mathrm{Si} \approx$ 3.48) is around 500 nm \cite{naik2013alternative}.\@ SPP modes offer a way to overcome this limit by achieving subwavelength confinement at metal-dielectric interfaces.\@ SPPs have a shorter effective wavelength than light in free space or a dielectric medium, leading to this tighter confinement.\@ This is due to the interaction of light with free electrons at the metal-dielectric interface, creating a wave that propagates along the interface with a shorter wavelength.\@ As mentioned earlier in Section \ref{SPP_guided_waves}, SPP-based waveguides inherently exhibit a trade-off between mode confinement and propagation loss.\@ Stronger confinement (smaller mode size) usually comes at the cost of increased propagation losses due to higher interaction with the metal and its inherent ohmic losses.\@ This trade-off is particularly critical for resonator devices like ring modulators, where low-loss propagation is essential for maintaining high \emph{Q}s and strong light-matter interaction is needed for efficient modulation.

CHPW structures offer a promising solution to the challenge of balancing mode confinement and propagation loss in plasmonic waveguides.\@ By carefully engineering the waveguide geometry and material properties, CHPWs can achieve nanoscale device footprints while maintaining microscale propagation lengths and high Purcell factors (\emph{F}$_\mathrm{P}$).\@ \emph{F}$_\mathrm{P}$, which quantifies the enhancement of spontaneous emission rates, is defined as \cite{purcell1946spontaneous}:
\begin{equation}
	F_\mathrm{P} = \frac{3Q_\mathrm{r}}{4\pi^{2}V_\mathrm{m}}\Big(\frac{\lambda_{0}}{n_\mathrm{r}}\Big)^{3},
\end{equation}

\noindent{where \emph{Q$_\mathrm{r}$} is the resonator's quality factor, \emph{V}$_\mathrm{m}$ is the mode volume, $\lambda_\mathrm{0}$ is the free-space wavelength of light, and \emph{n$_\mathrm{r}$} is the refractive index of the resonator material.\@ High \emph{F}$_\mathrm{P}$ values are particularly important for applications requiring low optical losses, such as resonators, where they enhance light-matter interaction and enable phenomena like cavity quantum electrodynamics (CQED).\@ While crucial for resonators, Purcell enhancement also plays a significant role in other optical devices, such as light-emitting diodes (LEDs) \cite{arneson2024color,lee2024top} and single-photon sources \cite{kaupp2023purcell,gritsch2023purcell}, by increasing spontaneous emission rates.}

\begin{figure*}[h]
	\centering
	\includegraphics[width=0.85\linewidth]{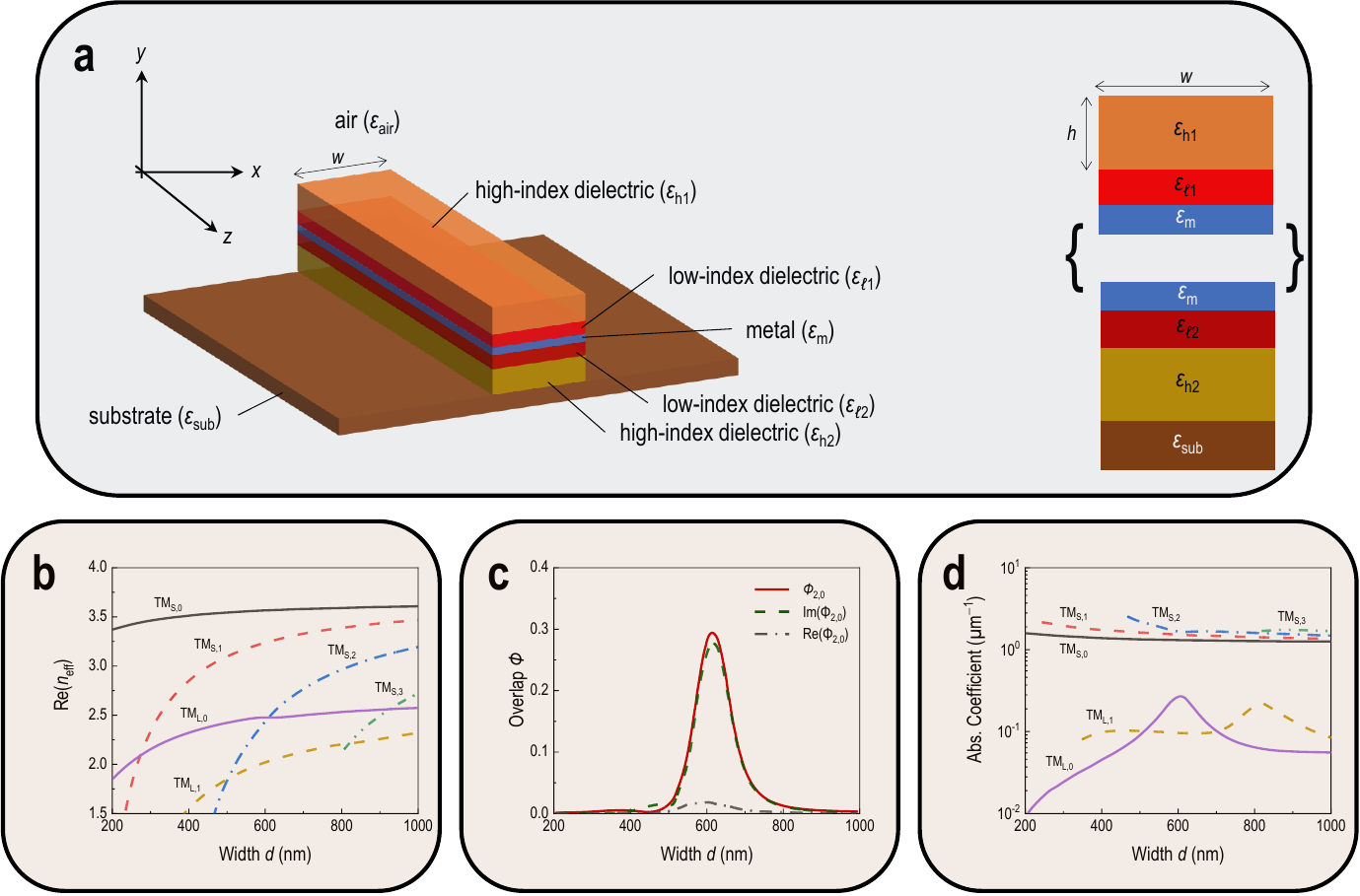}
	\caption{Generalized asymmetric waveguide topology analyzed as a system of two coupled plasmonic waveguides.\@ \textbf{a}, The figure illustrates a material-independent design approach for achieving efficient optical communications on a chip.\@ A CHPW structure, composed of a top high-index dielectric waveguide supporting SPP modes, coupled to a bottom HPW via a thin metal layer, is analyzed as a system of two coupled waveguides.\@ The HPW supports both SPP and TIR modes.\@ By manipulating the geometric and structural asymmetry of the waveguide, the field overlap within the metal layer can be for example maximized, leading to enhanced absorption and higher \emph{R}$_\mathrm{ph}$ levels for photodetection applications.\@ This approach minimizes device volume, reducing parasitic effects and enabling higher RC BW and speed with minimal energy consumption.\@ The design focuses on waveguide structural engineering rather than relying on specific materials, offering a versatile platform for on-chip optical communications.\@ \textbf{b}, \emph{n}$_\mathrm{eff}$ values, \textbf{c}, $\Phi$, and \textbf{d}, $\alpha$ of the supermodes guided by the asymmetrical CHPW structure as functions of the waveguide width.\@ Panels \textbf{b}--\textbf{d} are adapted with permission from Lin et al.\@ \cite{lin2020supermode}.\@ Copyright 2020, American Chemical Society.}\label{CHPW}
\end{figure*}

The dependence of the supermodes on the dimensions of the waveguide is a crucial aspect of CHPW design.\@ For example, varying the top Si thickness (\emph{h}, \figurename{ \ref{CHPW}\textbf{a}}) of the waveguide can tune the modal loss ($\gamma_{i}$) and other properties of the supermodes.\@ As \emph{h} increases, the SPP mode supported by the top Si-Al interface and the HPW mode supported by the bottom Al-SiO$_\mathrm{2}$-Si stack become phase-matched at around \emph{h} = 108 nm, leading to strong coupling between the modes \cite{lin2019photonic}.\@ This coupling results in the formation of two supermodes:\@ the short-range (SR) supermode (TM$_\mathrm{S}$) and long-range supermode (TM$_\mathrm{L}$).\@ The TM$_\mathrm{S}$ has an antisymmetric \emph{E}$_y$ field distribution, which leads to a stronger field overlap in the metal layer and, consequently, higher loss.\@ Conversely, the TM$_\mathrm{L}$ has a symmetric \emph{E}$_y$ field distribution, resulting in a reduced field overlap in the metal layer and lower loss.

The overlap of the \emph{E}$_z$ component with the different layers of the CHPW ($\Phi_\textit{j}$) can be quantified using the following equation \cite{lin2019photonic}:
\begin{equation}
	\Phi_{j}=\int_{t_{j}}\frac{\mathrm{Im}\{E_{z}\}}{P_{z}}dl,
\end{equation}

\noindent{where \emph{j} represents the different waveguide layers, and \emph{t}$_\textit{j}$ is the thickness of the layer.\@ The loss of the TM$_\mathrm{L}$ supermode is minimized when $\Phi_\mathrm{metal}$ is minimized.\@ At this point, the corresponding loss of the TM$_\mathrm{L}$ supermode is significantly smaller than that of the standalone HPW and SPP modes.}

Furthermore, the ability of CHPWs to minimize the transverse E-field component \emph{E}$_y$ within the metal layer (\figurename{ \ref{CHPW}}\textbf{a}), even in asymmetric structures, offers additional design freedom for tailoring the optical response and optimizing performance.\@ As described in equation \ref{field_sym}, the modal loss $\gamma_{i}$ heavily depends on the \emph{E}$_y$ field symmetry within the metal layer \cite{snyder1983optical}:
\begin{equation}\label{field_sym}
	\gamma_{i} = \frac{1}{2} \sqrt{\frac{\varepsilon_{0}}{\mu_{0}}} \frac{\int \varepsilon^{''}_{i} |E_{y}|^{2} \hspace{0.25em} dA}{\int P_{z} \hspace{0.25em} dA},
\end{equation}

\noindent{where $\varepsilon_{0}$ is the permittivity of free space, $\mu_{0}$ is the permeability of free space, $\sqrt{\frac{\varepsilon_{0}}{\mu_{0}}}$ $\approx$ 376.73 $\Omega$ is the impedance of free space, $\varepsilon^{''}_\mathrm{i}$ is the imaginary part of the metal permittivity, \emph{P}$_{z}$ is the Poynting vector in the $z$ axis of propagation of the wave, \emph{E}$_\mathrm{y}$ is the E-field component in the $y$ direction (transverse E-field component).\@ By minimizing \emph{E}$_y$, CHPW structures can achieve low propagation losses even when using an asymmetric geometry.\@ The design and analysis of CHPW structures can be approached through a simplified one-dimensional (1D) model using coupled mode theory and the transfer-matrix method.\@ This approach allows for the optimization of modal losses and other performance metrics, enabling the development of highly efficient and compact plasmonic devices \cite{lin2019photonic}}.

The unique properties of CHPWs have not only facilitated the development of low-loss LR SPP (LRSPP) propagation and ultra-compact photonic devices with high Purcell factors but also enabled the investigation of other intriguing phenomena and applications.\@ For example, Li et al.\@ theoretically calculated and analyzed the coupling modes in a composite HPW, demonstrating that the LR and SR supermodes in CHPW correspond to two different SPPs at two metal layer interfaces \cite{li2023narrow}.\@ The authors proposed a narrow bandwidth perfect absorber based on the CHPW structure, achieving a bandwidth of only 12.9 nm in the visible range and 6.67 nm in the NIR range.\@ The narrow bandwidth is attributed to the low-loss mode resulting from the coupling of localized surface plasmons (LSPs) in the absorber, similar to the coupling mechanism of SPPs in CHPW.\@ This work suggests that the CHPW configuration could facilitate the development of more advanced low-loss, high-performance plasmonic devices.

A crucial aspect of CHPW design is the choice of the mode area definition, which should be based on the intended device application.\@ For active devices, where strong light-matter interaction within localized regions is desired, a phenomenological measure of the mode area is more suitable.\@ Therefore, the effective mode area of the TM$_\mathrm{L}$ supermode ($A_\mathrm{L}$) can be defined as the ratio between the total mode energy density per unit length and the peak energy density \cite{oulton2008confinement}:
\begin{equation}
	A_\mathrm{L} = \frac{1}{\mathrm{max} \{ W(\textbf{r}) \}} \int_{A_{\infty}} W(\mathrm{\textbf{r}})dA,
\end{equation}

\noindent{where \emph{W}(\textbf{r}) is the mode energy density that can be calculated using}
\begin{equation}
	W(\mathrm{\textbf{r}}) = \frac{1}{2} \mathrm{Re}\Bigg\{\frac{d\big(\omega \varepsilon(\textbf{r})\big)}{d\omega}\Bigg\}|\textbf{E}(\textbf{r})|^{2} + \frac{1}{2} \mu_{0}|\textbf{H}(\textbf{r})|^{2}.
\end{equation}

\noindent{Here, $|\textbf{E}(\textbf{r})|^{2}$ and $|\textbf{H}(\textbf{r})|^{2}$ are the electric and magnetic fields, respectively.\@ This definition ensures that the mode area accurately reflects the spatial extent of the optical field, which is critical for achieving efficient light-matter interaction in devices such as EO modulators and photodiodes.\@ The mode area is also dependent on the dimensions of the waveguide.\@ For instance, assuming that the top high-index layer is Si (\figurename{ \ref{CHPW}\textbf{a}}), increasing its thickness, \emph{h}, from 100 to 150 nm can result in a significant reduction of the mode area.\@ This reduction, which can range from 0.01 to 0.002 $\mu$m$^\mathrm{2}$, can be attributed to the stronger confinement of the optical mode within the waveguide as the coupling between the SPP and HPW modes becomes stronger.\@ To gain a deeper understanding of these mode confinement effects, readers are encouraged to consult the references cited for a more in-depth analysis of the modeling results and a more extensive discussion \cite{lin2019photonic,oulton2008confinement,snyder1983optical}.}

The unique design of CHPWs, which can incorporate both low-loss and high-loss regions, makes them particularly suitable for applications requiring both efficient light transmission and strong light-matter interaction, such as modulators.\@ By engineering the field symmetry and exploiting the properties of different materials, CHPWs enable the creation of supermode hybridization, leading to enhanced performance and greater design flexibility.\@ This approach allows for optimizing waveguide losses and developing a versatile platform for various photonic applications, such as modulators and PDs.\@ The ability to achieve field symmetry in asymmetric structures further expands the design space and offers new possibilities for miniaturization and functional integration of photonic devices.

The versatility of CHPWs extends beyond low-loss applications.\@ By tuning the waveguide dimensions and materials, CHPWs can also be designed for high-loss operation, as required in PDs.\@ For instance, increasing the width of the waveguide can maximize the field overlap inside the metal layer, leading to enhanced absorption and higher \emph{R}$_\mathrm{ph}$ levels.\@ This capability enables a new paradigm for on-chip optical communications, where both low-loss (e.g., for modulators) and high-loss (e.g., for PDs) components can be realized within the same platform, facilitating monolithic integration and reducing complexity \cite{lin2020monolithic}.\@ Note that the width of the waveguide can be adjusted to fine-tune the absorption of the TM$_\mathrm{L}$ supermode.\@ Specifically, the absorption of the TM$_\mathrm{L}$ supermode is sensitive to the waveguide width (\emph{w}) and increases drastically with \emph{w}, reaching a peak at, say, \emph{w} = 620 nm \cite{lin2019photonic}.\@ This high absorption is due to a combination of factors, including the coupling of higher-order SPP and HPW modes and the hybridization between the TM$_\mathrm{L, 0}$ and TM$_\mathrm{S, 2}$ supermodes.

While high absorption is desirable for PDs, modulators require a different approach.\@ It is important to note that modulator operation requires both lower losses in the ON state (light passes) and higher losses in the OFF state (no light passes).\@ CHPWs can be exploited to create these desired effects by tailoring the mode hybridization and field confinement, thus enabling the design and fabrication of efficient modulators \cite{lin2015dynamically,lin2020monolithic,su2017highly}.\@ The ability to manipulate geometric and structural asymmetry in CHPWs, without being constrained by specific materials, opens up a vast design space for tailoring waveguide properties.\@ This supermode hybridization approach allows for the optimization of various device characteristics, including modal losses, propagation lengths, and field confinement, leading to the development of next-generation integrated photonic devices with enhanced performance, functionality, and versatility \cite{dhingra2023high,chatzitheocharis2022efficient}.

In summary, the CHPW architecture relies on the strategic coupling of different plasmonic modes at a single metal interface, a concept known as supermode hybridization.\@ This versatile design strategy allows for the engineering of both SR and LR modes, as depicted in \figurename{ \ref{CHPW_versatility}.\@ This flexibility can be leveraged to address the design challenges in integrated Schottky PDs and modulators, enabling a wide range of functionalities within a single platform.\@ As diagrammed in \figurename{ \ref{CHPW_versatility}}, the CHPW architecture, through the manipulation of physical dimensions and the resulting supermode hybridization, offers a pathway towards the monolithic integration of plasmonic optical circuits.\@ By tuning the waveguide dimensions and leveraging the field overlap within the metal layers, the CHPW platform can be tailored for both high-loss and low-loss applications.\@ This versatility enables the co-integration of diverse functionalities, such as photodetection and modulation, on a single chip, leading to compact and efficient optoelectronic systems.\@ Moreover, the CHPW's unique ability to support both SR and LR modes through supermode hybridization allows for optimizing device performance across various applications.\@ This design flexibility not only addresses the limitations of traditional plasmonic waveguides but also opens up new avenues for the development of next-generation integrated photonic devices with enhanced functionality and reduced complexity.

\begin{figure*}[h]
	\centering
	\includegraphics[width=0.85\linewidth]{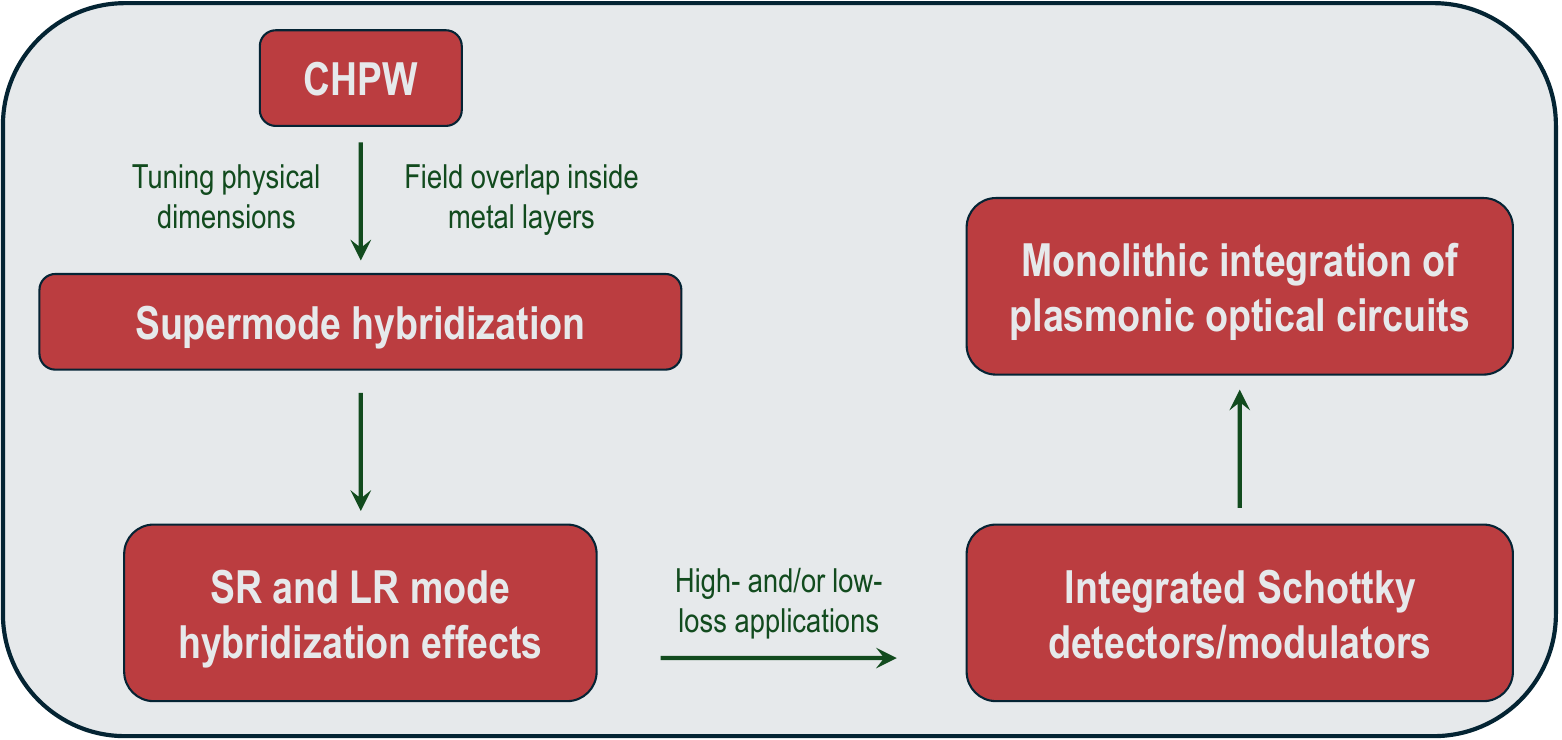}
	\caption{CHPW as a versatile platform for integrated photonics---The diagram illustrates the key design features and potential applications of CHPWs.\@ By tuning physical dimensions, CHPWs can be engineered to support supermode hybridization, enabling both low-loss and high-loss applications.\@ The resulting SR and LR mode hybridization effects can be leveraged for the monolithic integration of various photonic components, including Schottky PDs and modulators, on a single chip.}\label{CHPW_versatility}
\end{figure*}

\section{Optical Mode Hybridization}\label{hybridization}

This section delves deeper into the area of optical mode hybridization, an approach that merges the strengths of both plasmonic and dielectric waveguides to overcome limitations such as high propagation losses in plasmonics and weak confinement in dielectrics.\@ By strategically manipulating the interaction between these distinct modes, researchers can achieve precise control over light, paving the way for a new generation of high-performance, Si-compatible photonic devices.\@ We will explore the strategies in this domain, specifically the utilization of CHPWs.\@ This is achieved through the strategic engineering of field symmetry in asymmetric waveguide structures.\@ Both approaches offer unique advantages and design flexibilities, enabling the realization of diverse functionalities with enhanced optical properties.

\subsection{Engineering field symmetry in asymmetric waveguides}

To overcome the limitations of traditional plasmonic materials and device architectures, researchers have explored innovative optical mode hybridization approaches, aiming to leverage the strengths of both plasmonic and dielectric waveguide modes.\@ By strategically engineering the interaction between these modes, it is possible to tailor the optical properties of waveguides and realize a new class of high-performance, Si-integrable plasmonic devices.\@ In this section, we delve into two key approaches that have shown significant promise in achieving this goal: the utilization of CHPWs and the engineering of field symmetry in asymmetric waveguide structures.

CHPWs, as depicted in \figurename{ \ref{SPP_CHPW}\textbf{a}}, consist of two distinct plasmonic waveguides vertically coupled via a thin metal layer.\@ The top waveguide, often made of a high-index dielectric material, primarily supports SPP modes, while the bottom waveguide exhibits hybrid behavior, supporting both SPP and TIR modes.\@ The interaction between these waveguides results in the formation of coupled supermodes, which exhibit characteristics distinct from the individual SPP and HPW modes.\@ The key advantage of CHPWs lies in the constructive interference between the LR HPW mode and the SR SPP mode, leading to a hybrid mode with significantly reduced propagation losses.\@ Simultaneously, the coupling enhances the E-field intensity within the low-index dielectric layer, enabling subwavelength field confinement crucial for miniaturization.\@ Thus, CHPWs offer a pathway to achieve both low-loss propagation and high field confinement, overcoming the limitations of traditional plasmonic waveguides.

\begin{figure*}[h]
	\centering
	\includegraphics[width=0.90\linewidth]{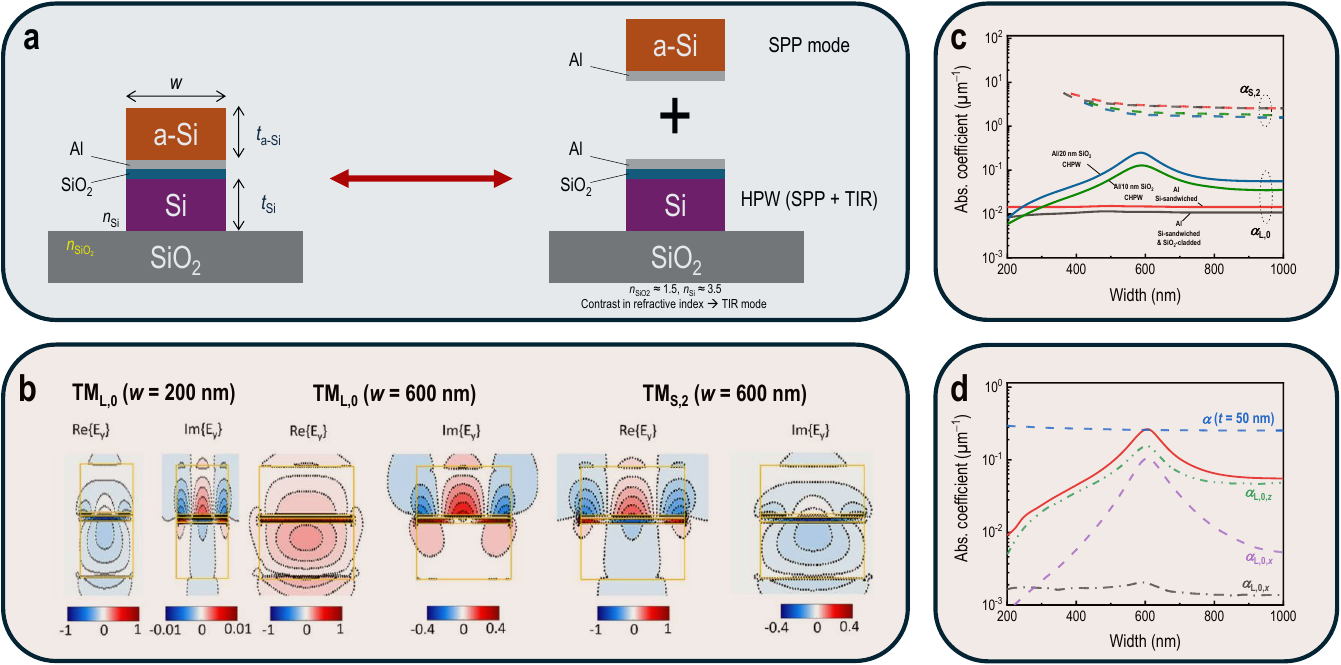}
	\caption{CHPW architecture.\@ \textbf{a}, Cross-sectional schematic illustration of a CHPW architecture.\@ The CHPW consists of a top waveguide (a high-index dielectric material) supporting SPP modes, and a bottom HPW supporting both SPP and TIR modes.\@ The two waveguides are coupled via a thin metal layer, enabling the formation of coupled supermodes with reduced propagation losses and enhanced field confinement.\@ \textbf{b}, Modal properties of a CHPW depicting field distributions of the two guided supermodes, TM$_{\mathrm{L,0}}$ and TM$_{\mathrm{S,2}}$, at $\lambda$ = 1.55 $\mu$m.\@ \textbf{c}, The impact of structural asymmetry $\alpha$ of the TM$_{\mathrm{L,0}}$ and TM$_{\mathrm{S,2}}$ supermodes.\@ The absorption coefficients are plotted for different configurations of the CHPW structure, including a completely symmetrical structure and asymmetrical structures with varying waveguide widths.\@ \textbf{d}, Breakdown of the absorption coefficient ($\alpha$) of the TM$_{\mathrm{L,0}}$ mode into contributions from its different E-field components (E$_x$, E$_y$, and E$_z$).\@ The absorption coefficient of a representative Schottky PD with a 50 nm metal layer on a Si nanowire is also shown for comparison.\@ Panel \textbf{b}--\textbf{d} are adapted with permission from Lin et al.\@ \cite{lin2020supermode}.\@ Copyright 2020, American Chemical Society.}
	\label{SPP_CHPW}
\end{figure*}

The CHPW architecture, as illustrated in \figurename{ \ref{SPP_CHPW}\textbf{a}}, offers a high degree of versatility and tunability.\@ By manipulating the waveguide dimensions and leveraging the interplay between different plasmonic modes, CHPWs can be tailored to exhibit specific modal characteristics for a wide range of applications.\@ Notably, the CHPW platform exhibits the unique capability to support both low-loss and high-loss modes through the coupling and hybridization of different supermodes.

Furthermore, \figurename{ \ref{CHPW}\textbf{c}--\textbf{e}} show how the effective indices (\emph{n}$_\mathrm{eff}$), field overlap strength ($\Phi$), and absorption coefficients ($\alpha$) of the TM$_{\mathrm{L,0}}$ and TM$_{\mathrm{S,2}}$ modes can be controlled by varying the width of the waveguide.\@ This tunability allows for achieving specific modal properties and optimizing the performance of CHPWs for various applications.\@ For instance, by maximizing the field overlap strength between the TM$_{\mathrm{L,0}}$ and TM$_{\mathrm{S,2}}$ modes, one can create a highly efficient PD with enhanced absorption due to the increased interaction with the metal.\@ Conversely, by minimizing the overlap and optimizing the structural asymmetry, one can achieve low-loss propagation suitable for other functional on-chip device applications.

\figurename{ \ref{SPP_CHPW}\textbf{c}} shows how varying the waveguide width allows for engineering both the low-loss (TM$_{\mathrm{L,0}}$) and high-loss (TM$_{\mathrm{S,2}}$) modes.\@ The absorption coefficients of these modes can be tuned by adjusting the degree of structural asymmetry in the CHPW, as depicted in \figurename{ \ref{SPP_CHPW}\textbf{c}}.\@ For instance, increasing the asymmetry by introducing a 10 nm-thick SiO$_\mathrm{2}$ layer at the bottom Schottky interface results in a 13-fold enhancement of the absorption coefficient $\alpha_{\mathrm{L,0}}$ for the TM$_{\mathrm{L,0}}$ mode.\@ This enhancement arises from increased mode hybridization, leading to a stronger interaction with the metal layer.\@ Further increasing the SiO$_\mathrm{2}$ thickness to 20 nm amplifies $\alpha_{\mathrm{L,0}}$ even more.\@ This tunability allows for tailoring the optical properties of the CHPW to match specific application requirements.\@ The low-loss TM$_{\mathrm{L,0}}$ mode exhibits a remarkably low propagation loss due to its reduced overlap with the metal layer, as depicted in \figurename{ \ref{SPP_CHPW}\textbf{b}}.\@ The absorption coefficient of the TM$_{\mathrm{L,0}}$ mode can be further understood by examining the contributions from its different E-field components, as shown in \figurename{ \ref{SPP_CHPW}\textbf{d}}.\@ This analysis reveals that the \emph{E}$_\mathrm{y}$ component, which is dominant in the TM$_{\mathrm{L,0}}$ mode, contributes minimally to the absorption.\@ In contrast, the \emph{E}$_\mathrm{x}$ and \emph{E}$_\mathrm{z}$ components, which have a stronger presence near the metal layer, lead to increased absorption.\@ By engineering the waveguide structure to minimize the \emph{E}$_\mathrm{x}$ and \emph{E}$_\mathrm{z}$ components, one can further reduce the propagation loss of the TM$_{\mathrm{L,0}}$ mode and enhance the overall performance of the CHPW.

So far, we have discussed a CHPW design optimized for photodetection, where maximizing optical losses is crucial.\@ By tuning the waveguide width ($w$), the field overlap within the metal layer ($\varepsilon_\mathrm{m}$) can be maximized, enhancing absorption and therefore the \emph{R}$_\mathrm{ph}$ levels of the PD (\figurename{ \ref{CHPW}\textbf{a}}).\@ The CHPW architecture's ability to minimize device volume translates to reduced parasitic effects, resulting in higher RC BW and speed while maintaining minimal energy consumption, a critical advantage for high-performance photodetection applications.\@ The specific width of 600 nm, mentioned in the previous case, highlights an example where this optimization strategy leads to significantly increased optical losses within the metal layer, ideal for efficient light detection.

Nanophotonics research has recently focused on leveraging asymmetric waveguide structures to achieve LR plasmonics, which offer the benefits of both nanoscale device footprints and microscale wave propagation.\@ This approach enables the realization of high Purcell factors (F$_\mathrm{P}$).\@ One promising approach in this area is the development of CHPWs, which leverage the concept of engineering field symmetry by controlling light confinement through careful material choices and geometric design.\@ This approach enables CHPWs to sustain wave propagation over several hundred microns, overcoming the typical limitation of high losses associated with traditional plasmonic waveguides.

While conventional wisdom suggests that higher degrees of waveguide geometric symmetry are required for achieving field symmetry conditions, CHPWs demonstrate that this is not necessarily the case.\@ The Helmy group at the University of Toronto has pioneered the development of asymmetric waveguide architectures that achieve field symmetry through a coupled system approach.\@ In this paradigm, the entire waveguide structure is treated as a coupled system, allowing for field symmetry even in the absence of strict geometric symmetry.\@ This opens up a greater degree of freedom in waveguide design, enabling engineers to optimize modal losses by manipulating parameters such as the thickness of a-Si layers.\@ The design topology of these asymmetric waveguides revolves around simultaneously maintaining nanoscale device footprints and microscale wave propagation while achieving high Purcell factors for low-loss applications.\@ This approach has led to the development of ultra-LR plasmonic waveguides that offer a promising platform for various nanophotonic applications, including ring resonators, modulators, and PDs.

\begin{figure}[h]
	\centering
	\includegraphics[width=\linewidth]{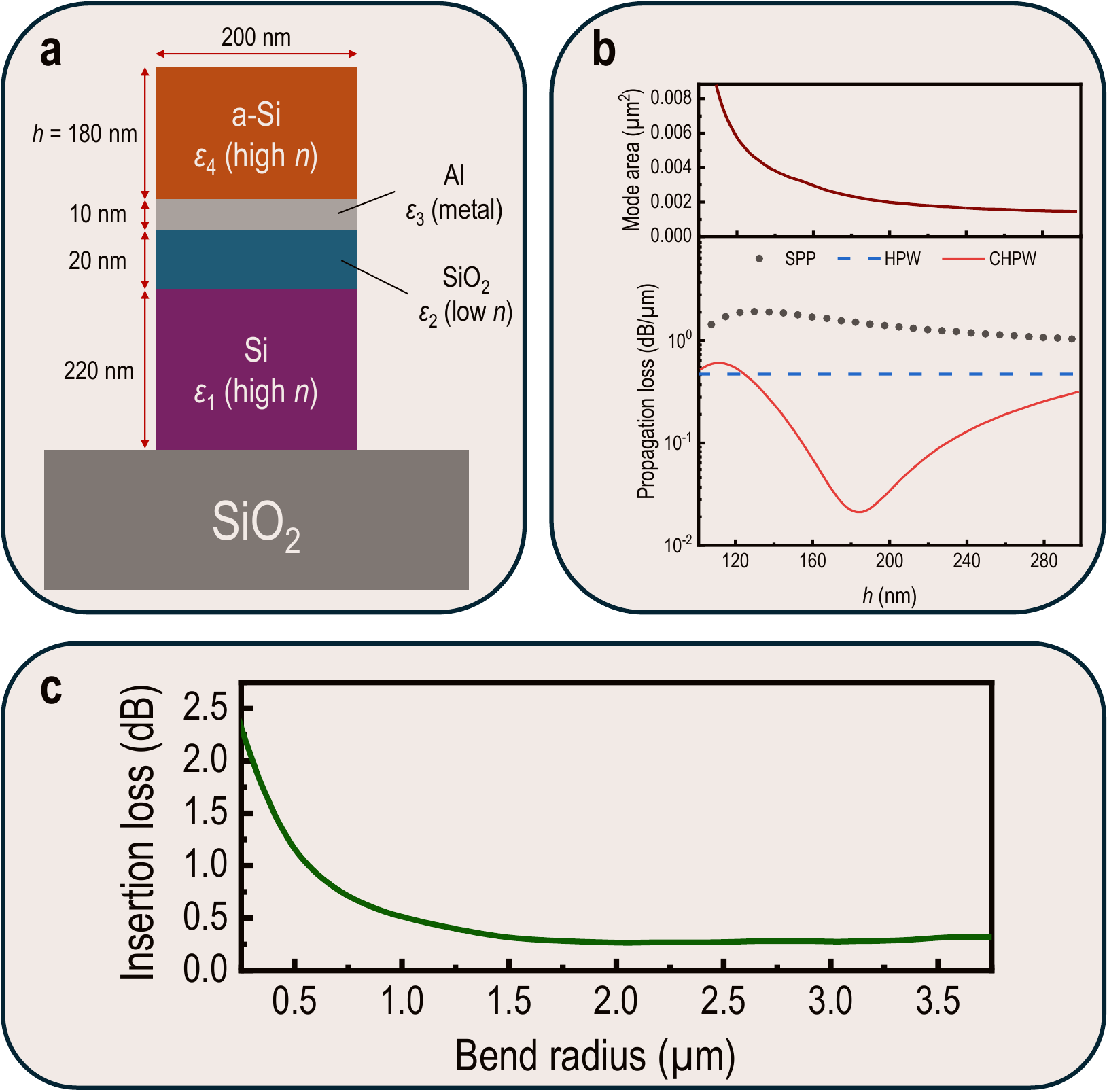}
	\caption{Modal properties of a CHPW.\@ \textbf{a}, Schematic cross-section of a CHPW core comprising a four-layer stack of Si/SiO$_\mathrm{2}$/Al/a-Si.\@ The overlaid modal E-field intensity profile highlights the confinement of the LR mode within the low-index SiO$_\mathrm{2}$ layer.\@ \textbf{b}, Calculated modal area and propagation loss of the LR mode as a function of the a-Si layer thickness.\@ The optimal thickness for minimizing loss corresponds to the minimum flux within the metal layer.\@ \textbf{c}, Simulated IL levels for 90$^\circ$ bends in the CHPW, indicating an optimal bend radius range.\@ All panels are adapted from Su et al.\@ \cite{su2019record} in accordance with a Creative Commons Attribution NonCommercial License 4.0 (CC BY-NC) (https://creativecommons.org/licenses/by-nc/4.0/).\@ Copyright 2019, The Authors, published by American Association for the Advancement of Science.}
	\label{Su_2DHPW}
\end{figure}

In our work published in Science Advances (Su et al.) \cite{su2019record}, we demonstrated a novel CHPW architecture that leverages the strategic coupling of dissimilar plasmonic modes at a single metal interface.\@ \figurename{ \ref{Su_2DHPW}\textbf{a}} illustrates the design and properties of a CHPW consisting of a four-layer stack of Si (high index), SiO$_\mathrm{2}$ (low index), Al (metal), and a-Si (high index), while \figurename{ \ref{Su_2DHPW}\textbf{b}} presents the calculated modal area and propagation loss of the LR mode as a function of the high-index a-Si layer thickness.\@ The optimal a-Si thickness (\emph{h} = 185 nm) for minimizing propagation loss corresponds to the point where the optical power flux within the metal layer is minimized.\@ This approach allows for significant reduction in the propagation losses of modes in a wide range of CHPW structures, without imposing any symmetry constraints, while maintaining a highly localized effective mode area of 0.002 $\mu$m$^2$.\@ The LR mode in this waveguide is primarily confined within the low-index SiO$_\mathrm{2}$ layer.\@ \figurename{ \ref{Su_2DHPW}\textbf{c}} presents simulations of the IL for 90$^\circ$ bends in the CHPW, indicating an optimal bend radius range between 1.5 and 3 $\mu$m, and demonstrates the critical coupling condition for a CHPW ring resonator in an all-pass filter configuration.\@ These results collectively highlight the potential of CHPWs for achieving low-loss, highly confined optical modes in Si-integrable plasmonic devices.\@ This design effectively reduces the overlap of the optical field with the lossy metal while simultaneously confining the power within a subwavelength area.\@ The CHPW platform does not require any specific structural or modal symmetry, thereby providing significantly relaxed fabrication tolerances.\@ This unique characteristic allows for the realization of LR plasmonic modes in any platform, free from material or structural constraints.\@ Experimental validation of the proposed CHPW structures has yielded record-low propagation losses and record-high \emph{Q}/\emph{V}$_\mathrm{eff}$ values.\@ As a proof of concept, we have only investigated how the thickness of the top a-Si layer can influence the supermode attributes.\@ Table 1 in the paper compares the experimental attributes of various traveling-wave optical microcavities.\@ It is evident that CHPW rings with a radius of 2.5 $\mu$m outperform their plasmonic and dielectric counterparts of similar radii, achieving \emph{Q}/\emph{V}$_\mathrm{eff}$ and extinction ratios (6507 and 29 dB, respectively) that are an order of magnitude higher compared with ultrahigh-\emph{Q} dielectric cavities \cite{frankis2019high,yang2018bridging,spencer2014integrated,zhu2014submicron,tseng2013study,zhu2012performance}.\@ This achievement effectively mitigates the detrimental effects of ohmic losses inherent to plasmonic waveguides.\@ We anticipate that further optimization of the waveguide layers or an extension to a whispering-gallery disk structure could lead to additional performance enhancements.

In summary, while plasmonic waveguides facilitate the design of highly compact photonic components for ICs due to their subwavelength confinement capabilities, they also suffer from increased losses compared to traditional TIR waveguides.\@ The choice between these two waveguide types depends on the specific application requirements.\@ If miniaturization and high-field confinement are paramount, plasmonic waveguides are the preferred choice.\@ However, if low-loss transmission over long distances is the primary concern, TIR waveguides are more suitable.\@ The development of Si-integrable plasmonic materials and device architectures aims to bridge this gap by combining the benefits of both worlds, offering a path towards miniaturized, low-loss, and highly integrated photonic devices.

\section{Trade-Offs in Si-Integrated Plasmonic Devices}\label{tradeoffs}

Si-integrable plasmonics offer the potential to integrate the unique capabilities of plasmonic devices, such as high-speed operation and miniaturization, with the mature and cost-effective CMOS fabrication platform \cite{burla2019500,ummethala2019thz,heni2020ultra}.\@ However, this integration comes with inherent trade-offs that need to be carefully considered and addressed in device design and optimization.\@ This section delves deeper into these trade-offs, exploring the advantages and challenges associated with Si-based plasmonic devices.

One of the primary challenges in Si-integrable plasmonics is the relatively high optical losses associated with CMOS-compatible metals like Al.\@ While noble metals such as Au and Ag exhibit low optical losses due to their unique electronic structure and high conductivity, their incompatibility with standard CMOS fabrication processes limits their integration with Si photonics.\@ At telecommunication wavelengths (around $\lambda$ = 1.55 $\mu$m), Al waveguides exhibit relatively high propagation losses, typically exceeding several dB/cm, rendering them impractical for most applications requiring long-distance signal transmission.\@ This high loss is primarily attributed to the metal's significant absorption coefficient in this wavelength range.\@ While precise loss values can vary based on waveguide geometry and fabrication techniques, Al's high conductivity contributes to significant signal attenuation.\@ In contrast, dielectric waveguides, relying on TIR for light confinement, offer significantly lower propagation losses, typically on the order of 0.1 dB/km or less.\@ However, these waveguides have limited ability to achieve subwavelength confinement, with modal areas typically in the range of 0.2–0.5 $\mu$m$^2$.\@ Non-Si-compatible plasmonic waveguides, utilizing noble metals, can achieve both relatively low losses (e.g., well-fabricated Ag waveguides at $\lambda$ = 1.55 $\mu$m can exhibit losses on the order of 0.1 dB/$\mu$m) and excellent field confinement, with modal areas readily reaching below 0.01 $\mu$m$^2$.\@ However, it is important to note that this performance is sensitive to factors such as waveguide geometry (e.g., width, height, and cladding materials) and fabrication techniques (e.g., lithography resolution, sidewall roughness, and deposition precision).\@ Achieving such strong confinement often involves a trade-off with propagation loss, as increased confinement can lead to higher scattering and absorption \cite{corato2024absorption,kharintsev2024photon}.\@ This is further exacerbated by the possible presence of surface states and oxygen vacancies \cite{alaloul2021low,alfaraj2020time}, which can act as scattering centers and increase material absorption, particularly in silver.\@ Furthermore, the material properties of Ag, particularly its susceptibility to oxidation \cite{mcmillan1962higher}, can influence long-term performance and necessitate protective measures like passivation layers.

Plasmonic devices, due to their inherent ohmic losses and the need for active modulation or switching mechanisms, can also have higher power consumption compared to their purely dielectric counterparts.\@ This poses a challenge for applications where energy efficiency is critical.\@ Additionally, the material properties of Si-integrable plasmonic materials can lead to limitations in the operating BW of devices.\@ Material dispersion, where the refractive index varies with wavelength, can cause pulse broadening and limit the achievable data rates in optical communication systems.\@ In contrast, noble metals offer broader operating BW due to their favorable dispersion characteristics.

Besides performance considerations, Si-based plasmonic devices also face challenges related to fabrication and design complexity.\@ Integrating plasmonic structures with CMOS technology requires careful consideration of material compatibility, process temperatures, and potential interactions between different device layers.\@ This can increase the complexity of fabrication processes and necessitate additional steps, such as the use of barrier layers or specialized deposition techniques.\@ Furthermore, optimizing the performance of Si-based plasmonic devices involves navigating a complex trade-off space between various factors, including material properties, device dimensions, and fabrication constraints.\@ For example, reducing the device footprint to achieve higher integration density can lead to increased optical losses or higher power consumption.\@ Efficient thermal management strategies are also essential to ensure stable operation and prevent device degradation due to the heat generated by ohmic losses.

\tablename{ \ref{tab:waveguide_comparison}} summarizes the key performance metrics and trade-offs associated with Si-integrable plasmonic waveguides, traditional dielectric waveguides, and non-Si-integrable plasmonic waveguides.\@ It highlights the strengths and weaknesses of each approach, emphasizing the need for careful consideration of the specific application requirements when choosing the appropriate technology.

\begin{table*}[h]
	\centering
	\caption{Comparison of key performance metrics and trade-offs associated with Si-integrable plasmonic waveguides, traditional dielectric waveguides, and non-Si-integrable plasmonic waveguides.\label{tab:waveguide_comparison}}
    \renewcommand{\arraystretch}{1} 
	\begin{tabular*}{\textwidth}{@{\extracolsep\fill}llll@{\extracolsep\fill}}
		\toprule
		\textbf{Feature} & \textbf{Si-integrable plasmonic} & \textbf{Dielectric waveguides} & \textbf{Non-Si-integrable plasmonic} \\
		\midrule
		Propagation loss (dB/$\mu$m) & 0.02--0.03 (CHPWs) \cite{su2019record,lin2020monolithic}, > 0.1 (Al) & < 0.00001 (SiO$_\mathrm{2}$) & $\sim$0.1 (Ag) \\
		Modal area ($\mu$m$^\mathrm{2}$) & 0.002--0.02 (CHPWs) \cite{su2019record,lin2020monolithic} & 0.2--0.5 & < 0.01 \\
		BW & Sub-1 THz  & Up to 100 THz & Up to 500 THz \\
		CMOS integration & Seamless (MEOL, BEOL) or full (Al, TiN) & Seamless & Challenging \\
		Fabrication complexity & Low to moderate & Low & High \\
		Cost & Low & Low & High \\
		Scalability & High & High & Limited \\
		Material losses & Moderate (ITO, TiN), High (Al) &      Very low (Si, SiO$_\mathrm{2}$) & Low (Ag, Au) \\
		Confinement-loss trade-off & Significant & Less significant & Significant \\
		Sensitivity to fabrication & Moderate & Low & High \\
		Long-term stability$^\dagger$ & Material degradation risk & High & Material degradation risk \\
		\bottomrule
	\end{tabular*}
	\begin{tablenotes}
		\item $^\dagger$Oxidation of Ag in non-Si-integrable plasmonics, or changes in the properties of ITO in Si-integrable plasmonics due to environmental factors or operational stress.\@ Dielectric waveguides generally exhibit high long-term stability.
	\end{tablenotes}
\end{table*}

Our earlier research investigated the integration of ITO into Si-integrable plasmonic devices for modulation \cite{alfaraj2023facile}.\@ We utilized the ENZ behavior of ITO, where its permittivity approaches zero, to enhance light-matter interaction and achieve modulation.\@ Two device structures were fabricated and characterized.\@ The first was a Si-integrated Al/ITO/SiO$_\mathrm{2}$/TiN metal--insulator--semiconductor--metal (MISM) heterojunctions (\figurename{ \ref{facile_schematic_updated_01}\textbf{c}}), used to investigate the electrical properties of ITO layers and their response to applied bias voltages.\@ The polycrystalline TiN (poly-TiN) layer served as a conductive template and bottom electrode, enabling the growth of ITO and SiO$_\mathrm{2}$ layers compatible with standard Si foundry and CMOS fabrication processes.\@ The second structure was a SiO$_\mathrm{2}$/ITO-based CHPW modulator on SOI (\figurename{ \ref{facile_schematic_updated_01}\textbf{d}}).\@ This design, incorporating an upon-bias graded-index ITO layer, enabled dynamic reconfiguration for amplitude, phase, or 4-quadrature amplitude modulation.\@ Focusing on the performance of the second structure, the device exhibited an ER of 1 dB/$\mu$m and an IL of 0.128 dB/$\mu$m for a 10 $\mu$m waveguide length.\@ It is important to consider that this relatively high IL value highlights the inherent trade-off between achieving strong light-matter interaction for modulation and minimizing propagation losses in plasmonic devices.\@ The CHPW leveraged the hybridization of plasmonic modes, which exhibit combined dielectric and plasmonic characteristics, allowing for optical field confinement and enhanced light-matter interaction.\@ However, as discussed earlier, the confinement achievable in these hybrid modes is still limited in comparison to conventional plasmonic modes achievable with noble metals.\@ By applying a bias voltage, the carrier concentration in the ITO layer was modulated, leading to a change in its permittivity and a shift in the effective index, enabling modulation of the optical signal.

This research demonstrated the feasibility of integrating ITO into Si-integrable plasmonic devices for light modulation \cite{alfaraj2023facile}.\@ The focus was on exploiting the ENZ properties of ITO to enhance light-matter interaction.\@ However, it is crucial to acknowledge that the performance of ITO-based plasmonic devices, particularly in terms of propagation loss, can be significantly affected by factors such as material quality, surface states, and oxygen vacancies.\@ The two fabricated device structures highlighted the potential of ITO in this context.\@ The first structure, the Si-integrated Al/ITO/SiO$_\mathrm{2}$/TiN MISM heterojunction, served to investigate the electrical properties of ITO layers.\@ The use of a poly-TiN layer facilitated the growth of ITO and SiO$_\mathrm{2}$ layers compatible with CMOS processes.\@ The second structure, the SiO$_\mathrm{2}$/ITO-based CHPW modulator, demonstrated plasmonic modulation using oxide materials.\@ The CHPW design enabled dynamic reconfiguration for various modulation formats.\@ While the achieved ER of 1 dB/$\mu$m is noteworthy, it is essential to consider the accompanying IL of 0.128 dB/$\mu$m, which can lead to significant signal attenuation over longer waveguide lengths.\@ This underscores the ongoing challenge of balancing modulation efficiency with propagation loss in ITO-based plasmonic devices.\@ The efficient modulation was attributed to the hybridization of plasmonic modes, which offer a compromise between field confinement and propagation loss compared to pure plasmonic and pure dielectric modes.\@ By applying a bias voltage, the carrier concentration within the ITO layer was modulated, affecting its permittivity and the effective index of the supermodes.

\begin{figure*}[h]
	\centering
	\includegraphics[width=0.85\linewidth]{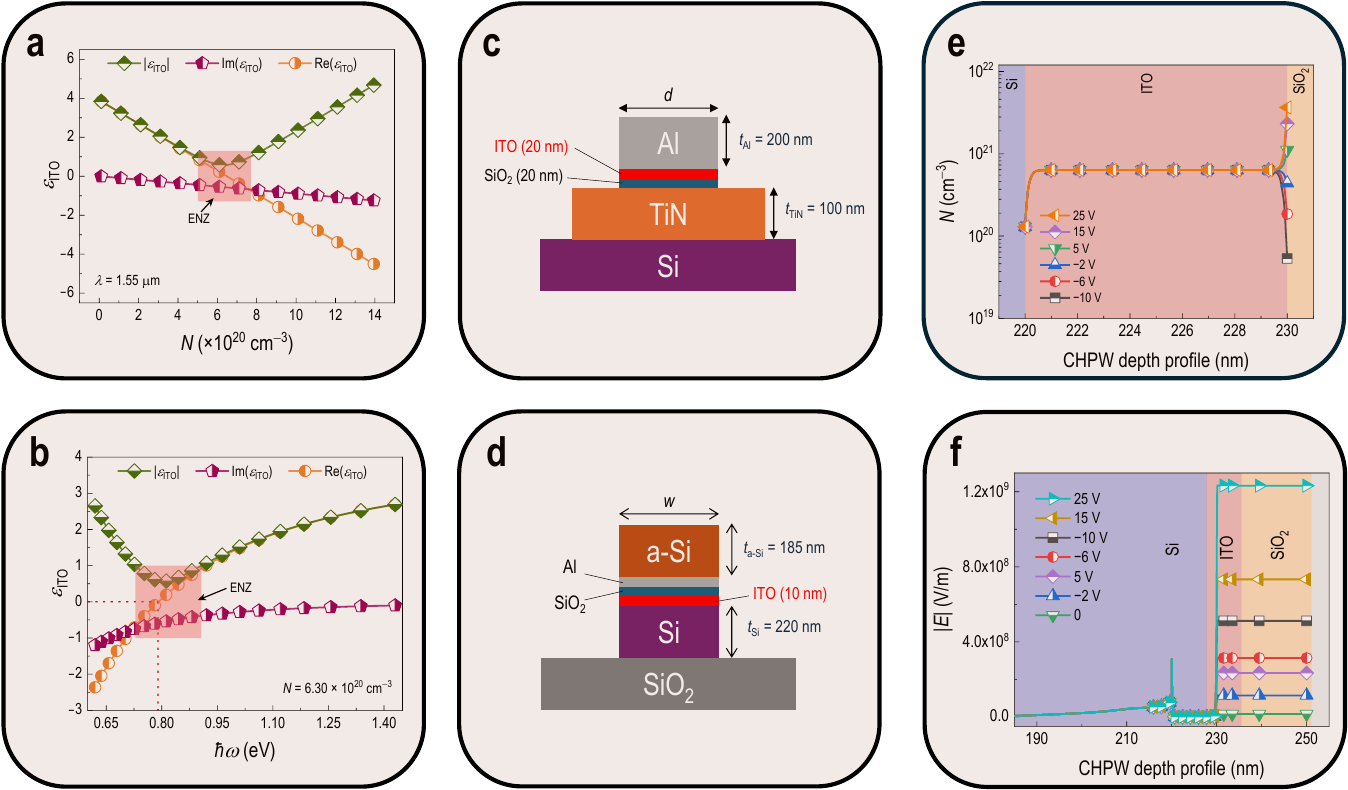}
	\caption{ITO optoelectronic properties and SOI-integrated CHPW modulator.\@ Panels \textbf{a},\textbf{b} illustrate the optical properties of ITO, calculated using the Drude model:\@ \textbf{a}, The real and imaginary parts, alongside the magnitude, of ITO permittivity as a function of accumulation layer carrier density at $\lambda = 1.55$ $\mu$m.\@ \textbf{b}, ITO permittivity as a function of light energy, with the ENZ regime highlighted with shading.\@ An optical modulator device platform and test structures are shown:\@ \textbf{c}, A schematic cross-section of a Si-integrated Al/ITO/SiO$_\mathrm{2}$/TiN MISM structure, and \textbf{d}, a representation of a multilayer SiO$_\mathrm{2}$/ITO-based CHPW modulator integrated on SOI.\@ \textbf{e}, The carrier concentration distribution within the ITO layer of the CHPW modulator under different bias voltages.\@ \textbf{f}, The E-field magnitude $|E|$ distribution across the CHPW modulator under different bias voltages.\@ Panels \textbf{c}--\textbf{f} are adapted from Alfaraj et al.\@ \cite{alfaraj2023facile} in accordance with a Creative Commons Attribution 4.0 International License (https://creativecommons.org/licenses/by/4.0/).\@ Copyright 2023, The Authors, published by Light Publishing Group.}
	\label{facile_schematic_updated_01}
\end{figure*}

\figurename{ \ref{facile_schematic_updated_02}} provides further insights into the performance and experimental validation of the ITO-based modulator devices.\@ \figurename{ \ref{facile_schematic_updated_02}\textbf{a}} shows the measured frequency response curves of fabricated SiO$_\mathrm{2}$/ITO MISM test structures with different ITO carrier concentrations (\emph{N}$_\mathrm{ITO}$) as functions of bias voltage, demonstrating modulation capabilities at different carrier concentrations and bias levels.\@ \figurename{ \ref{facile_schematic_updated_02}\textbf{b}} illustrates the simulated change in propagation loss (PL) levels of the waveguide as a function of bias voltage, highlighting operation within the ENZ regime.\@ The plot reveals a significant change in PL levels with varying bias, indicating potential for efficient modulation.\@ \figurename{ \ref{facile_schematic_updated_02}\textbf{c}} presents the simulated change in both the real and imaginary parts of the effective refractive index of the waveguide as a function of bias voltage, a key mechanism behind the modulator's operation.\@ The CHPW unbiased response is depicted in \figurename{ \ref{facile_schematic_updated_02}\textbf{d}}, showcasing the extracted IL and coupling efficiency (CE) as functions of wavelength, providing insights into the modulator's performance in terms of optical losses and fiber coupling efficiency.\@ The simulated ER is 0.85 dB/$\mu$m, and the minuscule overlap between the accumulation layer and the plasmonic modal cross-section in the plasmonic device necessitates an extended device length to obtain reasonable ER levels, which inevitably leads back to substantial IL levels.

\begin{figure*}[h]
	\centering
	\includegraphics[width=0.85\linewidth]{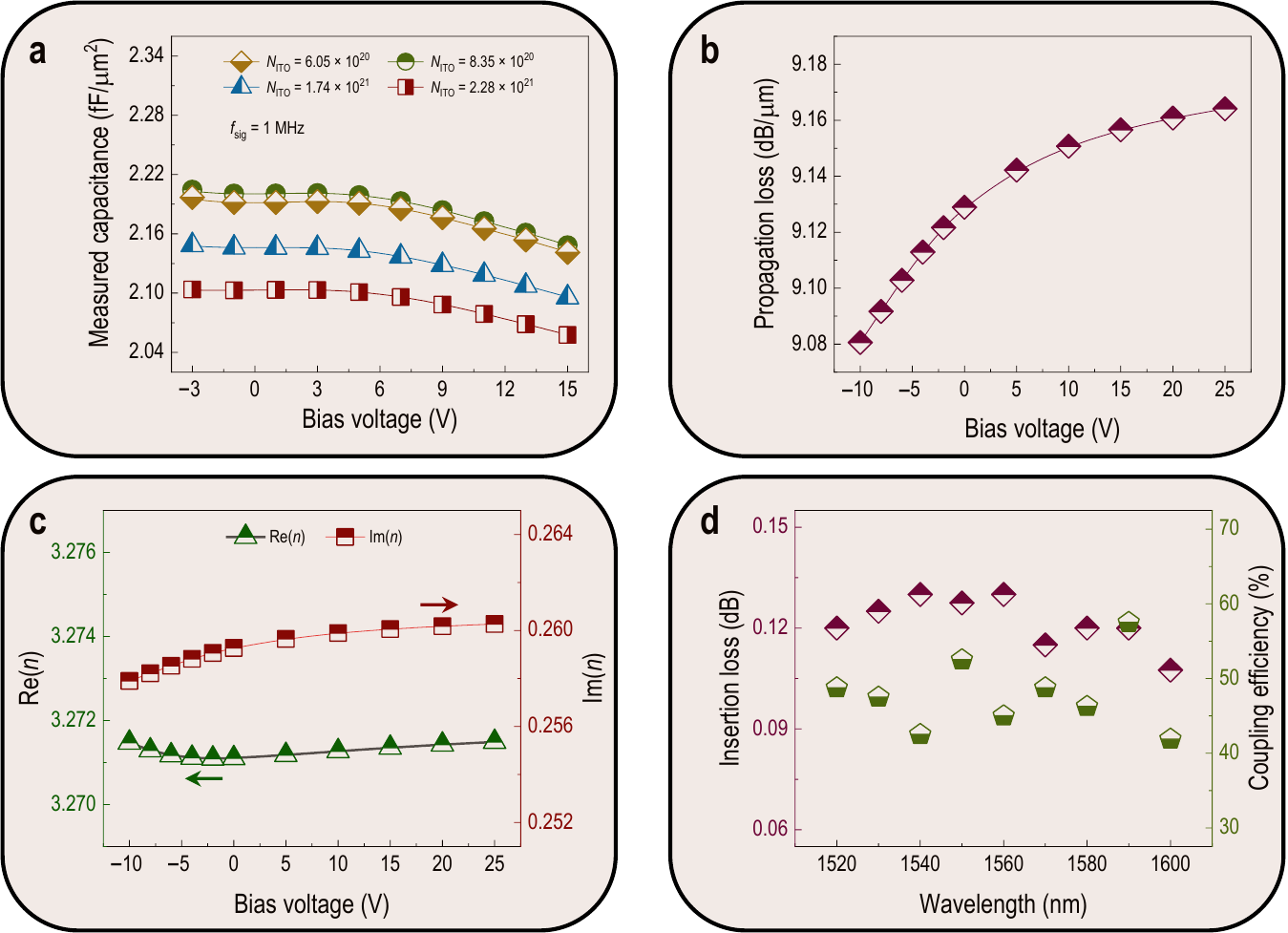}
	\caption{Experimental and numerical results related to the SiO$_\mathrm{2}$/ITO-based CHPW modulator.\@ \textbf{a}, Frequency response curves of fabricated SiO$_\mathrm{2}$/ITO MISM test structures with varying ITO carrier concentration \emph{N}$_\mathrm{ITO}$ levels as functions of bias voltage.\@ \textbf{b}, Numerical simulations of a CHPW, illustrating the evolution of effective propagation loss PL levels with bias voltage, highlighting operation within the ENZ regime.\@ \textbf{c}, Simulated refractive index of the CHPW waveguide as a function of bias voltage, again emphasizing ENZ operation.\@ \textbf{d}, Unbiased response of the CHPW, showing extracted IL and CE as functions of wavelength.\@ All panels are adapted from Alfaraj et al.\@ \cite{alfaraj2023facile} in accordance with a Creative Commons Attribution 4.0 International License (https://creativecommons.org/licenses/by/4.0/).\@ Copyright 2023, The Authors, published by Light Publishing Group.}
	\label{facile_schematic_updated_02}
\end{figure*}

Beyond demonstrating the successful integration and modulation capabilities of ITO, the study also provided valuable insights into the carrier dispersion properties of ITO and their impact on device performance.\@ A key finding was the validation of the graded-index layer (GIL) model for accurately predicting the accumulation layer width in the ITO-based MISM structure, particularly in the ENZ regime, where it outperformed the uniform accumulation layer (UAL) model.\@ Furthermore, the analysis of carrier distribution within the CHPW modulator underscored the critical role of ITO's carrier dispersion in achieving efficient modulation.\@ The ability to dynamically tune the carrier concentration in ITO, and thus its permittivity, through applied bias voltages proved essential for modulating the effective index of the hybrid plasmonic modes and achieving the desired modulation functionality.\@ Numerical simulations based on the GIL model were employed to gain deeper insights into the modulation mechanism within the CHPW structure (\figurename{ \ref{facile_schematic_updated_01}\textbf{e},\textbf{f}}).\@ These simulations revealed the dynamic changes in carrier density and E-field distribution across the waveguide under varying bias conditions.\@ As depicted in \figurename{ \ref{facile_schematic_updated_01}\textbf{e}}, applying a negative bias at the insulator side resulted in a depletion of free electrons, while a positive bias led to an accumulation of free electrons, facilitating higher current flow levels that increased with the positive bias.\@ The E-field magnitude across the SiO$_\mathrm{2}$ layer exhibited a linear dependence on the applied bias, directly influenced by the carrier accumulation/depletion at the SiO$_\mathrm{2}$/ITO interface (\figurename{ \ref{facile_schematic_updated_01}\textbf{f}}).\@ This observation highlighted the pivotal role of ITO's carrier dispersion properties in enabling the operation of the CHPW modulator.

The paper also addressed several challenges associated with the use of ITO in device fabrication, including its high production cost, mechanical brittleness, poor adhesion to certain materials, and concerns about long-term durability.\@ These challenges highlight the need for further research and development to fully realize the potential of ITO in practical plasmonic devices \cite{shan2024hemispherical}.\@ Additionally, the paper discussed the optimization of ITO deposition conditions to achieve optimal conductivity and modulator performance.\@ By carefully controlling the deposition parameters, it was possible to tailor the electrical and optical properties of ITO, leading to improved device performance and enhanced modulation efficiency.

In conclusion, Si-integrable plasmonics offer a promising pathway for realizing high-performance, integrated photonic devices.\@ However, it is crucial to acknowledge and address the inherent trade-offs between material properties, device performance, fabrication complexity, and cost.\@ Ongoing research in this field is focused on developing novel materials, device architectures, and fabrication techniques to overcome these challenges and unlock the full potential of Si-integrable plasmonics for a wide range of applications.

\subsection{Design trade-offs in plasmonic ITO-based MZMs}

Gui et al. investigated essential design trade-offs in their ITO-based MZM, focusing on optimizing ER and IL for high-performance, compact devices.\@ By employing an asymmetric MZI structure, they achieved fine control over ER and IL through careful adjustments to the power-splitting ratio in the Y-junction.\@ This structure is illustrated in \figurename{ \ref{gui_SOI_ITOmodulator}\textbf{a}-\textbf{d}}, which collectively show the device design and its performance metrics.

\begin{figure*}[h]
	\centering
	\includegraphics[width=0.85\linewidth]{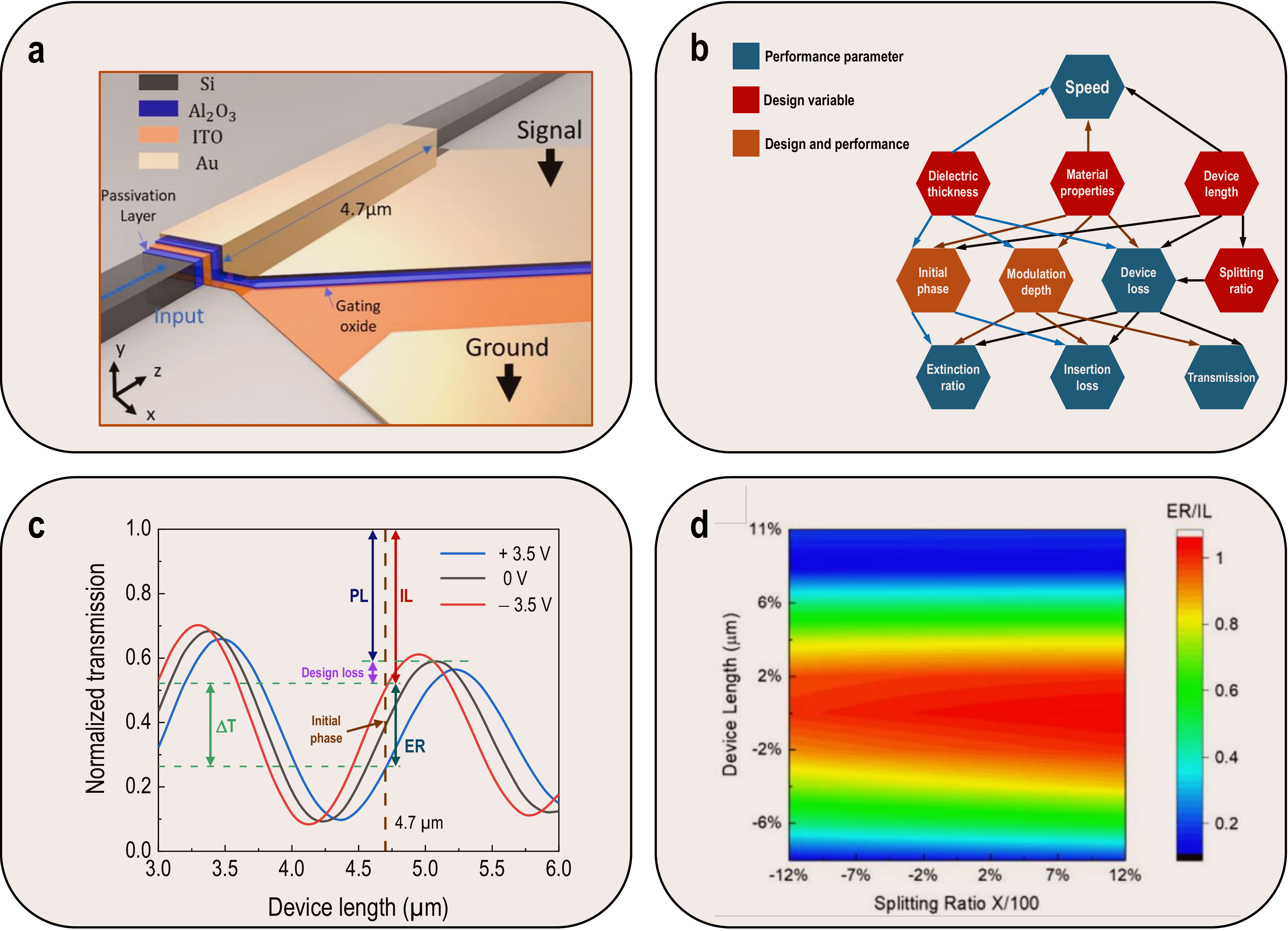}
	\caption{Structure, design variables, and performance analysis of an ITO-based plasmonic MZM.\@ \textbf{a}, Layer structure of an ITO-based MZM.\@ The device was fabricated on an SOI substrate and features a thin layer of ITO for plasmonic waveguiding and EO modulation.\@ The ITO layer is sandwiched between a passivation layer and an Au electrode which serves as the top electrode for applying the modulation voltage.\@ \textbf{b}, Relationship between MZM device design variables and performance parameters.\@ \textbf{c}, Normalized transmission as a function of device length for the asymmetric plasmonic mode ITO-based MZM with 15 nm-thick Al$_\mathrm{2}$O$_\mathrm{3}$ under ON ($-$3.5 V) and OFF states (+3.5 V).\@ \textbf{d}, Performance map illustrating the fabrication tolerance of device length and splitting ratio for the proposed design.\@ Panels \textbf{a},\textbf{d} are reproduced and \textbf{b},\textbf{c} are adapted from Gui et al.\@ \cite{gui2022100} in accordance with a Creative Commons Attribution 4.0 International License (https://creativecommons.org/licenses/by/4.0/).\@ Copyright 2022, The Authors, published by De Gruyter.}
	\label{gui_SOI_ITOmodulator}
\end{figure*}

\figurename{ \ref{gui_SOI_ITOmodulator}\textbf{b}} highlights the relationship between the power-splitting ratio and performance, illustrating how directing more power to the active arm (e.g., a 60:40 split) enhances ER by increasing modulation depth.\@ This higher power allocation to the active arm, however, also leads to elevated IL, as greater optical intensity results in increased propagation losses in the active region.\@ This configuration is thus suited for applications where strong modulation contrast is essential, though some increase in IL can be tolerated.\@ On the other hand, a lower splitting ratio (such as 37:63) reduces IL by directing less power to the active arm, thereby minimizing propagation loss but also yielding a reduced ER.\@ The 50:50 ratio offers a balanced approach, achieving moderate ER and IL, making it adaptable for scenarios that require a compromise between high modulation depth and minimal loss.

\figurename{ \ref{gui_SOI_ITOmodulator}\textbf{c}} and \figurename{ \ref{gui_SOI_ITOmodulator}\textbf{d}} provide additional insights into device performance as a function of design parameters.\@ \figurename{ \ref{gui_SOI_ITOmodulator}\textbf{c}} schematically illustrates normalized transmission as a function of device length, under ON and OFF bias states, for the asymmetric plasmonic mode MZM with a 15 nm-thick Al$\mathrm{2}$O$\mathrm{3}$ layer.\@ This plot underscores the influence of device length on modulation efficiency and loss:\@ shorter device lengths generally minimize IL, whereas longer lengths may improve modulation but at the expense of increased IL.\@ Gui et al.\@ demonstrate that a carefully chosen device length can provide a practical balance between efficient modulation and manageable IL.

\figurename{ \ref{gui_SOI_ITOmodulator}\textbf{d}} further explores the design space by presenting a performance map that evaluates the impact of device length and splitting ratio on ER and IL.\@ This performance map offers a detailed visualization of how these variables affect overall modulation efficiency, allowing designers to select optimal configurations based on specific application requirements.\@ For instance, the plot shows that high ER values can be achieved with a higher power-splitting ratio and longer device length, while low IL is associated with shorter device lengths and lower splitting ratios.\@ This approach underscores the flexibility of the ITO-based MZM design, where performance can be finely tuned by adjusting these parameters to suit particular operational needs, such as low-loss or high-modulation applications.

In summary, Figures \ref{gui_SOI_ITOmodulator}\textbf{b}-\textbf{d} collectively illustrate the complex trade-offs in ER and IL that can be optimized through strategic design choices in power splitting, device length, and asymmetric configuration. Gui et al.'s work exemplifies how careful modulation of these design parameters enables high-speed, energy-efficient performance in silicon-compatible plasmonic modulators, supporting their potential as versatile components for next-generation, high-density optical communication systems.

\section{Fabricating Plasmonic Devices on Si}\label{fabrication}

The integration of plasmonic components with Si platforms holds immense potential for realizing compact, high-performance optical devices.\@ However, the fabrication of such devices presents unique challenges due to the distinct properties of plasmonic materials and the need for compatibility with existing CMOS and Si foundry fabrication processes.\@ This section delves into the fabrication techniques employed in creating Si-based plasmonic devices, highlighting the specific considerations and trade-offs involved in achieving efficient and scalable manufacturing.\@ We begin by examining the integration of ITO, a CMOS-friendly TCO, into SOI devices, followed by a comparative analysis of various fabrication approaches employed in other notable works.

\subsection{Basic fabrication techniques and considerations}

The fabrication of CHPWs involves the deposition and patterning of various materials with significantly different properties.\@ As such, the process requires careful selection of deposition techniques and optimization of processing parameters.\@ In our previous work \cite{su2019record}, the fabrication of CHPWs began with the deposition of a 20 nm-thick SiO$_\mathrm{2}$ layer onto 220 nm-thick SOI wafers using plasma-enhanced chemical vapor deposition (PECVD).\@ This process occurred at a plasma temperature of 400 $^\circ$C using SiH$_\mathrm{4}$ and N$_\mathrm{2}$ gases.\@ The SiO$_\mathrm{2}$ layer served as a low-index spacer within the CHPW structure, which enabled strong confinement of the optical field for enhanced light-matter interaction.\@ Following the SiO$_\mathrm{2}$ deposition, a 10 nm-thick layer of Al was deposited using sputtering at room temperature in an Ar plasma.\@ This Al layer served as the metal layer within the CHPW, which supported the propagation of SPPs and enabled the coupling of SPP and HPW modes for the formation of LR, low-loss supermodes.

To complete the CHPW structure, a layer of a-Si was deposited using PECVD at a substrate temperature of 180 $^\circ$C using SiH$_\mathrm{4}$ plasma.\@ The thickness of this a-Si layer is a critical parameter that determines the coupling strength between the SPP and HPW modes and hence dictates the overall loss of the propagating supermode.\@ To achieve optimal loss reduction, the thickness of this layer was varied across different samples, ranging between 0 and 250 nm.\@ The variation allowed for systematic investigation of the impact of a-Si thickness on the modal properties of the CHPW, enabling the identification of the optimal thickness for LR propagation.\@ With all the material layers in place, EBL was used to define the patterns for the waveguides and other components.\@ The high resolution of EBL is crucial for accurately defining the nanoscale features of the CHPWs, ensuring the desired optical properties and device performance.\@ After the lithography step, RIE was employed to transfer the patterns into the Si and oxide layers, while a wet etching process was used to selectively remove the Al layer.\@ The use of RIE for the dielectric materials ensures anisotropic etching and maintains the fidelity of the defined patterns, while the wet etching process for Al provides selectivity and avoids damage to the other material layers \cite{su2019record}.

To showcase the capabilities of the CHPW platform, 200-nm-wide waveguides, and rings were fabricated, as shown in \figurename{ \ref{Su_CHPW_ring}\textbf{a}}.\@ For cutback measurement, CHPWs with lengths ranging from 10 to 400 $\mu$m were fabricated.\@ Light was coupled into and out of the CHPW devices using 800 nm-wide Si nanowires, as depicted in \figurename{ \ref{Su_CHPW_ring}\textbf{b}}.

\begin{figure}[h]
	\centering
	\includegraphics[width=\linewidth]{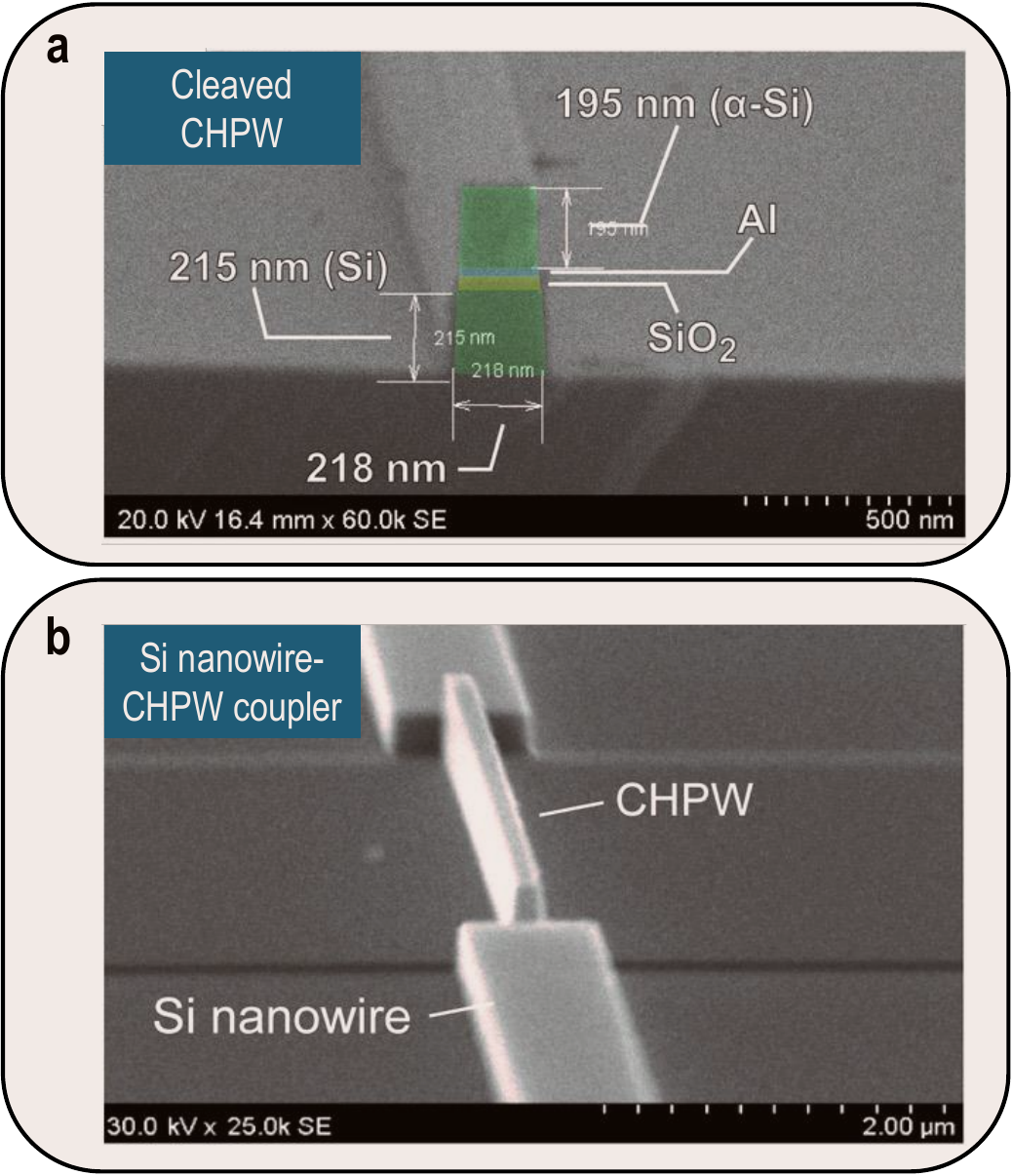}
	\caption{\textbf{a}, SEM image of a fabricated CHPW.\@ The image shows a cross-section of the waveguide, revealing the different material layers and their thicknesses.\@ \textbf{b}, SEM image of a Si nanowire--CHPW end-butt coupler.\@ This image illustrates how light is coupled into and out of the CHPW device using Si nanowires.\@ Both panels are reproduced from Su et al.\@ \cite{su2019record} in accordance with a Creative Commons Attribution NonCommercial License 4.0 (CC BY-NC) (https://creativecommons.org/licenses/by-nc/4.0/).\@ Copyright 2019, The Authors, published by American Association for the Advancement of Science.}
	\label{Su_CHPW_ring}
\end{figure}

\subsection{Fabrication of ITO-based CHPWs on SOI}

Here we discuss the fabrication techniques and practical considerations associated with CMOS- and Si-integrable materials and architectures, addressing the trade-offs and challenges in device performance.\@ TCOs like ITO, while not fully CMOS-integrable due to limitations imposed by FEOL processing, are amenable to integration at both the middle-end-of-line (MEOL) and BEOL stages due to their CMOS-friendly processing, which typically involves low-temperature deposition and patterning techniques that are compatible with other CMOS- and Si-friendly materials.

In our previous work \cite{alfaraj2023facile}, we demonstrated this compatibility by integrating ITO into an SOI-based CHPW modulator (\figurename{ \ref{facile_schematic_updated_01}} in section \ref{tradeoffs}).\@ Our fabrication process began with cleaning of the SOI substrate to eliminate contaminants that could negatively impact the performance of the nanoscale waveguides (\figurename{ \ref{WG_fab}}\textbf{a}).\@ This was followed by the deposition of a 5 nm-thick SiO$_\mathrm{2}$ passivation layer using sputtering (\figurename{ \ref{WG_fab}}\textbf{b}).\@ This SiO$_\mathrm{2}$ layer served several critical purposes.\@ First, it enhanced the adhesion of the subsequently deposited W alignment marks to the Si layer (\figurename{ \ref{WG_fab}}\textbf{c}).\@ Second, it acted as a protective barrier against further contamination, ensuring the integrity of the substrate and the proper operation of the waveguide.\@ Third, the controlled sputtering process provided a clean and uniform interface crucial for device performance.\@ Additionally, the SiO$_\mathrm{2}$ layer helped to passivate the Si surface, minimizing dangling bonds and surface states that could hinder device operation \cite{alfaraj2017photoinduced}.\@ Finally, it mitigated electron backscattering during the EBL process, improving the resolution and accuracy of the fabricated patterns.

\begin{figure*}[h]
	\centering
	\includegraphics[width=0.85\linewidth]{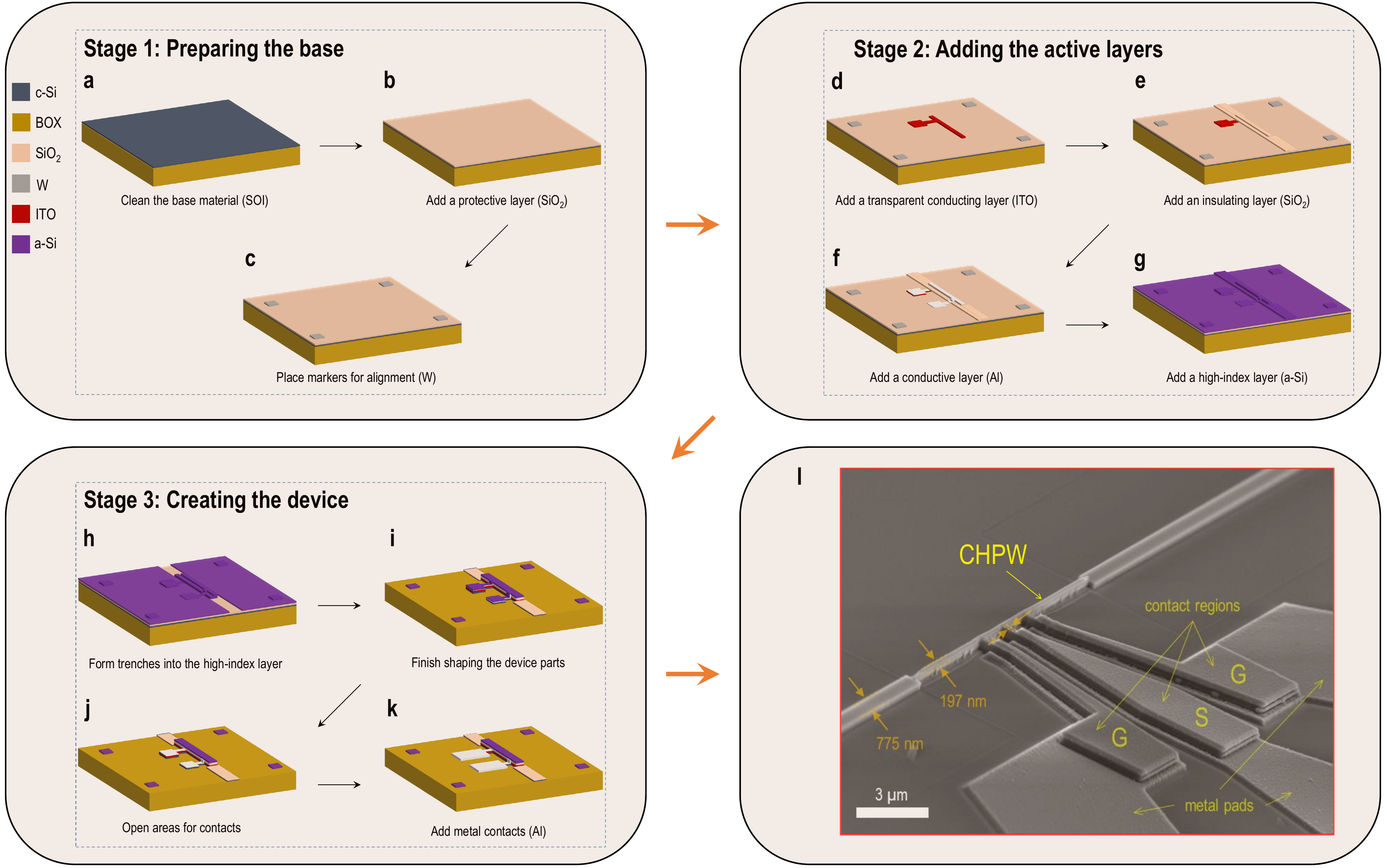}
	\caption{Fabrication of an a-Si/Al/SiO$_\mathrm{2}$/ITO CHPW.\@ The fabrication process includes the following steps:\@ \textbf{a} SOI wafer preparation;\@ \textbf{b} substrate passivation;\@ \textbf{c} deposition of W alignment marks;\@ \textbf{d} deposition of an ITO EO layer;\@ \textbf{e} deposition of a SiO$_\mathrm{2}$ insulating layer;\@ \textbf{f} deposition of an Al layer;\@ \textbf{g} deposition of an a-Si layer;\@ \textbf{h} creating trenches in the a-Si layer;\@ \textbf{i} etching the waveguide stack to form Si nanowires and CHPW components;\@ \textbf{j} selective-area etching of the a-Si layer;\@ and \textbf{k} formation of Al contact pads;\@ \textbf{l} inset:\@ Bird's-eye SEM image of a fabricated SiO$_\mathrm{2}$/ITO CHPW modulator.\@ All panels are adapted from Alfaraj et al.\@ \cite{alfaraj2023facile} in accordance with a Creative Commons Attribution 4.0 International License (https://creativecommons.org/licenses/by/4.0/).\@ Copyright 2023, The Authors, published by Light Publishing Group.}
	\label{WG_fab}
\end{figure*}

Following substrate preparation, the fabrication process proceeded with the deposition of a 100 nm-thick ITO layer using selective-area magnetron sputtering (\figurename{ \ref{WG_fab}}\textbf{d}).\@ A 10 nm-thick SiO$_\mathrm{2}$ layer was then selectively deposited on top of the ITO using electron-beam evaporation (\figurename{ \ref{WG_fab}}\textbf{e}).\@ This layer acted as an insulator between the ITO and the subsequent Al layer, which was also selectively deposited using electron-beam evaporation (\figurename{ \ref{WG_fab}}\textbf{f}).\@ The final step in the EO active layer deposition stage involved the deposition of a 330 nm-thick a-Si layer using magnetron sputtering (\figurename{ \ref{WG_fab}}\textbf{g}).\@ In the CHPW modulator, the a-Si layer plays a multifaceted role that is crucial to the device's functionality.\@ It serves as the primary platform for the SPP mode at the a-Si/Al interface, which is essential for the formation of the low-loss, LR supermode through coupling with the HPW mode.\@ This supermode carries the optical signal with minimal energy dissipation.\@ Additionally, the thickness of the a-Si layer is precisely controlled to minimize the field overlap within the metal layer, reducing propagation loss and enabling a relatively efficient plasmonic modulation.\@ The a-Si layer also contributed to the dynamic reconfigurability of the modulator, allowing it to be adapted for various modulation formats.\@ This multifaceted role highlights the intricate design of the CHPW modulator, where the a-Si layer is integral to achieving efficient light manipulation and modulation \cite{lin2020monolithic,lin2020supermode}.

With the active EO layers in place, the device formation process commenced.\@ A 50 nm-deep trench was etched into the a-Si layer to define the waveguide (\figurename{ \ref{WG_fab}\textbf{h}}).\@ The remaining components were then etched to their final dimensions (\figurename{ \ref{WG_fab}\textbf{i}}).\@ Contact areas were opened, and an Al contact was deposited to complete the device fabrication (\figurename{ \ref{WG_fab}\textbf{j},\textbf{k}}).\@ This plasmonic device fabrication process highlights the compatibility of ITO with standard Si processing techniques, making them attractive candidates for integrated photonic devices, as the entire device stack was deposited and processed at room temperature.

To achieve efficient modulation, careful consideration was given to the design of the electrical contacts.\@ To facilitate efficient electrical contact and modulation, the device incorporates a specific contact design.\@ A bird's-eye view of the fabricated 10-$\mu$m long SiO$_\mathrm{2}$/ITO CHPW modulator, captured using an SEM, is shown in \figurename{ \ref{WG_fab}\textbf{l}}.\@ In this image, contacts were formed by extending the Al and ITO layers away from the waveguide region via fingers that are 475 nm-wide and separated by 2 $\mu$m gaps.\@ This design choice ensures minimal disturbance to the optical mode.\@ The anode (S) and cathodes (G), consisting of 200-nm-thick Al layers, are sputtered onto the extended Al and ITO regions.\@ The Al pads are designed to be larger than the existing contact region to facilitate probing.\@ Under an external bias, carriers are injected through these contacts, and the formation of a voltage-induced accumulation layer at the SiO$_\mathrm{2}$/ITO interface enables optical modulation.

The EO SiO$_\mathrm{2}$/ITO interface is the core element of CHPW modulator.\@ As shown in \figurename{ \ref{norm_trans_dB}}, we compared the performance of two CHPW modulators, with and without an ITO layer, to demonstrate their functionality.\@ For both scenarios, no modulation was observed without ITO, whereas with ITO, modulation was observed with an extinction ratio of approximately 1 dB/$\mu$m.\@ The MD varied with device length, and for a 15 $\mu$m-long modulator, the normalized transmission could reach $-$12 dB at 26 V.\@ Beyond $\lambda$ = 1.53 $\mu$m, the MD decreased as the permittivity of ITO moved away from the ENZ regime.\@ However, an extinction ratio of about 1 dB/$\mu$m was maintained between $\lambda$ = 1.53 $\mu$m and 1.56 $\mu$m.

\begin{figure}[h]
	\centering
	\includegraphics[width=\linewidth]{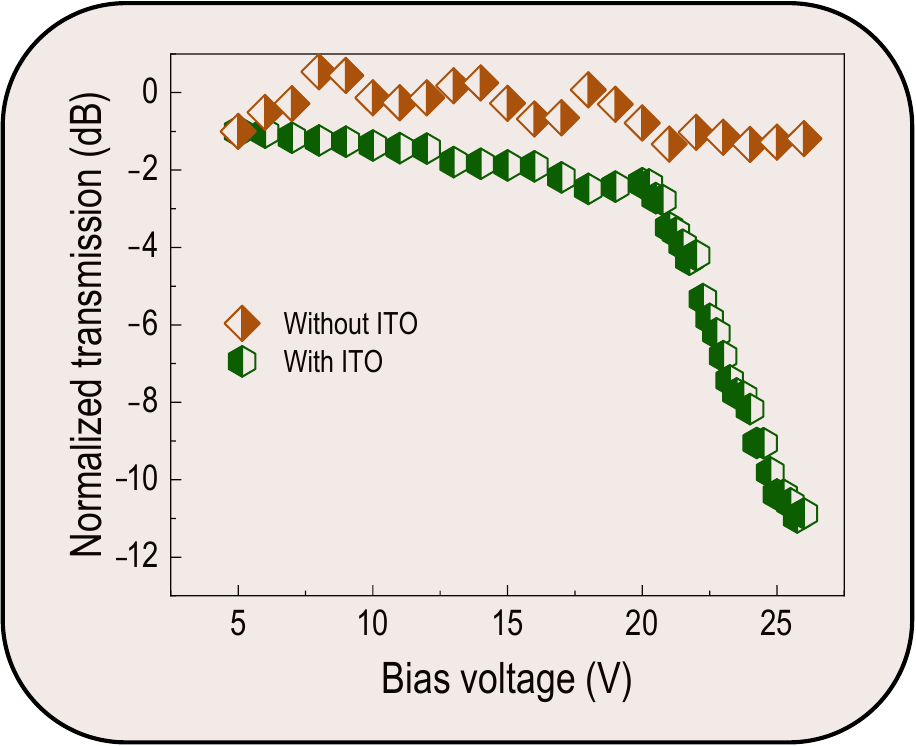}
	\caption{Transmission of a 10-$\mu$m SiO$_\mathrm{2}$/ITO-based CHPW modulator as a function of bias voltage.\@ The behavior of the modulator with and without an ITO layer is compared to demonstrate the functionality of the ITO layer.\@ For the modulator with ITO, modulation was observed with an extinction ratio of approximately 1 dB/$\mu$m.\@ Adapted from Alfaraj et al.\@ \cite{alfaraj2023facile} in accordance with a Creative Commons Attribution 4.0 International License (https://creativecommons.org/licenses/by/4.0/).\@ Copyright 2023, The Authors, published by Light Publishing Group.}
	\label{norm_trans_dB}
\end{figure}

The CHPW modulator incorporates a Schottky junction, creating a unique architecture when combined with the MOS capacitor.\@ This configuration facilitates the excitation of both SPP and HPW modes, as illustrated in the device's energy band diagram \cite{alfaraj2024cmos}.\@ The Al/Si interface at the top of the device supports a tightly confined SPP mode, while the bottom Al/SiO$_\mathrm{2}$/ITO/Si structure guides an HPW mode.\@ The thin Al layer facilitates spatial overlap and hybridization between these modes, resulting in two TM modes:\@ a LR mode with exceptionally low propagation losses (around 0.02 dB/$\mu$m at $\lambda$ = 1.55 $\mu$m) and a SR mode.\@ The integration of the Schottky junction and the subsequent reduction in IL contribute to a higher figure of merit for the modulator, making it more efficient.

While Gui et al.\@ did not detail their fabrication process \cite{gui2022100}, Amin et al.\@ focused on the heterogeneous integration of ITO on an SOI platform for their plasmonic MZM using the same device structure \cite{amin2021heterogeneously}.\@ Their fabrication process involved ion beam deposition (IBD) of ITO and ALD of the gate oxide, followed by EBL patterning.\@ These techniques offer good control over the material properties and device dimensions, but using ALD may increase fabrication complexity given the relatively high-temperature processing (100 $^\circ$C).\@ In contrast, our fabrication process, detailed earlier in this subsection, was carried out entirely at room temperature, which could provide a simpler and potentially more scalable alternative.\@ The fabrication process presented by Gui et al., depicted in \figurename{ \ref{gui_SOI_ITOmodulator}\textbf{a}}, involved the deposition of various materials at elevated temperatures.\@ Specifically, the Al$_\mathrm{2}$O$_\mathrm{3}$ layers were deposited using ALD at 100 $^\circ$C.\@ While this temperature remains within the thermal budget of BEOL processes, it is higher than room temperature, which could be a consideration for certain CMOS fabrication flows, especially when Au is involved at a later stage.\@ The use of Au, a non-Si-compatible material, is a significant concern due to potential integration challenges with CMOS fabrication.\@ However, Gui et al. highlighted that the Au layer served a dual purpose:\@ it acted as an electrical contact and contributed to the plasmonic behavior of the modulator by enhancing light-matter interaction and enabling strong modulation of the ITO layer.

Their fabrication process included the deposition of a 5 nm-thick Al$_\mathrm{2}$O$_\mathrm{3}$ passivation layer using ALD at 100 $^\circ$C to protect the Si waveguide and improve the adhesion of subsequent layers.\@ A 30 nm-thick Au layer was then sputtered onto the substrate to act as the top electrode for the capacitor structure, followed by a 10 nm-thick ITO layer deposited using sputtering, which serves as the active EO material responsible for phase modulation.\@ The final step involved the deposition of a 15 nm-thick Al$_\mathrm{2}$O$_\mathrm{3}$ layer using ALD to function as the gate dielectric for the capacitor.\@ The use of ALD for depositing the Al$_\mathrm{2}$O$_\mathrm{3}$ layers offers precise thickness control and excellent uniformity, crucial for achieving consistent device performance.\@ However, the incorporation of Au may limit the process's full compatibility with CMOS technology and Si foundry processes.

Despite this limitation, Gui et al.\@ demonstrated the successful operation of their ITO-based modulator with CMOS-compatible voltage levels, indicating its potential for integration into Si platforms.\@ A critical aspect of CMOS compatibility is the use of CMOS-compatible bias voltages, typically in the range of 0.5 V to 3.3 V \cite{chen2007output}.\@ Gui et al.\@ demonstrated the operation of their ITO-based modulator with a bias voltage of $\pm$3.5 V, which, while slightly exceeding the typical CMOS voltage range, remains within the acceptable limits for many CMOS technologies.\@ The ability to operate at such voltage levels underscores the potential of ITO-based modulators for CMOS integration, enabling the development of compact and efficient optical interconnects for next-generation computing systems.

The choice of materials and the fabrication process significantly influence the device's performance.\@ The high EO coefficient of ITO, the optimized thickness of the Al$_\mathrm{2}$O$_\mathrm{3}$ layer, and the plasmonic enhancement from the Au layer enabled a reasonable ER of 3 dB with a 4.7 $\mu$m-long active region.\@ The device exhibited a high modulation BW of 100 GHz (switching rate), highlighting its potential for high-speed optical communication.\@ Gui et al.\@ reported an IL of 2.9 dB for their modulator, which appears to be for the entire 1 mm device, including the grating couplers and the access waveguides, based on their description of the device dimensions (33 $\mu$m in length for the MZI structure, with the total modulator including grating couplers being 1 mm long).\@ This suggests a low loss per unit length for the modulator itself.\@ Furthermore, they achieved a high ER-to-IL ratio of around 1, indicating efficient modulation with minimal signal attenuation.\@ The compact size of the ITO-based MZM (33 $\mu$m long) contributes to a significant increase in packing density.\@ According to Gui et al., this design allowed for the integration of more than 3500 of these modulators within the same chip area as a single Si MZM, highlighting the potential for high-density integration offered by this technology.\@ The device also demonstrated low energy consumption, with a switching energy of 380 fJ/bit.

\subsection{Other fabrication approaches}

A comparison between the fabrication approach employed by us\cite{alfaraj2023facile} and those used in the other referenced works reveals key similarities and differences, particularly regarding the specifics of the techniques used, their advantages and disadvantages, and the degree of CMOS and Si integration achieved.\@ Bisang et al.\@ also utilized an SOI-based platform for their plasmonic modulators, but their fabrication process involved more complex steps, including Au deposition for the plasmonic slot waveguides \cite{bisang2024plasmonic}.\@ While their fabrication approach allows for precise control over the device dimensions, it may pose challenges for large-scale CMOS and Si device manufacturing given the contamination challenges.\@ In contrast, our process relied on room-temperature, CMOS-compatible techniques, such as room-temperature sputtering of conventional oxide structures, which are simpler and potentially more cost-effective for mass production.

Valdez et al.\@ explored a hybrid integration approach, combining TFLN with a SiN waveguide on an SOI substrate (\figurename{ \ref{valdez_alaloul}\textbf{a}--\textbf{c}}) \cite{valdez2023buried}.\@ This heterogeneous integration scheme involves bonding the TFLN layer onto the SOI platform using hydrophilic bonding, adding complexity to the fabrication process.\@ After bonding, the LNOI substrate is removed to expose the LN layer for subsequent processing.\@ While this approach leverages the benefits of both materials, it may introduce challenges related to bonding quality and compatibility.\@ Our integration of ITO on SOI \cite{alfaraj2023facile}, on the other hand, offers a simpler fabrication process and potentially better compatibility between the materials.

\begin{figure*}[h]
	\centering
	\includegraphics[width=0.85\linewidth]{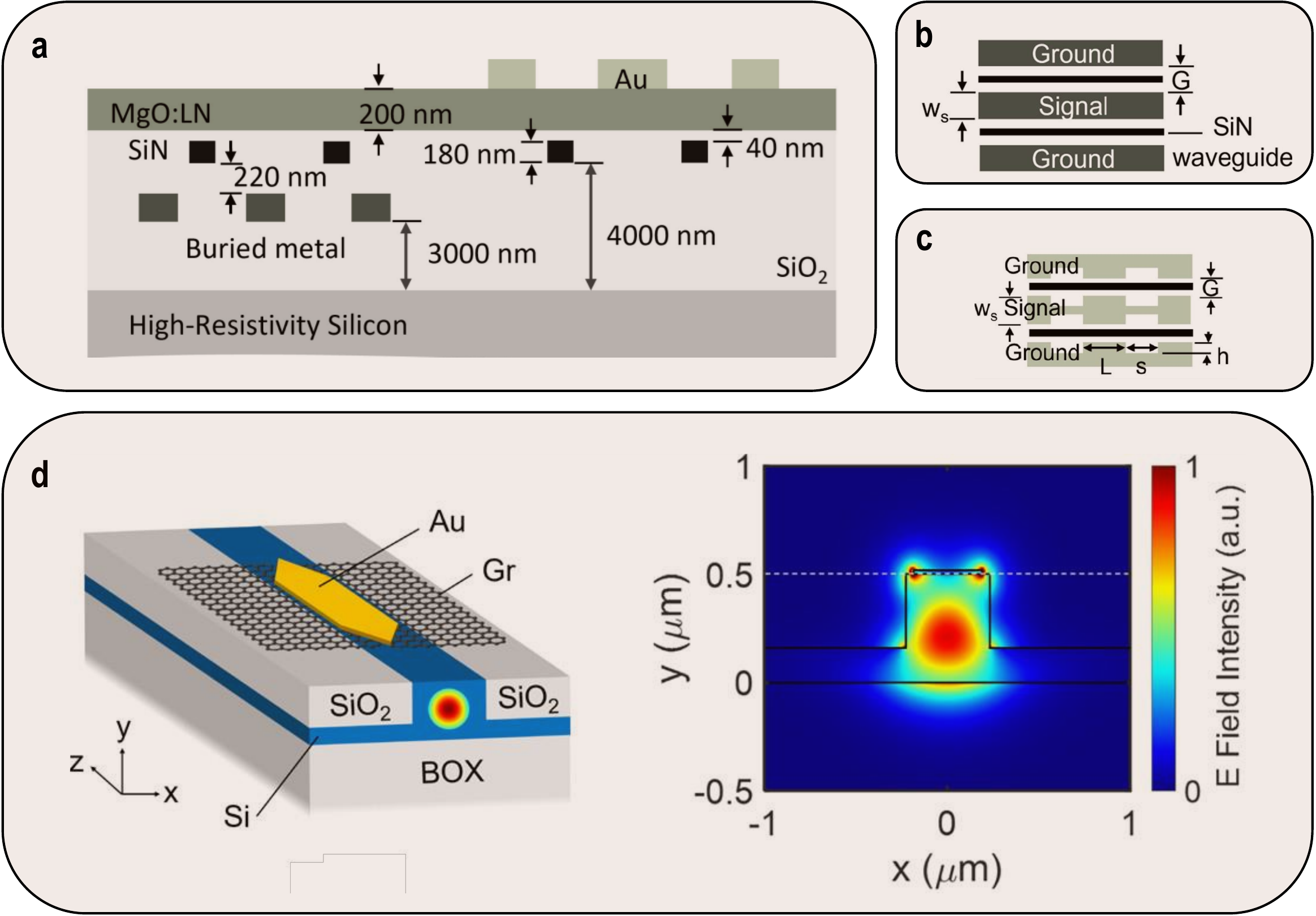}
	\caption{\textbf{a}, Cross-sectional diagram of a SiN-LN hybrid platform, illustrating two modulator designs:\@ one with buried metal electrodes and another with top Au electrodes positioned above the bonded LN film.\@ \textbf{b}, Schematic of a SiN-LN MZM device with buried electrodes in a coplanar waveguide (CPW) configuration.\@ \textbf{c}, Schematic of a SiN-LN MZM device with top Au electrodes in a slow-wave electrode (SWE) structure.\@ \textbf{d}, An all-optical modulator, utilizing a Si rib waveguide structure, is shown in this figure.\@ The corresponding visualization presents the E-field distribution of the quasi-TM mode as it propagates through the waveguide.\@ The planar position of the graphene sheet is identified by the dashed white line.\@ Panels \textbf{a}--\textbf{c} are reproduced with permission from Valdez et al.\@ \cite{valdez2023buried}.\@ Copyright 2023, IEEE.\@ Panel \textbf{d} is reproduced from AlAloul et al.\@ \cite{alaloul2021low} in accordance with a Creative Commons Attribution 4.0 International License (https://creativecommons.org/licenses/by-nc-nd/4.0/).\@ Copyright 2021, The Authors, published by American Chemical Society.}
	\label{valdez_alaloul}
\end{figure*}

The fabrication process began with the deposition and patterning of a SiN waveguide layer on the SOI substrate.\@ Then, a thin film of LN was bonded on top of the SiN waveguide layer.\@ After bonding, the top LNOI substrate was removed to expose the LN layer for subsequent processing.\@ The final steps involved the fabrication of the electrodes, which can be achieved using two different approaches:\@ either by burying them in the SOI substrate or by placing them on top of the bonded LN film.\@ The choice of electrode placement affected the device's performance characteristics, such as its BW and IL.

\figurename{ \ref{valdez_alaloul}\textbf{a}} shows cross-sections of the final fabricated devices with the two different electrode placements.\@ In terms of CMOS compatibility, the use of buried electrodes (\figurename{ \ref{valdez_alaloul}\textbf{b}}) offers a potential advantage.\@ The buried electrodes can be fabricated using standard CMOS processes, making the integration of the modulator with other CMOS-based components potentially easier.\@ However, the use of Au in the top-electrode design (\figurename{ \ref{valdez_alaloul}\textbf{c}}) poses a challenge to full CMOS compatibility.\@ Additionally, the bonding process used to integrate the LN film with the SiN waveguide may introduce complexities and compatibility issues with certain CMOS fabrication flows.

AlAloul et al.\@ investigated SOI-based plasmon-enhanced graphene modulators, as shown in \figurename{ \ref{valdez_alaloul}\textbf{d}} \cite{alaloul2021low}.\@ Their fabrication process involved placing an Au stripe on top of a graphene layer, which was then integrated onto a Si waveguide.\@ This approach requires careful alignment and placement of the Au stripe and graphene layer, potentially adding complexity to the fabrication process.\@ Our process, while not involving graphene, utilized simpler deposition and patterning techniques, potentially offering a more straightforward fabrication route.\@ Haffner et al.\@ demonstrated low-loss plasmon-assisted EO modulators on an SOI platform \cite{haffner2018low}.\@ Their fabrication process involved creating an Au MIM slot waveguide ring coupled to a buried Si bus waveguide.\@ This approach requires precise control over the slot dimensions and alignment with the Si waveguide, potentially increasing fabrication complexity.\@ Our process employed simpler deposition and patterning techniques, potentially offering a more streamlined fabrication process without incorporating Au.\@ Finally, Dionne et al.\@ presented a Si-based plasmonic modulator in an MOS geometry \cite{dionne2009plasmostor}.\@ Their fabrication process involved creating a suspended Si membrane, followed by thermal oxidation and deposition of Ag layers on each side.\@ This approach incorporates Ag and requires careful control over the membrane thickness and uniformity, potentially increasing fabrication complexity.\@ Our process, while also utilizing an SOI platform, employed simpler deposition and patterning techniques, potentially offering a more straightforward fabrication route.

In summary, the fabrication of plasmonic devices on Si platforms compatible with CMOS technologies requires careful consideration of various factors, including material compatibility and process complexity.\@ Our room-temperature fabrication approach for integrating ITO on an SOI platform offers advantages in terms of simplicity and potential scalability \cite{alfaraj2023facile}.\@ It utilizes standard deposition and patterning techniques, potentially leading to a more cost-effective and manufacturable solution for integrated photonic devices.\@ Additionally, the demonstrated stability of our device's EO response at elevated temperatures further highlights the robustness and potential for practical applications of our approach.\@ However, other approaches also present unique strengths.\@ Bisang et al.'s approach, which involves using EBL to pattern plasmonic slot waveguides with widths that are close to only 100 nm, allows for precise control over device dimensions \cite{bisang2024plasmonic}.\@ This precise control is crucial for achieving specific performance targets and minimizing optical losses.\@ The hybrid integration approach employed by Valdez et al., while potentially introducing complexities related to the bonding process, successfully combines the relatively high EO coefficient of LN \cite{xu2023attojoule,xu2022dual} with the low optical loss of SiN \cite{zabelich2024silicon}, enabling the realization of high-performance modulators \cite{valdez2023buried}.\@ AlAloul et al.'s investigation of graphene-based modulators highlights the potential of 2D materials for plasmonic applications, offering unique advantages such as ultrafast carrier dynamics and broadband optical response \cite{alaloul2021plasmon}.\@ Ultimately, the choice of fabrication approach depends on the specific requirements of the application, considering factors such as performance targets, CMOS compatibility, and manufacturing scalability.

\section{Si- vs. Non-Si-Integrable Plasmonics}\label{comparison}

Having detailed the fabrication of Si-compatible plasmonic devices in the previous section, we now broaden the discussion by comparing the capabilities and trade-offs of Si-integrable and non-Si-integrable plasmonic approaches, providing a comprehensive overview to guide material and architecture selection for various applications.

CMOS-integrable plasmonic approaches, as summarized in \tablename{ \ref{Si_comparison}}, offer distinct advantages in terms of integration, scalability, and cost-effectiveness by leveraging the mature infrastructure and well-established processes of CMOS technology.\@ This facilitates the seamless co-integration of plasmonic and electronic components on a single chip, potentially leading to the realization of compact, high-performance, and multifunctional optoelectronic systems.\@ Moreover, the utilization of Si-integrable materials like Al, TiN, and TCOs can significantly reduce fabrication complexity and cost, making plasmonic devices more accessible for large-scale production.\@ For example, the use of Al, TiN, and ITO allows for the fabrication of plasmonic devices using standard CMOS processes, eliminating the need for specialized and potentially costly fabrication techniques.\@ While the optical losses in these materials are typically higher than those in noble metals, advancements in hybrid mode engineering, such as the use of CHPWs, have demonstrated the potential to achieve propagation losses as low as 0.02 dB/$\mu$m, making them competitive for various on-chip applications.

In contrast, non-Si-integrable plasmonic approaches, typically employing noble metals like Au and Ag, exhibit superior optical properties.\@ These materials offer lower optical losses (e.g., Ag at $\lambda$ = 1.55 $\mu$m: on the order of 0.1 dB/$\mu$m) and broader operating BWs, enabling high-performance waveguides and facilitating applications such as plasmonic lasers and nonlinear optics.\@ They also excel in achieving superior field confinement, with plasmonic nanoantennas capable of localizing light to areas smaller than 0.01 $\mu$m$^\mathrm{2}$ \cite{savage2012revealing,novotny2011antennas,schuller2010plasmonics,curto2010unidirectional,cubukcu2006plasmonic,schuck2005improving}.\@ However, their integration with CMOS electronics often necessitates complex and costly fabrication techniques, primarily limited to BEOL processing, which can hinder their scalability and practical implementation in ICs.

\begin{table*}[h]
	\centering
	\caption{Comparison of Si-integrable and non-Si-integrable plasmonic approaches.\@ This comparison highlights the key trade-offs to consider when selecting a plasmonic approach for specific applications.\label{Si_comparison}}
    \renewcommand{\arraystretch}{1} 
	\begin{tabular*}{\textwidth}{@{\extracolsep\fill}lll@{\extracolsep\fill}}
		\toprule
		\textbf{Feature} & \textbf{Si-integrable plasmonics} & \textbf{Non-Si-integrable plasmonics}\\
		\midrule
		\textbf{Materials} & 
		\begin{tabular}[c]{@{}l@{}}
			Al, TCOs, TiN, SiN, Cu.
		\end{tabular}
		& 
		\begin{tabular}[c]{@{}l@{}}
			Au, Ag. 
		\end{tabular}
		\\
		\addlinespace 
		\textbf{Optical losses} &
		\begin{tabular}[c]{@{}l@{}}
			Typically higher.\\
			Mitigable with CHPWs\\(0.02--0.03 dB/$\mu$m at $\lambda$ = 1.55 $\mu$m) \cite{su2019record,lin2020monolithic}.
		\end{tabular}
		&
		\begin{tabular}[c]{@{}l@{}}
			Lower (e.g., Ag at $\lambda$ = 1.55 $\mu$m: $\sim$0.1 dB/$\mu$m) \cite{johnson1972optical}.\\
			But still significant for LR applications.
		\end{tabular}
		\\
		\addlinespace
		\textbf{Field confinement} &
		\begin{tabular}[c]{@{}l@{}}
			Comparable or slightly lower.\\
			Subwavelength confinement with\\CHPWs (0.002 $\mu$m$^\mathrm{2}$ modal area) \cite{su2019record,lin2020monolithic}.
		\end{tabular}
		& 
		\begin{tabular}[c]{@{}l@{}}
			Superior\\(e.g., nanoantennas: < 0.01--0.03 $\mu$m$^\mathrm{2}$) \cite{firby2024enhanced}.
		\end{tabular}
		\\
		\addlinespace
		\textbf{Operating BW} &
		\begin{tabular}[c]{@{}l@{}}
			Can be limited by material dispersion.\\
			Careful design and material choice can extend BW.
		\end{tabular}
		& 
		\begin{tabular}[c]{@{}l@{}}
			Broader, especially for noble metals \cite{anquillare2016efficient}.
		\end{tabular}
		\\
		\addlinespace
		\textbf{CMOS integration} &
		\begin{tabular}[c]{@{}l@{}}
			Seamless (MEOL, BEOL) for most materials.\\
			Full CMOS compatibility for Al, TiN \cite{alfaraj2021silicon}.\\
			Potential for 2D materials (e.g., MoS$_\mathrm{2}$)\\with ongoing research.
		\end{tabular}
		&
		\begin{tabular}[c]{@{}l@{}}
			Challenging, requires additional processing steps.\\
			Primarily limited to BEOL integration.
		\end{tabular}
		\\
		\addlinespace
		\textbf{Fabrication complexity} & 
		\begin{tabular}[c]{@{}l@{}}
			Lower, leverages existing CMOS processes.
		\end{tabular}
		& 
		\begin{tabular}[c]{@{}l@{}}
			Higher (often requires specialized techniques).
		\end{tabular}
		\\
		\addlinespace
		\textbf{Cost} & 
		\begin{tabular}[c]{@{}l@{}}
			Lower (benefits from economies of scale in CMOS).
		\end{tabular}
		& 
		\begin{tabular}[c]{@{}l@{}}
			Higher (non-standard materials and processes).
		\end{tabular}
		\\
		\addlinespace
		\textbf{Scalability} & 
		\begin{tabular}[c]{@{}l@{}}
			High, inherent to CMOS technology.
		\end{tabular}
		& 
		\begin{tabular}[c]{@{}l@{}}
			Limited (challenging for large-scale production).
		\end{tabular}
		\\
		\addlinespace
		\textbf{Applications} &
		\begin{tabular}[c]{@{}l@{}}
			On-chip optical interconnects \cite{liu2016integrated}.\\
			Optical modulators and PDs \cite{lin2020monolithic}.
		\end{tabular}
		&
		\begin{tabular}[c]{@{}l@{}}
			Plasmonic lasers \cite{wang2023recent}.\\
			Biosensors \cite{minopoli2022nanostructured}.\\
			High-performance waveguides.
		\end{tabular}
		\\
		\bottomrule
	\end{tabular*}
\end{table*}

To further illustrate the trade-offs between Si-integrable and non-Si-integrable plasmonic approaches, \figurename{ \ref{bar_chart}} presents a visual comparison of their key features based on the information summarized in \tablename{ \ref{Si_comparison}}.\@ The bar chart quantifies the relative performance of each approach across various aspects, including optical losses, field confinement, operating BW, CMOS integration, fabrication complexity, cost-effectiveness, and scalability.\@ The chart underscores the inherent strengths and weaknesses of each approach.\@ While non-Si-integrable plasmonics (typically using Au and Ag) excel in optical performance, they face challenges in CMOS integration, fabrication complexity, cost, and scalability.\@ This trade-off highlights the challenges associated with incorporating these materials into mainstream semiconductor manufacturing processes.

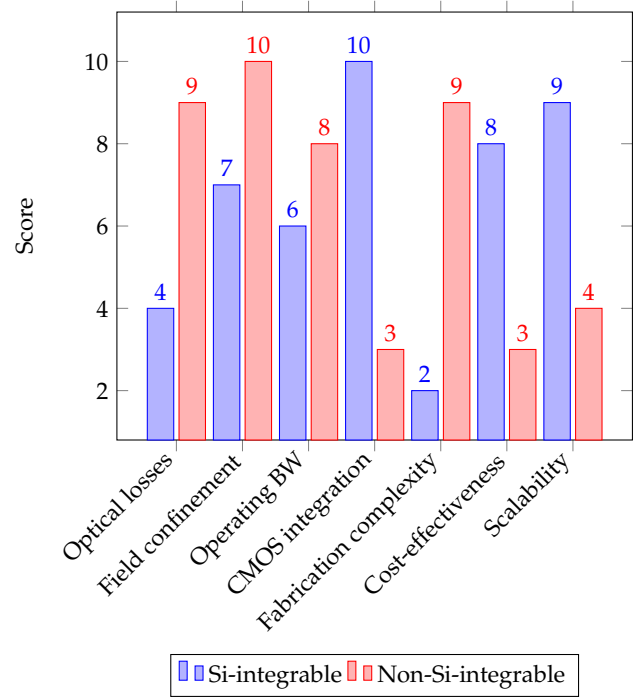
\begin{figure}[h]
	\centering
	\begin{tikzpicture}
		\begin{axis}[
			ybar,
			enlargelimits=0.15,
			legend style={at={(0.5,-0.5)}, 
				anchor=north,legend columns=-1}, 
			ylabel={Score},
			symbolic x coords={Optical losses, Field confinement, Operating BW, CMOS integration, Fabrication complexity, Cost-effectiveness, Scalability},
			xtick=data,
			nodes near coords,
			nodes near coords align={vertical},
			x tick label style={rotate=45,anchor=east},
			]
			\addplot coordinates {(Optical losses, 4) (Field confinement, 7) (Operating BW, 6) (CMOS integration, 10) (Fabrication complexity, 2) (Cost-effectiveness, 8) (Scalability, 9)};
			\addplot coordinates {(Optical losses, 9) (Field confinement, 10) (Operating BW, 8) (CMOS integration, 3) (Fabrication complexity, 9) (Cost-effectiveness, 3) (Scalability, 4)};
			\legend{Si-integrable,Non-Si-integrable}
		\end{axis}
	\end{tikzpicture}
	\caption{Comparative analysis of Si-integrable and non-Si-integrable plasmonic approaches.\@ The bar chart illustrates the relative performance of each approach based on key features relevant to integrated photonics.\@ Non-Si-integrable plasmonics, typically utilizing noble metals, excel in optical losses, field confinement, and operating BW, but face challenges in CMOS integration, fabrication complexity, cost, and scalability.\@ In contrast, Si-integrable plasmonics, employing materials like Al and TCOs, offer advantages in integration, fabrication simplicity, cost-effectiveness, and scalability, while recent advancements in hybrid mode engineering have mitigated their limitations in optical losses.\@ This comparison highlights the trade-offs involved in selecting the appropriate plasmonic approach for specific applications.}
	\label{bar_chart}
\end{figure}

The scores in \figurename{ \ref{bar_chart}} were determined based on a qualitative assessment of the relative performance of each approach across various aspects, with higher scores indicating better performance.\@ For example, non-Si-integrable plasmonics received a higher score for optical losses due to the inherently lower losses of noble metals like Au and Ag.\@ Si-integrable plasmonics received a higher score for CMOS integration due to the compatibility of materials like Al and TiN with CMOS processes.

In contrast, Si-integrable plasmonics, utilizing materials such as Al, TiN, and TCOs, demonstrate significant advantages in integration, fabrication simplicity, cost-effectiveness, and scalability, aligning well with the requirements of large-scale IC manufacturing.\@ Although these materials may exhibit higher optical losses compared to their non-CMOS counterparts, ongoing research in hybrid mode engineering, as exemplified by CHPWs, has shown promising results in mitigating these losses, making them increasingly competitive for on-chip applications.

The choice between Si-integrable and non-Si-integrable plasmonic approaches hinges on the specific application requirements and the priorities placed on different performance metrics.\@ Non-Si-integrable plasmonics offer superior optical performance in certain aspects, while Si-integrable plasmonics presents a compelling alternative for applications where integration, scalability, and cost-effectiveness are critical factors.

\section{Outlook}\label{outlook}

The pursuit of Si-integrable plasmonics has yielded significant progress in overcoming the limitations of traditional photonic devices \cite{xiao2023recent,min2009high}.\@ The advent of hybrid mode integration, exemplified by the CHPW architecture, has opened up new avenues for designing and fabricating compact, efficient, and multifunctional photonic components.\@ The ability to engineer the field symmetry in asymmetric waveguides, combined with the strategic coupling of different plasmonic modes, has led to record-breaking performance in various applications.\@ For instance, the CHPW platform has enabled the realization of ultra-compact microring resonators with unprecedented Purcell factors \emph{Q}/\emph{V}$_\mathrm{eff}$, approaching 10$^\mathrm{4}$ \cite{su2019record,tseng2013study,xu2008silicon}, demonstrating the potential for highly efficient light-matter interactions.

Moreover, the unique properties of CHPWs have facilitated the development of high-performance modulators and PDs.\@ By leveraging the ENZ effect in Si-integrable and CMOS-friendly materials integrable at the BEOL, such as ITO, researchers have achieved impressive MD levels with minimal IL levels \cite{jafari2023innovative}.\@ Specifically, the MD exhibited dependence on the device length.\@ For instance, a Si-integrable 15 $\mu$m-long modulator achieved a normalized transmission of $-$12 dB at 26 V.\@ Beyond $\lambda$ = 1.55 $\mu$m, the MD decreased due to the permittivity of ITO deviating from the ENZ regime.\@ Despite this, an ER of approximately 1 dB/$\mu$m was maintained within the 1.53 $\mu$m to 1.56 $\mu$m wavelength range \cite{alfaraj2023facile}.\@ The integration of metal-semiconductor Schottky junctions within the CHPW structure has led to PDs with high sensitivity and IQE, reaching levels of $-$55 dBm and 3.1\%, respectively, for a 5 $\mu$m CHPW PD at 10 V \cite{lin2020supermode}.\@ The versatility of the CHPW platform is further exemplified by its ability to be electrically reconfigured to support various passive and active functionalities, including resonators, modulators, and PDs, within the same waveguide stack \cite{lin2020monolithic}.\@ These advancements have the potential to revolutionize the optoelectronics field through ``More than Moore'' disruptive system architectures and integration solutions, where the co-integration of electronics and photonics on a single chip can lead to unprecedented levels of performance and functionality.

Despite these remarkable achievements, significant challenges remain in the field of Si-integrable plasmonics.\@ One of the primary challenges lies in optimizing the trade-off between modal confinement and propagation losses, particularly for LR applications.\@ While CHPWs have demonstrated significant improvements in reducing losses, further research is needed to push the boundaries of propagation distance and minimize signal attenuation.\@ Additionally, integrating complex plasmonic devices with existing CMOS electronics poses fabrication and design challenges.\@ Ensuring seamless compatibility and efficient coupling between plasmonic and electronic components remains an ongoing area of investigation.\@ Addressing these challenges will require innovative materials and device designs and the development of advanced modeling and simulation tools.\@ In the long term, integrating deep neural artificial intelligence (AI) networks for device modeling could prove invaluable in accelerating the design process and optimizing device performance, potentially eliminating some of the time-consuming and computationally intensive EM and device junction simulation steps.

Looking ahead, the future of Si-integrable plasmonics holds immense promise.\@ The continued exploration of novel materials, device architectures, and fabrication techniques is expected to lead to even more compact, efficient, and multifunctional photonic devices.\@ The integration of plasmonic components with emerging technologies like Si photonics and quantum photonics could open up new avenues for applications in areas such as high-speed communication, sensing, and quantum information processing.\@ By addressing the aforementioned challenges and embracing new tools and methodologies, researchers can unlock the full potential of Si-integrable plasmonics and usher in a new era of integrated photonics.

\section*{Acknowledgments}
The authors gratefully acknowledge the Natural Sciences and Engineering Research Council of Canada (NSERC) for their support of relevant work conducted by the Helmy Group at the University of Toronto.\@ We also thank Sherif Nasif and M.\@ Raquib Ehsan for helpful discussions and insights.\@ N.\@ A.\@ acknowledges the support of the Ibn Rushd Postdoctoral Fellowship Program, administered by the King Abdullah University of Science and Technology (KAUST).

\newpage
\bibliography{sample}

\bibliographyfullrefs{sample}


\ifthenelse{\equal{\journalref}{aop}}{%
\section*{Author Biographies}
\begingroup
\setlength\intextsep{0pt}
\begin{minipage}[t][6.3cm][t]{1.0\textwidth} 
  \begin{wrapfigure}{L}{0.25\textwidth}
    \includegraphics[width=0.25\textwidth]{john_smith.eps}
  \end{wrapfigure}
  \noindent
  {\bfseries John Smith} received his BSc (Mathematics) in 2000 from The University of Maryland. His research interests include lasers and optics.
\end{minipage}
\begin{minipage}{1.0\textwidth}
  \begin{wrapfigure}{L}{0.25\textwidth}
    \includegraphics[width=0.25\textwidth]{alice_smith.eps}
  \end{wrapfigure}
  \noindent
  {\bfseries Alice Smith} also received her BSc (Mathematics) in 2000 from The University of Maryland. Her research interests also include lasers and optics.
\end{minipage}
\endgroup
}{}

\end{document}